\documentclass{aa} 
\usepackage{graphicx}
\usepackage[varg]{txfonts}
\usepackage{bbm}
\usepackage{tabu}
\usepackage{tablefootnote}
\begin{document} 

   \title{Magnetic and tidal migration of close-in planets}
   \subtitle{Influence of secular evolution on their population}
   \author{J. Ahuir
          \inst{1}, A. Strugarek \inst{1}, A.-S. Brun \inst{1} \and S. Mathis \inst{1}
          }
   \institute{D\'epartement d’Astrophysique-AIM, CEA/DRF/IRFU, CNRS/INSU, Universit\'e Paris-Saclay, Universit\'e de Paris, F-91191 Gif-sur-Yvette, France\\
             \email{jeremy.ahuir@cea.fr}}

   \date{Received XXX; accepted XXX}

 
  \abstract
   {Over the last two decades, a large population of close-in planets has been detected around a wide variety of host stars. Such exoplanets are likely to undergo planetary migration through magnetic and tidal interactions.}
   {We aim to follow the orbital evolution of a planet along
the structural and rotational evolution of its host star, simultaneously taking into account tidal and magnetic torques, in order to explain some properties of the distribution of observed close-in planets.}
   {We rely on a numerical model of a coplanar circular star--planet system called ESPEM, which takes into account stellar structural changes, wind braking, and star--planet interactions. We browse the parameter space of the star--planet system configurations and assess the relative influence of magnetic and tidal torques on its secular evolution. We then synthesize star--planet populations and compare their distribution in orbital and stellar rotation periods to \textit{Kepler} satellite data.}
   {Magnetic and tidal interactions act together on planetary migration and stellar rotation. Furthermore, both interactions can dominate secular evolution depending on the initial configuration of the system and the evolutionary phase considered. Indeed, tidal effects tend to dominate for high stellar and planetary masses as well as low semi-major axis; they also govern the evolution of planets orbiting fast rotators while slower rotators evolve essentially through magnetic interactions. Moreover, three populations of star--planet systems emerge from the combined action of both kinds of interactions. First, systems undergoing negligible migration define an area of influence of star--planet interactions. For sufficiently large planetary magnetic fields, the magnetic torque determines the extension of this region. Next, planets close to fast rotators migrate efficiently during the pre-main sequence (PMS), which engenders a depleted region at low rotation and orbital periods. Then, the migration of planets close to slower rotators, which happens during the main sequence (MS), may lead to a break in gyrochronology for high stellar and planetary masses. This also creates a region at high rotation periods and low orbital periods not populated by star--planet systems. We also find that star--planet interactions significantly impact the global distribution in orbital periods by depleting more planets for higher planetary masses and planetary magnetic fields. However, the global distribution in stellar rotation periods is marginally affected, as around 0.5 \% of G-type stars and 0.1 \% of K-type stars may spin up because of planetary engulfment. More precisely, star--planet magnetic interactions significantly affect the distribution of super-Earths around stars with a rotation period higher than around 5 days, which improves the agreement between synthetic populations and observations at orbital periods of less than 1 day. Tidal effects for their part shape the distribution of giant planets.}
   {}
   \keywords{planet-star interactions -- stars: evolution -- stars: solar-type -- stars: rotation}
   \maketitle

\section{Introduction}
Since the detection of 51 Pegasi b by \citet{mayor}, more than 4000 exoplanets have been detected. The observed populations show a wide variety of host stars, orbital architectures, planetary sizes, and masses. Moreover, because of the biases of the most successful detection methods, namely, the transit and radial velocity techniques, a majority of the discovered planets orbit close to their host stars, whether they are of mass comparable to that of Jupiter \citep[forming the population of hot Jupiters, e.g.,][]{mayor,henry2000,charbonneau2000} or slightly larger than that of the Earth (the so-called super-Earths, such as 55 Cnc e; see \citealt{dawson}). Close-in planets orbit in a dense and magnetized medium, which leads to the emergence of star--planet interactions that can affect the dynamics and evolution of the orbital systems \citep{cuntz}. In particular, angular momentum exchanges can occur between the planetary orbit and the stellar spin, leading to migration of the planet. Potential signatures of these interactions may have been identified in individual systems (e.g., HD 189733 ; see \citealt{dowling,cauley18}) as well as in the distribution of some planetary populations. More precisely, \citet{MMA13} estimated the rotation period of 737 stars hosting Kepler objects of interest using an auto-correlation method, and  identified a possible dearth of planets with orbital periods shorter than 2-3 days around fast rotators (with a rotation period shorter than 10 days). \citet{teitler} first proposed that such a phenomenon may be attributed to the engulfment of close-in planets by their host stars through tidal interactions. \citet{lanzashkolnik} suggested an alternative scenario based on secular perturbations in multiplanet systems. 
These latter authors showed that remote planets which are excited on a sufficiently eccentric orbit around old stars may be tidally circularized on shorter orbits. Furthermore, \citet{walkowicz} found a concentration of massive planets with an orbital period equal to either  the rotation period of their host star or half that period, which could be the signature of tidal interactions. In view of these different aspects, understanding how compact systems form and evolve is a key astrophysical question to be addressed. 

The role of the protoplanetary disk in shaping the observed structure of planetary systems is strongly emphasized in the literature. Indeed, the disk structure and evolution have a significant influence on the mass and semi-major axis distribution of the young planets \citep[e.g.,][]{mordasini09a,mordasini09b}. In particular, planet migration in the disk through Lindblad resonances is thought to be efficient in shaping planetary systems \citep{baruteau14,bouviercebron,heller}. Moreover, population synthesis models have been developed to better understand the interplay between the properties of the disk and the different processes shaping planetary systems \citep[we refer the reader to][for an extended review]{mordasini18}. The predicted distributions were then compared to \textit{Kepler} observations in order to constrain models of planetary formation and evolution \citep{mulders}. For a large range of multi-planet system properties (e.g., orbital period ratios, mutual inclination, position of the innermost planet), these synthetic populations are in good agreement with \textit{Kepler} global distributions, if multiple interacting seed planet cores per disk are taken into account.\\

 \begin{figure*}[h]
 \centering
 \includegraphics[scale=0.4]{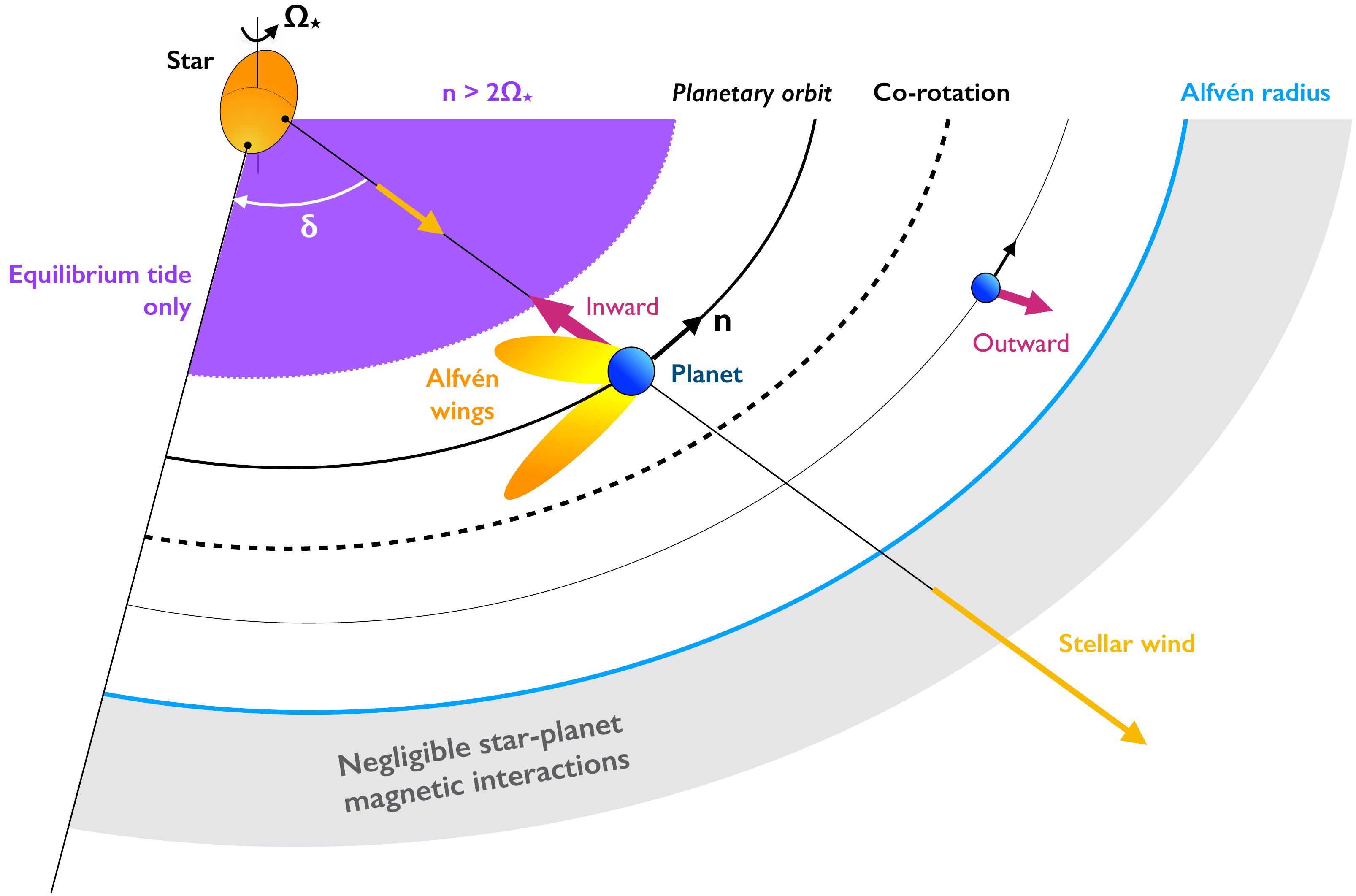}
 \caption{\label{SPI} Sketch of the main features and locations of interest involved in a star--planet system undergoing tidal and magnetic interactions. The planet (in blue) orbits the star (in orange) with an orbital angular velocity \(n\). As a result of the presence of a planet, the star presents a bulge misaligned with the line joining the centers of the two celestial bodies, which induces a lag angle \(\delta\). This angle has been greatly exaggerated here for visualization purposes (it is indeed much smaller than 1 degree in almost all cases). If the planetary mean motion is greater than twice the stellar rotation rate, inertial waves cannot be excited in the stellar convective zone and no dynamical tide is raised within the convective envelope of the star (see the purple area). The relative motion between the planet and the ambient wind (represented with orange arrows) leads to the formation of Alfv\'en wings (orange lobes around the planet) if the planet is below the Alfv\'en radius (in blue). Beyond this distance, no Alfv\'en wings can connect the star and the planet (see the gray area). For both tidal and magnetic interactions, a planet situated below the co-rotation radius (see the black dashed line) undergoes an inward migration, and a planet situated beyond this distance migrates outwards. The relative position of the different orbits of interest may vary depending on the initial configuration of the system considered.}
\end{figure*}
After the dissipation of the disk, dynamical interactions, in particular Kozai oscillations, may occur in multiplanet systems, leading to an intricate evolution of their orbital architecture \citep{laskar12,bolmont15}. However, isolated close-in planets can already suffer efficient migration because of magnetic and tidal interactions with their host star. We consider this simpler case in the present work, where we aim to account for the variety of such star--planet interactions. We therefore consider a simplified system comprising a star and a single planet on a circular orbit perpendicular to the stellar rotation axis. One of the main physical processes acting in such a configuration are tidal interactions. These result from the gravitational response of a star to the presence of a planet and play a key role in the evolution of the orbital configuration of the system. Two components arise from the stellar response: the hydrostatic nonwavelike equilibrium tide, dissipated by turbulent friction \citep{zahn66,remus,ogilvie13}, and the dynamical tide, which consists in the excitation of waves inside the star by the tidal potential as well as their dissipation. A dynamical tide can exist in both radiative zones (through internal gravity waves, \citealt{zahn75,goldreich89, goodman, terquem}) and convective zones \citep{ogilvie2007,ogilvie13,mathis15}. We focus on the latter in this study and leave a detailed investigation of internal gravity waves to future work \citep{barker20, ahuir21}. In the stellar convective zone, if the orbital period of the planet is longer than half of the stellar rotation period, inertial waves restored by the Coriolis force are excited and dissipated in the envelope of solar-type stars. Otherwise, inertial waves cannot be excited and no dynamical tide is raised in the star (see the violet area in Fig. \ref{SPI}). The associated dissipation, which depends on stellar internal structure as it arises from the reflection of the waves on the radiative core \citep{ogilvie13,goodman09,mathis15}, can be several orders of magnitude higher than the dissipation of the equilibrium tide \citep{ogilvie2007,bolmont16}. When tidal dissipation is taken into account, the stellar response presents a delay and the associated bulge is misaligned with the line joining the centers of the two celestial bodies. This misalignment then induces a lag angle \(\delta\), which increases with dissipation magnitude, and is at the origin of an exchange of angular momentum between the star and the planet. Indeed, a net tidal torque is applied both on the planetary orbit and on stellar rotation to reduce the angle \(\delta\). The position of the planet with respect to the co-rotation radius (for which \(n = \Omega_\star\), see the black dashed line in Fig. \ref{SPI}) then determines the evolution of the system. If the planet is situated beyond this characteristic distance, an outward migration makes it move away from the host star, which then spins down. Otherwise the planet migrates inward, moving closer to the spinning-up star. In the latter case, the fate of the system is determined by the orbital angular momentum \(L_\text{orb}\) of the planet and the stellar angular momentum \(L_\star\): if \(L_\text{orb} \geq 3L_\star\), an equilibrium state is reached where the angular velocities \(\Omega_\star\) and \(n\) are synchronized. Otherwise, the planet migrates too efficiently and it spirals towards its host star until its disruption at the Roche limit \citep{hut80,hut81}. It is important to note that such a condition is based on the conservation of the total angular momentum of the star--planet system. \citet{damiani15} derived a similar criterion by taking into account magnetic braking. More generally, to provide a realistic and complete equilibrium criterion, it is necessary to take into account magnetic braking, angular momentum redistribution within the star, and the various star--planet interactions  simultaneously.

Tidal interactions are modulated by the structural and rotational evolution of the star. The stellar rotation rate generally exhibits a complex evolution due to the initial disk--star interaction, the internal redistribution of angular momentum within the star during contraction phases, and the angular momentum extraction by the stellar wind  \citep[e.g.][]{weberdavis,skumanich,kawaler,matt12,reville15a}. As this intricacy may significantly affect the evolution of a given star--planet system, it is necessary to take such processes into account as much as possible. This requires models calibrated on gyrochronology and rotational distributions in open clusters \citep{gallet15,matt15}, leading up to general frameworks combining stellar rotation, wind, and magnetism \citep{johnstone15a,ahuir20}. Past studies have taken these constraints into account to some extent. For instance, \citet{zhang} relied on the two-layer rotational model of \citet{macgregorbrenner} and a constant tidal dissipation to deal with the secular evolution of star--planet systems, and subsequently adopted a statistical approach to their numerical simulations in order to apply constrains to tidal theory. \citet{bolmont16} then first incorporated the effects of dynamical tide in stellar convection zones and studied their impact on the secular evolution of star--planet systems by considering a one-layer rotational model for the central star. More recently, \citet{benbakoura} performed a study based on the amalgamation of the two previous studies, taking a bi-layer structure for the star and both the equilibrium and dynamical tides into account. This allowed them to provide a criterion for planetary engulfment due to tidal effects, taking into account stellar evolution. They were also able to characterize the influence of such a phenomena on the rotation of the host star. In parallel, \citet{gallet18} and \citet{gallet19} relied on a similar model to investigate the rotational evolution of planet-hosting stars in open clusters (in particular in the Pleiades)  in more detail, which allowed them to assess some limits of gyrochronology \citep{barnes}. Finally, the evolution of star--planets system under tidal interactions during the red giant phase has also been extensively investigated \citep{privitera16a,privitera16b,meynet17,rao18}.

However, in past studies, tidal effects and magnetism have not been taken into account together systematically (apart from wind braking). \citet{bouviercebron} first explored the possibility that tidal and magnetic interactions may compete with accretion and contraction in the case of a close-in planet embedded in a disk. Furthermore, after the dissipation of the latter, star--planet magnetic interactions may occur because of the relative motion between the planet and the ambient wind at the planetary orbit (represented with orange arrows in Fig. \ref{SPI}). If the planet is below the Alfv\'en radius (at which the wind velocity is equal to the local Alfven speed; see the blue line in Fig. \ref{SPI}), the magnetic torque applied to the planet can lead to efficient transport of angular momentum between the planet and the star through the so-called Alfv\'en wings \citep[][see the orange lobes around the planet in Fig. \ref{SPI}]{neubauer}. Beyond the Alfv\'en radius, the wind becomes superalfv\'enic. In this case, Alfv\'en wings may still exist but do not connect back the star (see the gray region in Fig. \ref{SPI}). In this case, the planet may transfer energy and angular momentum to the ambient wind instead. In the context of close-in planets, we only consider here the subalfvenic scenario. Several regimes then appear depending on the star--planet configuration \citep{strugarek17hex}. If Alfv\'en waves have enough time to go back and forth between the star and the planet before the magnetic field lines slip through the planet, the interaction acts as a unipolar generator, leading to the so-called unipolar interaction \citep{laine,lainelin}. In the opposite case, magnetic interactions between the planet and the star still occur and the interaction becomes dipolar \citep{saur,strugarek15,strugarek16}. As the relative motion between the planet and the ambient wind is at the origin of the subsequent magnetic torques, these are then expected to act in the same way as tidal effects in most cases. The \textit{co-rotation} radius then plays a determining role in planetary migration in both cases. Many other notable effects, such as anomalous emissions or planet inflation, may result from star--planet magnetic interactions \citep[we refer the reader to][for a recent review]{lanza18}. \citet{strugarek17} performed a first study on planetary migration taking into account tidal and magnetic torques simultaneously. In particular, they computed the migration timescale of the planet for both contributions, finding that both effects could play a key role depending on the characteristics of the star--planet system considered. Thus, following the orbital evolution of a planet along the rotational and structural evolution of the host star by taking into account the coupled effects of tidal and magnetic torques is essential to better understanding the evolution of star--planet systems. Furthermore, synthesizing planetary populations by taking into account the whole variety of star--planet interactions to explain the observed distributions of exoplanets still has to be performed.

Following \citet{strugarek17} and \citet{benbakoura}, the main goal of this work is to assess the relative contribution of both tidal and magnetic interactions on the secular evolution of star--planet systems, and to investigate the role of the associated torques in shaping the distributions of planetary populations. In section 2, we present the hypotheses of our study, detail the interactions involved in our modeled star--planet systems, and describe the modeling approach used in this work. In section 3, we investigate the influence of the main characteristics of a star--planet system (e.g., stellar mass, stellar magnetism, semi-major axis, planetary type, etc.) on its secular evolution by assessing the relative contribution of magnetic and tidal torques. 
All these parameters are then taken into account simultaneously in section 4 in order to classify planetary populations emerging from the action of star--planet interactions and to highlight regions of interest resulting from their evolution.
Populations of star--planet systems are then synthesized in section 5 and are confronted with a statistical distribution obtained from \textit{Kepler} data. All of those results are then summarized,  discussed, and put into perspective in section 6.

\section{Star--planet interaction model}
\subsection{ESPEM: an overview}

\begin{figure}[h]
 \centering
\includegraphics[scale=0.4]{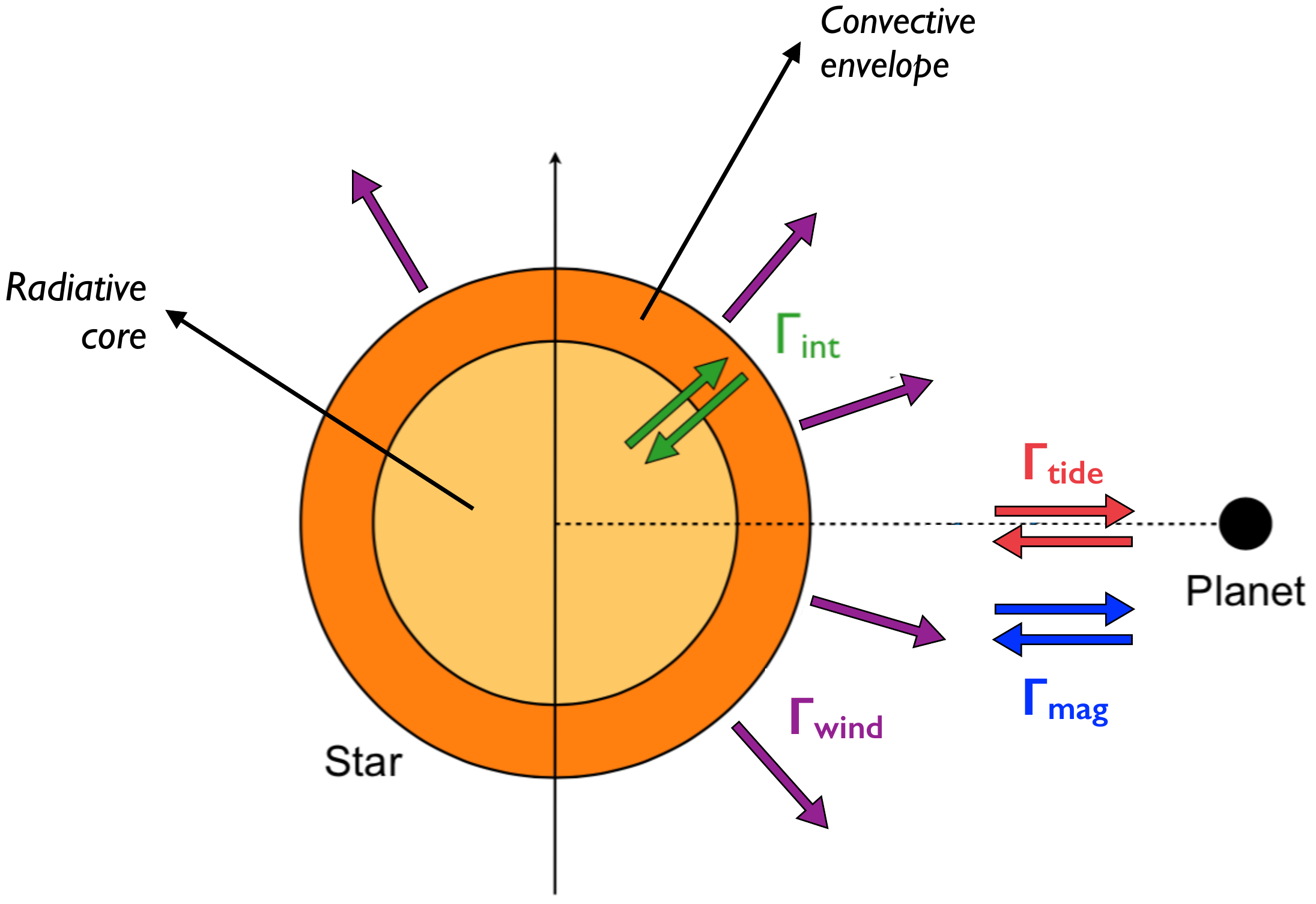}
 \caption{\label{ESPEM} Schematic view of the system and its interactions  \citep[adapted from][]{benbakoura}. The radiative core (in yellow) and the convective envelope (in orange) exchange angular momentum (green arrows). Stellar wind carries away angular momentum from the envelope and spins the star down (purple arrows). Stellar rotation and planetary orbit are coupled through tidal (red arrows) and magnetic effects (blue arrows).}
\end{figure}

ESPEM \citep[French acronym for Evolution of Planetary Systems and Magnetism ; see][]{benbakoura} is a numerical code computing the secular evolution of a star--planet system by following the semi-major axis of the planetary orbit as well as the stellar rotation rate. Furthermore, we assume here a coplanar and circular orbit, and a synchronized planetary rotation, as the reservoir of angular momentum of the planet is less important than the one in its orbit \citep{guillot96}. In this model, we consider a two-layer solar-type star composed of a radiative core and a convective envelope. Tidal dissipation is only considered in the stellar envelope in this work. The core is interacting with the envelope through internal coupling, and the latter exchanges angular momentum with the orbit through tidal and magnetic interactions (SPMI). Moreover, the whole system loses angular momentum through magnetic braking by the stellar wind. Hence, the angular momentum  of the planetary orbit, \(L_\text{orb}\), the stellar convective zone, \(L_c\), and radiative zone, \(L_r\), are evolved by the following system of equations:

\begin{equation}
\frac{dL_{\rm orb}}{dt} = -\Gamma_{\rm tide}-\Gamma_{\rm mag}
\end{equation}
\begin{equation}
\frac{dL_c}{dt} = \Gamma_{\rm int}+\Gamma_{\rm tide}-\Gamma_{\rm wind}+\Gamma_{\rm mag}
\end{equation}
\begin{equation}
\frac{dL_r}{dt} = -\Gamma_{\rm int},
\end{equation}
where \(\Gamma_\text{int}\) is the internal torque, coupling the core and the envelope of the star, \(\Gamma_\text{wind}\) is the wind-braking torque, and \(\Gamma_\text{tide}\) and  \(\Gamma_\text{mag}\) are the tidal and MHD torques between the star and the planet, respectively. A schematic global view of the system studied by ESPEM is provided in Figure \ref{ESPEM}.

\subsection{Stellar structure and evolution}

Stellar structure and evolution are taken into account in ESPEM during the pre-main sequence (PMS) and the main sequence (MS). The internal structure of the star, especially the radii, the masses, and the moments of inertia of radiative and convective zones are provided at each ESPEM time-step through grids precomputed with the stellar evolution model STAREVOL \citep{siess,palacios,amard16,amard19}.

The coupling between the radiative zone and the convective zone following the model proposed in \citet{macgregorbrenner} is taken into account as an exchange of angular momentum allowing the synchronization of their spins on a characteristic timescale \(\tau_{c-e}\), which is determined by internal transport processes in the radiative core \citep{brun06,mathis13, aerts} as well as the effective coupling between the radiative and the convective zones \citep{brun11,strugarek11}. The amount of angular momentum to be transferred between the two layers of the star to equilibrate their angular velocities can be expressed as

\begin{equation}
\Delta L = \frac{I_rI_c}{I_r+I_c}\left(\Omega_r-\Omega_c\right),
\end{equation}
where \(I_r, \Omega_r\) are the moments of inertia and the rotation rate of the core, and \(I_c,\ \Omega_c\) are the same quantities assessed in the envelope.

Moreover, the expansion of a radiative core during the PMS involves a rapid conversion of convective state to radiative state \citep{emeriau}. During this transition phase, a significant mass transfer occurs, which is accompanied by a transport of angular momentum. This way, the coupling between the core and the envelope can be modeled as a torque with two components applied to the convective zone:

\begin{equation}
\Gamma_{\rm int}=\frac{\Delta L}{\tau_{c-e}}-\frac{2}{3}{R_r}^2\Omega_c\frac{dM_r}{dt},
\end{equation}
where \(M_r\) and \(R_r\) are the mass and radius of the radiative core. The coupling timescale \(\tau_{c-e}\) is a free parameter of our model and has been calibrated with the \citet{gallet15} study as follows:
\begin{equation}
\tau_\text{c--e}\  [\text{Myr}] = 3.05\left(\frac{M_\star}{M_\odot}\right)^{-5.02}\left(\frac{P_\text{rot,c}}{P_{\text{rot,}\odot}}\right)^{0.67},
\end{equation}
where \(M_\star\) is the stellar mass and \(P_\text{rot,c}\) is the rotation period of the stellar convective zone.
Star–disk interaction is taken into account in a simplified way at the beginning of the PMS by assuming a constant surface rotation rate during the disk’s lifetime, which is also fixed by the \citet{gallet15} study as
\begin{equation}
\tau_\text{disk}\  [\text{Myr}] = 13.41\left(\frac{P_\text{rot,c}}{P_{\text{rot,}\odot}}\right)^{-0.56}.
\end{equation}
During this phase, the semi-major axis of the planet is assumed to be constant and the rotation of the radiative zone is only constrained through internal coupling. We focus here on the evolution after the disk dissipation. A more precise treatment of the early phase could be added in future works \citep{bouviercebron,galletzanni19}.

\subsection{Wind braking torque}
The wind braking torque is given in our model by \citet{matt15} :
\begin{equation}\label{eqn:matt15}
\Gamma_\text{wind} = \Gamma_\odot\left(\frac{R_\star}{R_\odot}\right)^{3.1}\left(\frac{M_\star}{M_\odot}\right)^{0.5} \left(\frac{Ro}{Ro_\odot}\right)^{-2}\left(\frac{\Omega_c}{\Omega_\odot}\right), \ Ro>Ro_\text{sat},
\end{equation}
\begin{equation}
\Gamma_\text{wind} = \Gamma_\odot\left(\frac{R_\star}{R_\odot}\right)^{3.1}\left(\frac{M_\star}{M_\odot}\right)^{0.5} \left(\frac{Ro_\text{sat}}{Ro_\odot}\right)^{-2}\left(\frac{\Omega_c}{\Omega_\odot}\right),\ Ro \leq Ro_\text{sat}
,\end{equation}
with \(\Gamma_\odot = 6.3\times10^{30}\ \text{erg}\). Such a prescription allows us to account for the mass and age dependencies of the distribution of stellar rotation periods in open clusters \citep{rodriguez, agueros} and in the sample of stars observed by the \textit{Kepler} satellite \citep{MMA14}. We use the stellar Rossby number for simplicity purposes, expressed as
\begin{equation}\label{eqn:def_rossby}
Ro = \frac{P_\text{rot,c}}{\tau_c},
\end{equation}
where \(P_\text{rot,c}\) is the rotation period of the stellar convective zone \citep[see][for a discussion on the various definitions found in the literature]{landin,brun17}. The convective turnover time \(\tau_{c}\) can be assessed with the \citet{ardestani} prescription: 
\begin{equation}\label{eqn:ardestani}
\begin{split}
\log_{10} \tau_c\ [\text{s}] =\  &8.79-2|\log_{10}(m_\text{CZ})|^{0.349}-0.0194\log_{10}^2(m_\text{CZ})\\
& - 1.62\ \text{min}\left[\log_{10}(m_\text{CZ})+8.55,0\right],
\end{split}
\end{equation}
where \(m_\text{CZ}\) is the mass of the convective envelope normalized to the stellar mass. The formulation obtained by these latter authors has the advantage of being valid during the pre-main sequence and the main sequence for metallicities ranging from [Fe/H] = -0.5 to 0.5 and was obtained with the CESAM stellar evolution code \citep{morel}. Their prescription leads to a solar value of \(Ro_\odot = 1.113\) and a saturation value of \(Ro_\text{sat} = 0.09\).\\

\begin{figure}[!h]
   \begin{center}
      \includegraphics[scale=0.47]{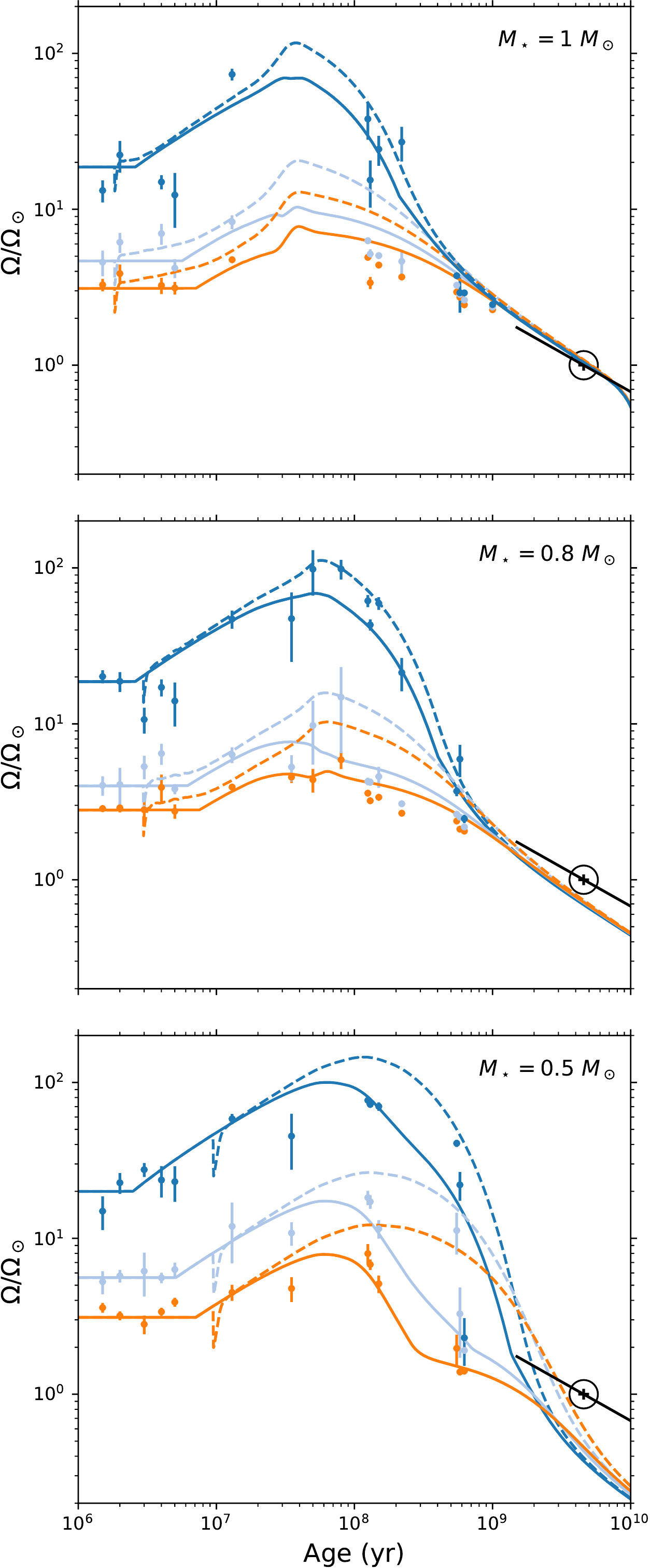}
   \end{center}
\caption{\label{GB15}
Secular evolution of the rotation rate of the convective envelope (solid lines) and of the radiative core (dashed lines) for stars with a mass \(M_\star =  \{1,\ 0.8,\ 0.5\}\ M_\odot\) (from top to bottom). Slow (orange), median (light blue), and fast (dark blue) rotators are considered. The  solar rotation rate at solar age is represented by a black circle in each panel. The dots with error bars correspond to the 25th, 50th, and 90th percentiles of rotational distributions of observed stellar clusters published by \citet{gallet15}. In black: Skumanich law normalized to the Sun.}
\end{figure}

Figure \ref{GB15} shows the typical rotational evolution obtained with our model for isolated stars. We show our results for three stellar masses, and for three initial rotation periods (1.4, 5, and 8 days) covering fast, median, and slow rotators from \citet{gallet15}. The initial rotation spread is reduced over the MS as all the stars converge towards a sequence where their rotation rate is fully determined by their age and mass \citep{barnes}. As seen in the top panel, solar-mass stars spin down to reach the solar rate at the solar age whereas less-massive stars reach a lower rotation at the same age. A steeper evolution of the stellar rotation rate compared to the Skumanich law occurs for each stellar mass because of the core-envelope coupling, in accordance with the \citet{gallet15} results, as the angular momentum stored in the radiative zone is redistributed on secular timescales.

\subsection{Planetary properties}
We consider in our model a planet of mass \(M_p\), ranging between 0.5 and 1589\ \(M_\oplus\) (corresponding to 5\ \(M_\text{Jup}\)), and a radius \(R_p\). It is assumed to be punctual and its rotation is synchronized with its orbit. Furthermore, we adopt the probabilistic mass–radius relations proposed by \citet{chen} based on a sample of well-constrained planets:
\begin{equation}\label{eqn:RMrel}
R_p \propto
\begin{cases}
M_p^{0.28},\ &M_p < 2.0\ M_\oplus\ (6.29\times10^{-3}\ M_\text{Jup})\\
M_p^{0.59},\ &2.0\ M_\oplus \leq M_p < 0.4\ M_\text{Jup}\\
M_p^{-0.04},\ &M_p \geq 0.4\ M_\text{Jup}.
\end{cases}
\end{equation}
Recent studies show that the distribution of planetary radii in the Kepler sample presents a gap between 1.5 $R_\oplus$ and 2 $R_\oplus$ \citep{fulton}. Such a bimodality in the distribution may be due to photoevaporation which may drive atmospheric mass loss on close-in planets. If so, the gap would originate from a discrepancy between planets with H/He envelopes of small mass and bare rocky cores. In this first statistical study, as such a feature does not appear in the star--planet sample we have considered (we refer the reader to the bottom panel of Fig. \ref{MMA1314}), the incorporation of this radius valley in our model is left for future work.

The equatorial field at the planetary surface \(B_p\), which is of prime importance in star--planet interactions, is by default assumed to be constant. Nevertheless, this is varied in \S 3.3.

\subsection{Tidal effects}
Tidal interactions lead to an angular momentum exchange between the star and the planet, which can be translated into a tidal torque applied to the stellar envelope \citep{murray,benbakoura}:
\begin{equation}\label{eqn:tidal_torque}
\Gamma_{\rm tide}=-{\rm sign}(\omega_{\rm tide})\frac {9}{4Q'}\frac{GM_p^2}{a^6}R_\star^5,
\end{equation}
where \(M_p\) is the planetary mass, \(R_\star\) the stellar radius, \(G\) the gravitational constant, \(a\) the semi-major axis, \(\omega_\text{tide} = 2(\Omega_c - n)\), which is the tidal frequency in the case of a planet with a circular orbit and a synchronized rotation, and \(n\) is the mean motion of the planetary orbit. The equivalent quality factor \(Q'\) takes into account the nature and the efficiency of tidal dissipation as a function of the internal structure and rotation of the star. Here, it is used to describe the so-called equilibrium \citep{zahn66, remus} and dynamical tides \citep{ogilvie13, mathis15}. We now summarize their treatment \citep[we refer the reader to][for a more detailed description]{benbakoura}.

The equilibrium tide is taken into account in ESPEM by following the \citet{hansen} prescription, relying on a constant value for the dimensionless dissipation factor \(\bar\sigma_\star\) along the evolution of the system. This quantity is calibrated using observations and leads to the quality factor \citep{bolmont16}
\begin{equation}
Q'_\text{eq} = \frac{3}{2}\frac{1}{\sigma_0\bar\sigma_\star}\frac{G}{R_\star^5|\omega_\text{tide}|},
\end{equation}
where \(\sigma_0 = \sqrt{G/(M_\odot R_\odot^7)}\). In this work, the value of \(\bar\sigma_\star\), which decreases with stellar mass, is taken from Fig. 3 of \citet{hansen}. Such a formulation provides the same order of magnitude as prescriptions derived from physical models \citep[for two different approaches, see][]{strugarek17,barker20}.

The dynamical tide, and more precisely the dissipation of tidal inertial waves (governed by the Coriolis acceleration) within the convective zone, is based on the prescription of \citet{ogilvie13} and \citet{mathis15}, who introduced a frequency-averaged effective constant tidal quality factor. Indeed, when computing the frequency dependence of the tidal torque due to tidal inertial waves in stellar convection zones \citep[see e.g.,][]{ogilvie2007}, its frequency dependence is highly resonant and erratic. This complex behavior relies on the physics of the friction applied by the turbulent convection on tidal inertial waves \citep[e.g.,][]{ogilvie04,auclair15}; works are ongoing to improve its complex modeling \citep{duguid20}. Therefore, as explained in Section 4.2 of \citet{benbakoura}, a consistent treatment of the frequency dependence of the torque induced by tidal inertial waves would require a coupling of ESPEM with 2D hydrodynamical computation of tidal inertial modes at each time-step, which is beyond the scope of this work and would also not allow us to explore the broad parameter space describing the diversity of star--planet systems.

To get an order of magnitude of this dissipation for a given stellar mass, age, and rotation, we perform the frequency average of the dissipation as proposed and described in \citet{ogilvie13}, \citet{mathis15}, and \citet{barker20} which provides us with good trends when compared with observational constraints. In the case of a two-layer star, the formulation  of the frequency-averaged tidal dissipation $\overline{Q'_\text{dyn}}$ \footnote{We refer the reader to Eq. (1) from \citet{mathis15} as well as the appendix B from \citet{ogilvie13} for an explicit definition of such an average.} provided by these latter authors gives
\begin{equation}\label{eqn:Qdyn}
\begin{split}
\frac{3}{2\overline{Q'_\text{dyn}}}&= \frac{100\pi}{63}\left(\frac{\Omega_c}{\Omega_\text{crit}}\right)^2\left(\frac{\alpha^5}{1-\alpha^5}\right)(1-\gamma)^2(1-\alpha)^4\\
&\times\frac{\left(1+2\alpha+3\alpha^2+\frac{3}{2}\alpha^3\right)^2\left[1+\left(\frac{1-\gamma}{\gamma}\right)\alpha^3\right]}{\left[1+\frac{3}{2}\gamma+\frac{5}{2\gamma}\!\left(1+\frac{1}{2}\gamma-\frac{3}{2}\gamma^2\right)\alpha^3-\frac{9}{4}(1-\gamma)\alpha^5\right]^2}, 
\end{split}
\end{equation}
where \(\displaystyle\alpha = \frac{R_r}{R_\star}\), \(\displaystyle\beta = \frac{M_r}{M_\star}\), and \(\displaystyle\gamma = \frac{\alpha^3(1-\beta)}{\beta(1-\alpha^3)}\). The latter quantity corresponds to the ratio of the density of the envelope to that of the core. \(\Omega_\text{crit}=\sqrt{GM_\star/R_\star^3}\) is the critical angular velocity of the star. Such a contribution will affect the secular evolution of the system if inertial waves are likely to be excited by the tidal potential, i.e., \(P_\text{orb} > \frac{1}{2} P_\text{rot}\) \citep{bolmont16}. The total quality factor \(Q'\), accounting for the sum of the tidal dissipation, is then given by:
\begin{equation}
\frac{1}{Q'} = \frac{1}{Q'_\text{eq}}+\frac{1}{\overline{Q'_\text{dyn}}}.
\end{equation}

\subsection{Magnetic star--planet interactions}
\label{sec:defSPMI}

When a planet orbits in a magnetized medium, MHD disturbances propagate away from the planet vicinity while transporting energy and angular momentum, forming the so-called Alfv\'en wings \citep{neubauer, saur}. Two Alfv\'en wings are always produced. Depending on the magnetic topology and the alfvenic Mach number of the interaction, one, both, or neither of the two may reach back to the star \citep[for more details, see][]{strugarek15}. If Alfv\'en waves do not have enough time to travel back and forth  between the star and the planet before the magnetic field lines slip through the planet, the magnetic interaction is dubbed dipolar \citep{saur,strugarek15,strugarek16}. Otherwise the interaction becomes unipolar \citep{laine,lainelin}.\\

The existence of Alfv\'en wings results in a magnetic torque applied to the planet. Assuming the planet possesses a magnetosphere, it can be written as a drag torque \citep{strugarek16}:

\begin{equation}
\Gamma_{\rm mag}=-{\rm sign}(\omega_{\rm mag})c_d\left(A_0M_a^\beta\Lambda_P^\alpha\ \pi R_p^2\right)p_{\rm tot}\ a,
\end{equation}
where \(c_d \approx {M_a}/\sqrt{1+M_a^2}\) is a drag coefficient representing the efficiency of the magnetic reconnection between the wind and the planetary magnetic fields, \(M_a\) is the alfvenic Mach number in the frame rotating with the planet, and \(\Lambda_P=(B_p^2/2\mu_0)/p_\text{tot}\) is the ratio between the planetary magnetic pressure and the total pressure of the ambient wind at the planetary orbit. $\omega_\text{mag}$ corresponds to the difference between the rotation rate of the ambient wind and the planetary mean motion. As the wind is in a first approximation co-rotating with the star below the Alfv\'en radius, one can assume that $\omega_\text{mag}$ and $\omega_\text{tide}$ have the same sign in the vast majority of cases. In the case of close-in planets, the total wind pressure can be approximated by the magnetic pressure of the wind \citep{reville15b}. The quantities \(A_0,\alpha,\beta\) are determined from a set of 3D MHD simulations in \citet{strugarek16}. Only the case of a planetary dipole aligned with the stellar magnetic field at the planetary orbit is considered from now on in our model. Such a configuration, which maximizes the torque, gives \(A_0 = 10.8,\ \alpha = 0.28,\ \beta = -0.56\). This way, the considered torque provides an upper bound of the influence of magnetic effects on planetary migration in the dipolar regime.

The dipolar interaction regime considered here is likely to be realized in most compact star--planet systems. Generally, for planets sustaining a magnetosphere against the ambient pressure, alfvenic perturbations do not have enough time to travel back and forth between the star and the planet before the magnetic field line slips around the planet, unless the planetary magnetosphere of the planet is sufficiently large \citep[comparable to the size of the Sun ; see][for an extensive review of star--planet magnetic interactions]{strugarek17hex}. In the case of a weakly magnetized planet, the time-dependent component of the stellar magnetic field is either dissipated in the planetary interior or screened by the magnetic field induced by large surface currents, depending on the planetary resistivity. Therefore, we only consider the steady component of the stellar magnetic field. In this configuration, if the planetary diffusivity is sufficiently high, the time-independent component of the stellar magnetic field is efficiently dissipated inside the planet, creating a true magnetic cavity in the planetary interior. The dipolar regime is found to generally hold in this case. Otherwise, if magnetic diffusivity is sufficiently low, the magnetic field lines are frozen in the planet interior and dragged with the orbital motion of the planet. In that case, propagating Alfv\'en waves can generally reach back to the planet, and the interaction becomes unipolar. Such a configuration has been extensively treated by \citet{laine}, \citet{lainelin} and is found to lead to far stronger magnetic torques than for the dipolar regime \citep[typically 4 or 5 orders of magnitude ; see][]{strugarek17}. More complex situations can occur depending on the conductive properties of the planet material and its degree of ionization. As the transition between the unipolar and the dipolar regimes is still poorly understood, we focus in this work on the dipolar regime, which is more likely to occur in exosystems, and leave the study of the unipolar regime to a future study.

In this context, if the planet is not able to sustain a magnetosphere, we consider in this work that it screens the surrounding wind magnetic field, which leads to a dipolar star--planet interaction. The effective area of the planetary obstacle then corresponds to the geometrical cross-section of the planet. Henceforth, we rely on the following prescription for the dipolar torque:\\
\begin{equation}
\Gamma_{\rm mag}=\begin{cases}
\displaystyle\frac{-\text{sign}(\omega_\text{mag})M_a}{\sqrt{1+M_a^2}}\left(10.8M_a^{-0.56}\Lambda_P^{0.28}\right)\pi R_p^2p_{\rm tot}a,\ \Lambda_P>1.\\
\displaystyle\frac{-\text{sign}(\omega_\text{mag})M_a}{\sqrt{1+M_a^2}}\pi R_p^2p_{\rm tot}a,\ \text{otherwise}.
\end{cases}
\end{equation}
Star--planet magnetic interactions occur because of the relative motion between the planet and the ambient wind at the planetary orbit. We must therefore estimate the radial profiles of the main characteristics of the wind, such as its velocity or its density. To this end, we incorporate a 1D isothermal magnetized wind model in ESPEM \citep[see][for an extensive description of the model]{lamers,preusse,johnstone17}. Such a modeling requires knowledge of the temperature \(T_c\) and density \(n_c\) at the base of the wind. For consistency with the observational constraints on stellar rotation, wind, and magnetism, we rely on the \citet{ahuir20} prescriptions for those quantities. More precisely, as the stellar magnetic field measured from Zeeman broadening and Zeeman-Doppler imaging \citep[see][]{montesinos,vidotto14,see17}
have only exhibited linear or super-linear dependencies between the large-scale magnetic field and the Rossby number, we consider for the sake of simplicity 
the following scaling law to assess the magnetic field at the stellar surface \(B_\star\) \citep{ahuir20}:
\begin{equation}\label{eqn:Bstar}
B_\star\ [\text{G}] = 2.0\left(\frac{Ro}{Ro_\odot}\right)^{-1}\left(\frac{M_\star}{M_\odot}\right)^{-1.76},\   Ro > Ro_\text{sat}.
\end{equation}
This leads to the following expressions for the coronal properties \citep{ahuir20}:
\begin{equation}\label{eqn:Tc}
T_c\ [\text{MK}] =1.5\left(\frac{Ro}{Ro_\odot}\right)^{-0.11}\left(\frac{M_\star}{M_\odot}\right)^{0.12},\  Ro > Ro_\text{sat},
\end{equation}
\begin{equation}\label{eqn:nc}
n_c\ [\text{cm}^{-3}] =  7.25\times 10^7\left(\frac{Ro}{Ro_\odot}\right)^{-1.07}\left(\frac{M_\star}{M_\odot}\right)^{1.97},\  Ro > Ro_\text{sat}.
\end{equation}
Stellar magnetic field as well as wind temperature and density are assumed to be independent of the Rossby number in the rotation-saturated regime (\(Ro \leq Ro_\text{sat}\)). 

For the sake of simplicity as well as to provide an upper bound on magnetic effects in the dipolar regime, we assume that the planet is located in an open field region. The stellar magnetic field is then assumed to be radial. For more complex topologies, the magnetic field decays faster with distance to the star, which reduces the efficiency of the star--planet magnetic interactions  accordingly.

\section{Tidal and magnetic interactions: an evolutive approach of star--planet systems}
\subsection{Outline of star--planet secular evolution}
\subsubsection{Planet migration: reference case}

We now aim to investigate the influence of the main properties of a star--planet system on its secular evolution by assessing the relative contribution of tidal and magnetic torques. Such an approach allows us to study star--planet magnetic interactions from a dynamical and evolutive point of view and to compare the associated results to the \citet{strugarek17} study, which relied on the instantaneous migration timescale of the planet. For the sake of simplicity, we use a reference case to investigate the influence of each parameter of our model on the secular evolution of star--planet systems. We consider a young star–planet system formed by a fast-rotating K star orbited by a strongly magnetized hot Neptune whose main features are presented in Table \ref{tab:refcase}. 

\begin{table}[!h]
\centering 
      \caption{\label{tab:refcase} Star--planet parameters of the reference case.}
      \begin{tabu}{cc}
            \hline
            \noalign{\smallskip}
            Star & Planet \\
            \noalign{\smallskip}
            \hline
            \noalign{\smallskip}
            $M_\star = 0.8\ M_\odot$ & $M_p = 0.1\ M_\text{Jup}$\\
            $P_\text{rot,ini} = 1.4\text{ d}$ & $a_\text{ini} = 0.035 \text{ AU}$\\
            &$B_p = 10\text{ G}$\\
            \noalign{\smallskip}
            \hline
         \end{tabu}
   \end{table}

\subsubsection{Secular evolution of a reference star--planet system and influence of initial semi-major axis}

\begin{figure}[!h]
   \begin{center}
      \includegraphics[scale=0.4]{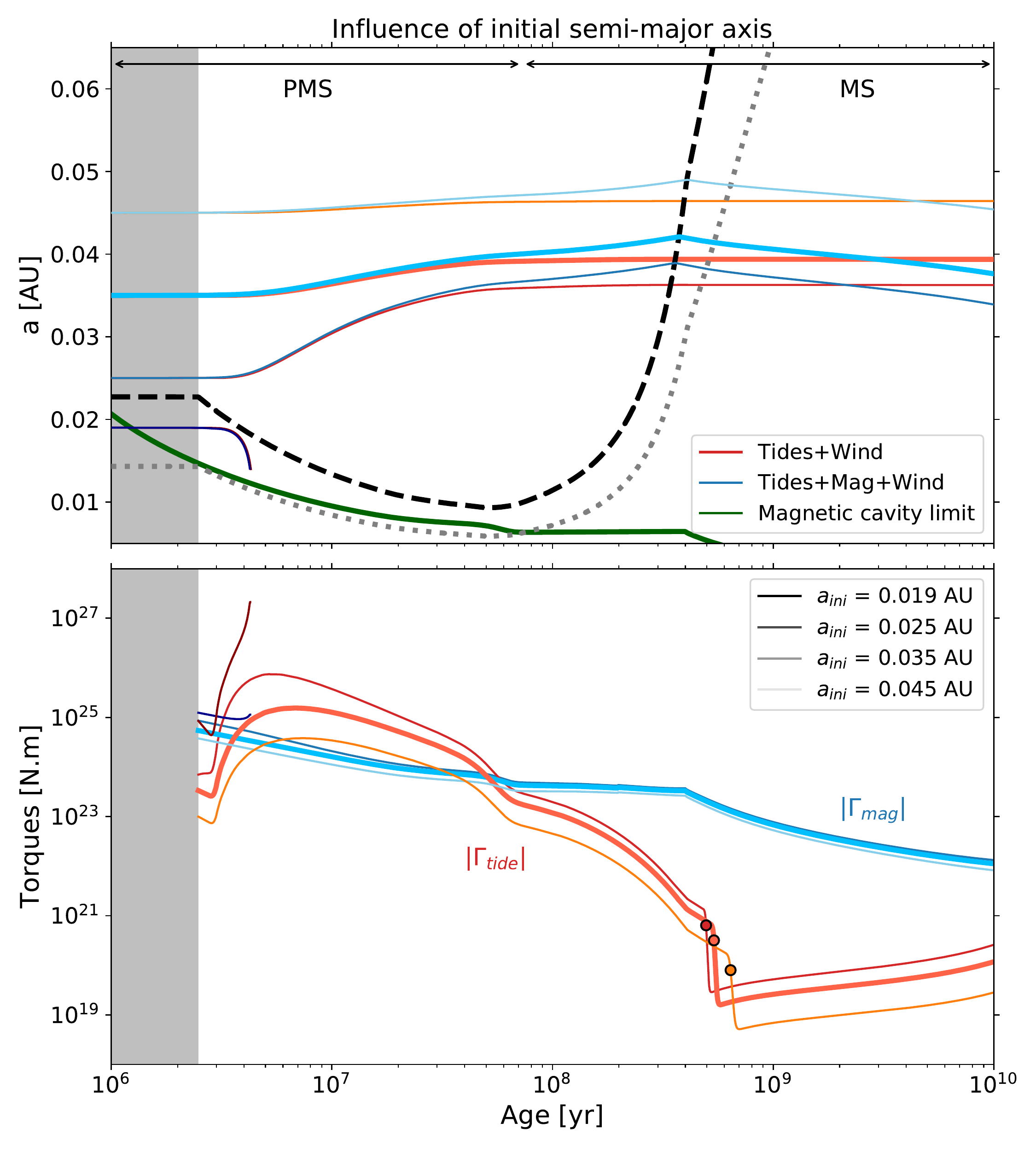}
   \end{center}
   \caption{\label{aini} Secular evolution of a star–planet system  formed by a fast-rotating K star (\(M_\star = 0.8\ M_\odot,\ P_\text{rot,ini} = 1.4\text{ d}\)) orbited by a strongly magnetized hot Neptune (\(M_p = 0.1\ M_\text{Jup},\ B_p = 10\text{ G}\)) for four different initial semi-major axes: \(a_\text{ini} = 0.019\text{ AU},\ 0.025\text{ AU},\ 0.035\text{ AU},\ 0.045\text{ AU}\) (dark to light colors). The thick curves correspond to our reference case discussed in \S 3.1.1 with \(a_\text{ini}=0.035\) AU. \textit{Top panel}: Semi-major axis (solid lines), co-rotation radius of the star (black dashed lines), and limit of excitation of the dynamical tide (black dotted lines). Wind braking + tides are shown in shades of red; wind braking +  tides + magnetic effects are shown in shades of blue. The gray bands on the left correspond to the disk-locking phase. The cavity formation limit, corresponding to \(\Lambda_p = 1\), is shown in green.
   \textit{Bottom panel}: Tidal (shades of red) and magnetic (shades of blue) torques in the case of an evolution with all the combined interactions. The red circles correspond to the crossing of the dynamical tide excitation limit by the planet.}
\end{figure}

We summarize the secular evolution of our reference system in Fig. \ref{aini}. The top panel shows the evolution of the semi-major axis of the planetary orbit and the bottom panel the tidal and magnetic torques applied to the planet (see the thick curves in the figure). Our reference model thus shows an outward migration of the planet after the disk dissipation (gray area on the left), followed by an inward migration after $t\sim 350$ Myr. This change occurs when the co-rotation radius \(r_\text{corot} = (GM_\star/\Omega_c^2)^{1/3}\) (for which the orbital period is equal to the rotation period of the stellar envelope; see the black dashed line in Fig. \ref{aini}) crosses the orbital distance. Indeed, the co-rotation radius varies throughout the life of the system in a similar way to the stellar rotation rate (see Fig. \ref{GB15}):  during the PMS, while the star is contracting, the induced spin-up leads to a decrease in the co-rotation radius; and after the ZAMS, as the stellar structure has stabilized, stellar wind spins the star down, leading to an increase in \(r_\text{corot}\). The limit of excitation of the dynamical tide, defined as \(P_\text{orb} = \frac{1}{2} P_\text{rot}\), evolves in the same way (see the black dotted line in Fig. \ref{aini}).

The initial outward migration in our reference model can be attributed to the dynamical tide. As the star--planet system reaches higher semi-major axis (near the ZAMS), the dissipation of inertial waves becomes less and less effective (see the red curves in the bottom panel). The secular evolution of the system is then driven by magnetic torques (in blue), which leads to a more efficient inward migration after 100 Myr compared to a system evolving through tidal effects only (in red in the top panel of Fig. \ref{aini}), no longer evolving significantly at older ages. Interestingly, the tidal torque drops by two orders of magnitude when the planet then crosses the dynamical tide excitation limit (see bottom panel of Fig. \ref{aini}) as the equilibrium tide  then remains the only contributor. However, because the magnetic torques already dominate during that phase, this does not affect the overall evolution of the system.\\

We finally also track the limit of formation of a planetary magnetosphere, corresponding to \(\Lambda_p = 1\), as a limit orbital distance \(a_\text{cav}\) below which, in our model, a magnetic cavity is formed around the planet. Within our model hypotheses (see \S \ref{sec:defSPMI}), this limit can be expressed as 
\begin{equation}\label{eqn:acav}
a_\text{cav} = R_\star\left(\frac{B_\star}{B_p}\right)^\frac{1}{2}.
\end{equation}
As we consider a constant planetary magnetic field, \(a_\text{cav}\) evolves in the same way as the magnetic field at the stellar surface. In the case of our reference system, the orbital semi-major axis is always beyond the cavity formation limit (green curve), which means that the planet is able to sustain a magnetosphere throughout the whole ESPEM simulation. This will be generally the case in what follows, and we highlight the special cases when a magnetic cavity is formed and influences the secular evolution.

We now focus on the influence of the initial semi-major axis on the secular evolution of the system. To this end, we vary the initial semi-major axis of our reference case by considering \(a_\text{ini} = 0.019, 0.025,\ 0.035,\text{ and } 0.045\text{ AU}\). For the three highest values of \(a_\text{ini}\), only outward migration occurs initially as they orbit outside the co-rotation radius. Remote planets migrate less efficiently because both tidal and magnetic torques decrease with higher semi-major axis (from dark to light colors in Fig. \ref{aini}). However, an increase in the initial semi-major axis  by a factor of 1.8 leads to a decrease in the magnetic torque  by a factor of 2.5 and a drop in tidal torque  by at most a factor of 30 when the dynamical tide dominates. Hence, for planets that are able to sustain a magnetosphere, the tidal torque presents a higher sensitivity to the orbital distance than the magnetic torque, which means that for remote planets the secular evolution will likely be dominated by magnetic torques for a higher fraction of the system's lifetime. In the case of \(a_\text{ini} = 0.019 \text{ AU}\), the planet is initially located below the co-rotation radius and beyond the tidal excitation limit. The  planet therefore migrates inward efficiently because of the rise in dynamical tide until it is engulfed very early on, after about 5 Myr (which explains why evolutionary tracks are so short  in that case). In this case, the dynamical tide is so efficient that the addition of magnetic torques does not change the already fast evolution of the star--planet system.

In what follows we assess the sensitivity of the secular evolution of a given system to the free parameters of the ESPEM model, namely \(a_\text{ini},\ P_\text{rot,ini},\ M_\star,\ M_p, \text{ and } B_p\). To this end, we focus on characterizing the time at which the co-rotation radius exceeds the semi-major axis of the orbit. This gives us as a first idea of the sensitivity of our model to the initial conditions and physical prescriptions we chose. For instance, we have seen that the crossing of the co-rotation radius can be delayed by hundreds of millions of years when the initial semi-major axis varies from 0.025 AU to 0.045 AU. We also found that the addition of magnetic torque to the tidal torques induces a delay of the order of 10 Myr in our reference case. Let us now characterize the sensitivity of our model to stellar (\S 3.2) and planetary (\S \ref{sec:planetparams}) parameters.

\subsection{Influence of stellar parameters on planet migration}
\subsubsection{Influence of initial stellar rotation and stellar mass}\label{sec:stellarparams}
We now investigate the influence of stellar rotation and stellar mass on the evolution of a star--planet system. The relative contribution of magnetic and tidal torques depending on the instantaneous stellar rotation is presented in Appendix B. To highlight the role of initial stellar rotation on the fate of the system, we consider models rotating initially slower than our reference case, that is \(P_\text{rot,ini} = 2.67\) and \(5\ \text{d}\).
\begin{figure}[!h]
   \begin{center}
    \includegraphics[scale=0.4]{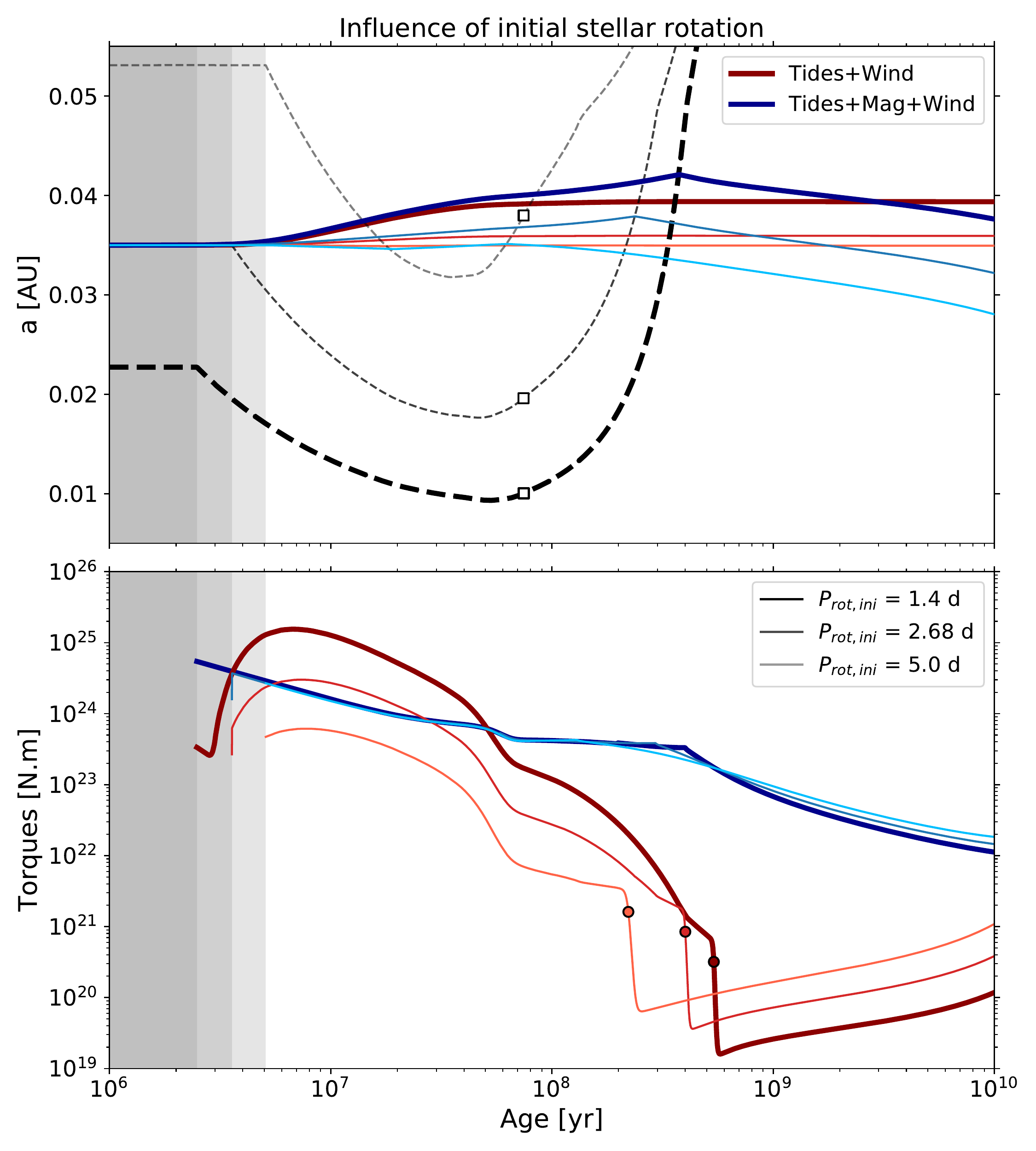}
   \end{center}
   \caption{\label{Protini} Secular evolution of a star–planet system formed by a K star (\(M_\star = 0.8\ M_\odot\)) orbited by a strongly magnetized hot Neptune (\(a_\text{ini} = 0.035\text{ AU},\ M_p = 0.1\ M_\text{Jup},\ B_p = 10\text{ G}\)) for three different initial stellar rotation periods: \(P_\text{rot, ini} = 1.4\text{ d},\ 2.67\text{ d},\ 5\text{ d}\) (dark to light colors). The thick curves correspond to our reference case discussed in \S 3.1.1.  \textit{Top panel}: Semi-major axis (solid lines), co-rotation radius of the star (black dashed lines). Wind braking + tides are shown in shades of red ; wind braking + tides + magnetic effects are shown in shades of blue. The gray bands on the left correspond to the disk locking phase. 
   \textit{Bottom panel}: Tidal (shades of red) and magnetic (shades of blue) torques in the case of evolution with all the combined interactions. The white squares correspond to the ZAMS and the red circles to the crossing of the dynamical tide excitation limit by the planet.}
\end{figure}  

The first striking effect of the initial stellar rotation period is that the planet is generally always closer to its host along the secular evolution for slower initial rotators. Indeed, our reference model shows a first phase of outward migration, followed by an inward migration after the crossing of the co-rotation radius. If a star rotates slowly initially, the first outward migration phase is very inefficient because both the tidal torque and the stellar magnetic field (and thus the magnetic torque) are small. On the contrary, the late inward migration phase is as efficient in all cases as stars converge on the same rotational tracks on the main sequence and therefore the magnetic torques that dominate the evolution here are of comparable amplitude. We note that planetary migration is negligible in the tidal case alone (in red in the upper panel) for the two slowest rotations. Indeed, as the dynamical tide is raised in this configuration, higher stellar rotation periods result in lower values of the tidal torque. However, in both cases,  the addition of the magnetic torque affects the secular evolution. We also considered the particular case where the planet is initially situated exactly at the co-rotation orbit (\(P_\text{rot,ini} = 2.67\) d); it weakly migrates outwards after the dissipation of the disk (see the gray shaded area in Fig. \ref{Protini}) as the star contracts and spins up, before the system evolves in the same way as our reference case, albeit with less efficient planetary migration. The tidal torque is indeed an order of magnitude lower, allowing the magnetic torque to dominate at all times. By taking magnetic interactions into account, a typical variation of \(P_\text{rot,ini}\) (from 1.4 to 5 d) leads to a delay in the crossing of the co-rotation radius of around 100 Myr.\\

\begin{figure}[!h]
   \begin{center}
      \includegraphics[scale=0.4]{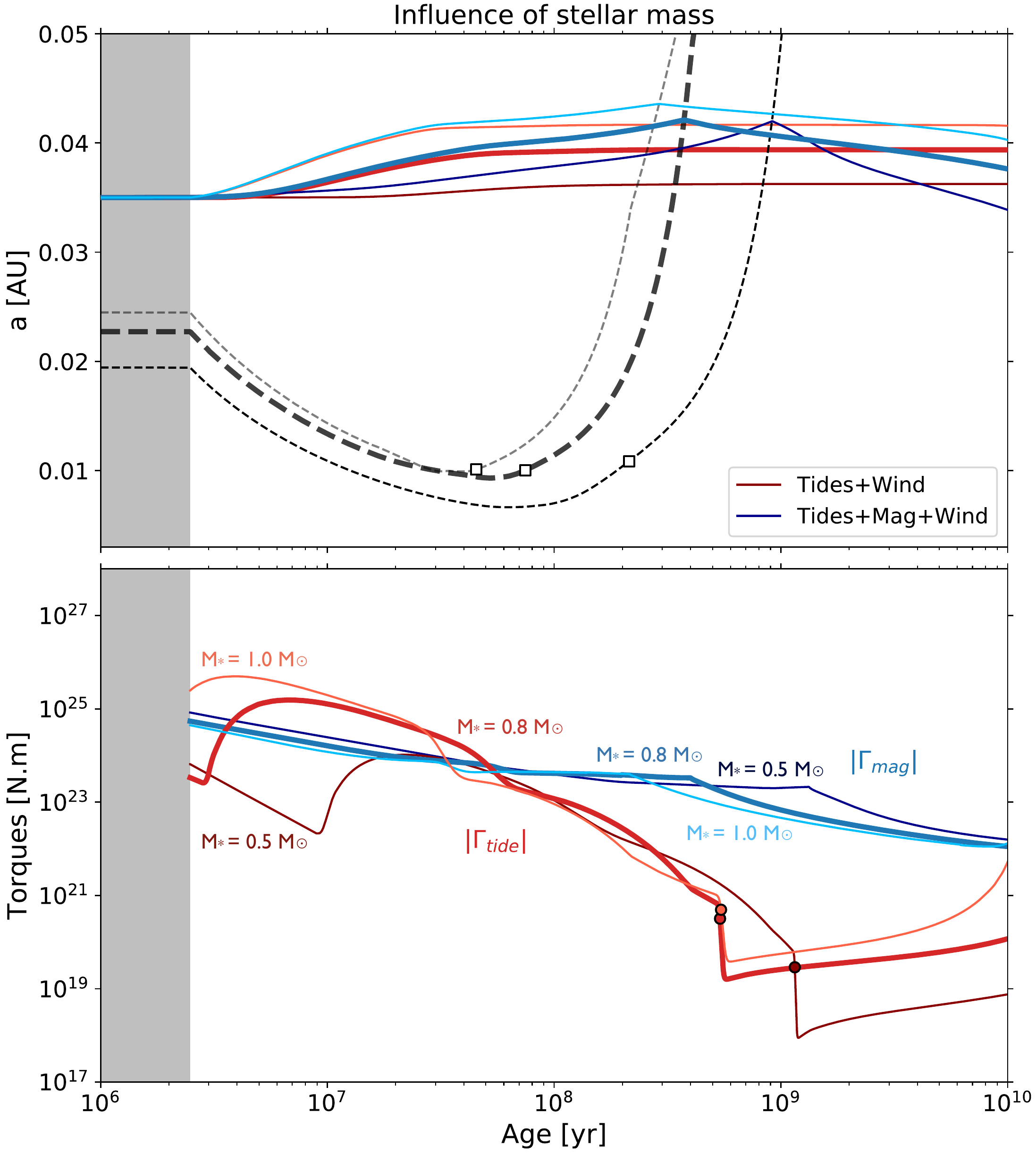}
   \end{center}
   \caption{\label{Mstar} Secular evolution of a star–planet system  formed by a fast-rotating star (\(P_\text{rot,ini} = 1.4\text{ d}\)) orbited by a strongly magnetized hot Neptune (\(M_p = 0.1\ M_\text{Jup},\ a_\text{ini} = 0.035\text{ AU},\ B_p = 10\text{ G}\)) for three different stellar masses: \(M_\star = \{0.5,\ 0.8,\ 1\}\ M_\odot\) (dark to light colors). The thick curves correspond to our reference case discussed in \S 3.1.1.  \textit{Top panel}: Semi-major axis (solid lines) and co-rotation radius of the star (black dashed lines). Wind braking + tides are shown in shades of red ; wind braking + tides + magnetic effects are shown in shades of blue. The gray bands on the left correspond to the disk locking phase. 
   \textit{Bottom panel}: Tidal (shades of red) and magnetic (shades of blue) torques in the case of evolution with all the combined interactions. The white squares correspond to the ZAMS and the red circles to the crossing of the dynamical tide excitation limit by the planet.}
\end{figure}

To investigate the influence of stellar mass on tidal and magnetic torques, we now consider our reference case for three different stellar masses: \(M_\star=0.5,\ 0.8, \text{ and } 1\ M_\odot\). As shown in the top panel of Fig. \ref{Mstar}, the co-rotation radius, the tidal excitation limit and the semi-major axis have an overall similar evolution as in \S 3.1.1 due to the initial conditions adopted. The planet, initially beyond the co-rotation radius, undergoes an outward migration through the dynamical tide in all cases. Near the ZAMS, the secular evolution of the system is driven by magnetic torques in all cases as well (in blue in the top panel of Fig. \ref{Mstar}).  

In the tidal case alone (in red in the top panel of Fig. \ref{Mstar}), the planet undergoes a more efficient migration around more massive stars at the beginning of the evolution. After dozens of millions of years, migration becomes negligible. Indeed, during the PMS, higher mass stars undergo a stronger tidal torque (in red in the bottom panel). Then, as the stellar structure has stabilized around the ZAMS, stellar spin-down leads to a continuous decrease in tidal dissipation towards the end of the evolution, varying weakly with stellar mass. In the presence of magnetic interactions,  planetary migration is found to be more and more efficient as stars are less massive. Indeed, those have a higher relative convective mass, a smaller Rossby number (Eqs. \eqref{eqn:def_rossby}-\eqref{eqn:ardestani}), and thus a stronger stellar magnetic field. This tends to enhance the stellar wind flow as well as star--planet magnetic interactions for low-mass stars. The magnetic torque then significantly affects the later evolution of the semi-major axis and tends to dominate the tidal torque over a longer phase (in blue in the top panel of Fig. \ref{Mstar}). More precisely, when magnetic interactions overcome their tidal counterparts (at \(t \sim 30\ \text{Myr}\) for \(M_\star = \{0.8,1\}\ M_\odot\), and immediately after the dissipation of the disk for \(M_\star = 0.5\ M_\odot\)), the three evolutions behave similarly until the crossing of the co-rotation radius. Indeed, the magnetic torque depends weakly on stellar mass during the PMS and the beginning of the MS. As less massive stars have a lower co-rotation radius, the planet undergoes an outward migration during a longer phase. This allows it to reach larger semi-major axes for low values of \(M_\star\).

A typical change in stellar mass (from 0.5 to 1 \(M_\odot\)) induces a delay of around 600 Myr of the crossing of the co-rotation radius, which makes it the most sensitive parameter of our model. For low stellar masses (in particular the case \(M_\star = 0.5\ M_\odot\) in the top panel of Fig. \ref{Mstar}), magnetic torques can lead to a migration delay of 100 Myr when compared to the case with only tidal effects.

In summary, the magnetic torque tends to dominate the evolution of the star--planet system during a longer fraction of its lifetime when the stellar magnetic field is stronger and when the dynamical tide is less efficient. This typically  occurs for lower mass stars, as was already pointed out by \citet{strugarek17}, and confirmed here with a fully dynamical evolution.

\subsubsection{Influence of stellar magnetism on planet migration}\label{sec:stellarparamsB}
We now aim to assess the influence of stellar magnetism on the secular evolution of star--planet systems. To this end, we first assess the dependency of \(\Gamma_\text{mag}\) on the stellar magnetic field heuristically. Indeed, the magnetic torque, as expressed in \S 2.6, presents the following dependencies for low alfvenic Mach numbers:
\begin{equation}\label{eqn:magtorque}
\Gamma_\text{mag}\propto
\begin{cases}
M_a^{0.44}p_\text{tot}^{0.72}a,\text{ if } \Lambda_p>1\\
M_a p_\text{tot}a,\ \text{otherwise (magnetic cavity)}.
\end{cases}
\end{equation}
By introducing the wind magnetic field at the planetary orbit, \(B_\text{wind}\), the alfv\'enic Mach number \(M_a = |\Omega_c - n|a/v_A\), where \(v_A\) is the Alfv\'en velocity, and scales as
\begin{equation}
M_a \propto B_\text{wind}^{-1}.
\end{equation}
Moreover, as we consider close-in exoplanets, the total wind pressure is dominated by the magnetic component close to the star \citep{preusse, reville15b}, which leads to
\begin{equation}
p_\text{tot} \propto B_\text{wind}^2.
\end{equation}
This results in the following dependency for the magnetic torque:
\begin{equation}
\Gamma_\text{mag} \propto B_\text{wind}.
\end{equation}
We now consider a multipolar topology of degree \(l\) for the magnetic field (\(B_\text{wind} \propto a^{-(l+2)}\), as the star is at the center of both the wind and the planetary orbit). The magnetic torque then becomes
\begin{equation}\label{eqn:Ptot}
\Gamma_\text{mag} \propto B_\star a^{-(l+2)} \mathcal{F}(a),
\end{equation}
where \(\mathcal{F}\) is a function of the semi-major axis, independent of the degree \(l\) at first order, and which is linked to wind acceleration. Even if those scaling laws are based on strong assumptions on the wind model, we can see that a more complex magnetic topology (corresponding to higher \(l\) values) implies a stronger dependency of the magnetic torque on the semi-major axis, which can make it more sensitive than the tidal torque itself. If the stellar magnetic field is dominated by small scales (large $l$) and its large-scale components are weak, the magnetic torque then also weakens efficiently and no longer affects the evolution of remote planets.

In addition, a change in the scaling law of \(B_\star\) (Eq. \ref{eqn:Bstar}) can affect the relative importance of the magnetic torque in our model. To illustrate this we consider the alternative prescription proposed by \citet{ahuir20}, which shows the steepest Rossby number dependency:
\begin{equation}\label{eqn:Bstar_max}
B_\star\ [\text{G}] = 2.0\left(\frac{Ro}{Ro_\odot}\right)^{-1.65}\left(\frac{M_\star}{M_\odot}\right)^{-1.04},\   Ro > Ro_\text{sat}. 
\end{equation}
To keep a consistent wind model,  the \(T_c\) and \(n_c\) prescriptions need to be updated as \citep{ahuir20}:
\begin{equation}\label{eqn:Tc_max}
T_c\ [\text{MK}] =1.5\left(\frac{Ro}{Ro_\odot}\right)^{-0.04}\left(\frac{M_\star}{M_\odot}\right)^{0.05},\  Ro > Ro_\text{sat},
\end{equation}
\begin{equation}\label{eqn:nc_max}
n_c\ [\text{cm}^{-3}] =  7.25\times 10^7\left(\frac{Ro}{Ro_\odot}\right)^{-0.64}\left(\frac{M_\star}{M_\odot}\right)^{1.49},\  Ro > Ro_\text{sat}.
\end{equation}
A steeper dependency on the Rossby number implies higher values of \(B_\star\) at young ages, and thus stronger star--planet magnetic interactions at the beginning of the evolution of the system compared to a scenario in which \(B_\star\) and \(Ro^{-1}\) scale linearly (Eqs. \eqref{eqn:Bstar}, \eqref{eqn:Tc}, and \eqref{eqn:nc}). Moreover, as \(Ro \propto M_\star R_\star^{1.2}\) during the MS in the \citet{ardestani} formulation, the stellar magnetic field in the present scenario is also more sensitive to \(M_\star\). Because of the solar normalization, a change in the prescription of \(B_\star\) then increases the stellar magnetic field of less massive stars, and emphasizes the significance of the magnetic torque on the secular evolution of the star--planet system.

\begin{figure}[!h]
   \begin{center}
    \includegraphics[scale=0.39]{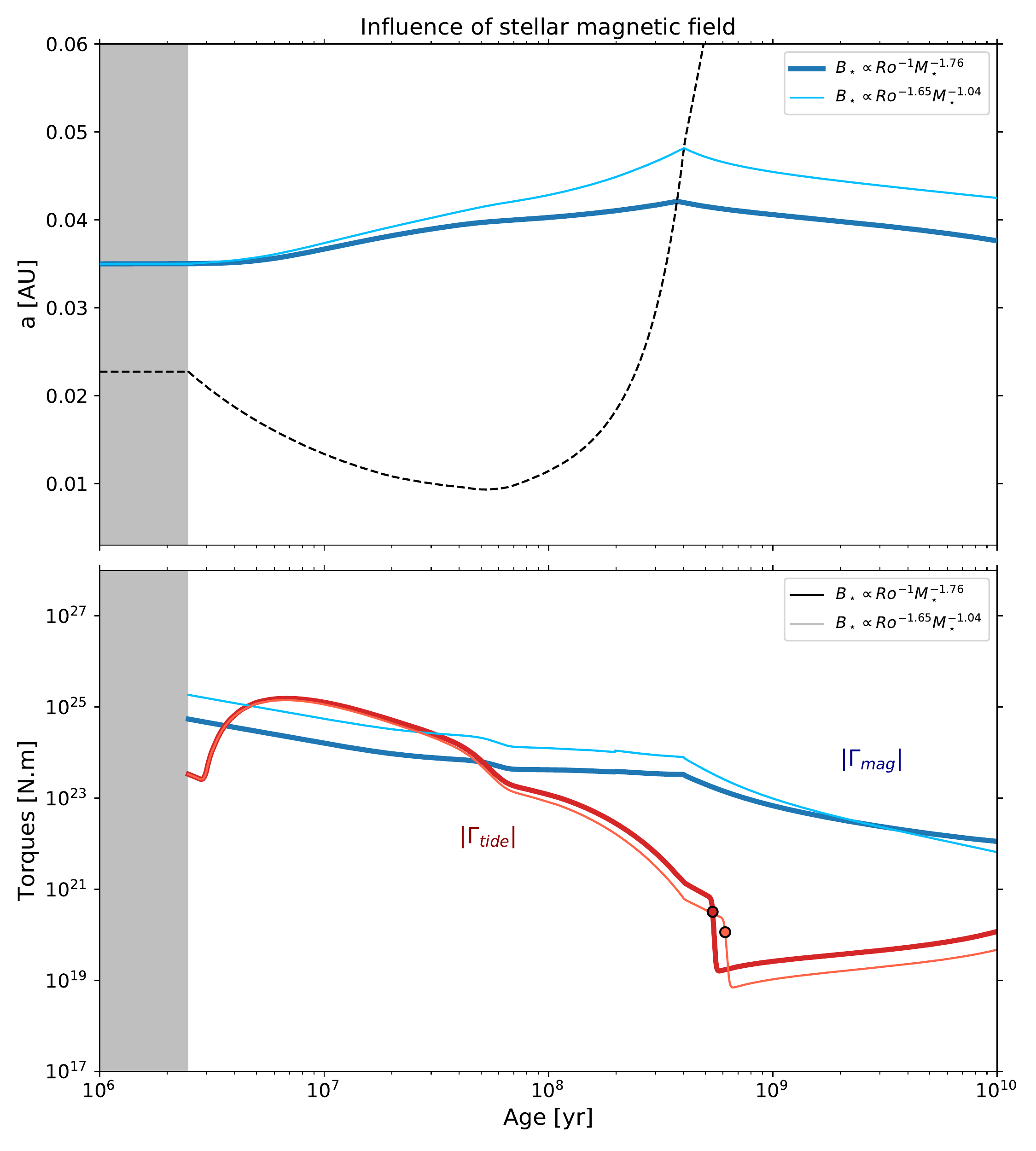}
   \end{center}
   \caption{\label{Bstar} Secular evolution of a star–planet system  formed by a fast-rotating K star (\(M_\star = 0.8\ M_\odot,\ P_\text{rot,ini} = 1.4\text{ d}\)) orbited by a strongly magnetized planet (\(a_\text{ini} = 0.035\text{ AU},\ B_p = 10\text{ G}\)) in the case of a linear \(B_\star-Ro^{-1}\) relationship (in dark colors; see Eqs. \eqref{eqn:Bstar}, \eqref{eqn:Tc} and \eqref{eqn:nc}) and a superlinear \(B_\star-Ro^{-1.65}\) relationship (in light colors; see Eqs. \eqref{eqn:Bstar_max}, \eqref{eqn:Tc_max} and \eqref{eqn:nc_max}). The thick curves correspond to our reference case discussed in \S 3.1.1.  \textit{Top panel}: Semi-major axis (solid lines) and co-rotation radius of the star (black dashed lines). The gray bands on the left correspond to the disk-locking phase. 
   \textit{Bottom panel}: Tidal (dark and light red) and magnetic (shades of blue) torques in the case of evolution with all the combined interactions. The red circles correspond to the crossing of the dynamical tide excitation limit by the planet.}
\end{figure}

The influence of the choice of a magnetic scenario on the evolution of our reference case is illustrated in Fig. \ref{Bstar}. Planetary migration through Eq. \eqref{eqn:Bstar_max} (upper panel) is more efficient (light blue curve) than for our reference case (dark blue curve) until \(t \sim 700\) Myr. Afterwards, both evolutions are similar. Indeed, as can be seen from the scaling laws, Eq. \eqref{eqn:Bstar_max} induces a more intense magnetic torque for systems younger than the Sun. Hence, at the beginning of the evolution, the magnetic torque is increased by nearly an order of magnitude (in blue in the bottom panel of Fig. \ref{Bstar}). The crossing of the co-rotation radius is delayed by about 10 Myr, showing a mild influence of the stellar magnetism scaling law on the system here. As the planet migrates further away from the star than in the case of a linear \(B_\star-Ro^{-1}\) scaling, the tidal torque becomes less efficient during the MS (in red in the bottom panel of Fig. \ref{Bstar}). However, this does not affect the planetary migration, as the magnetic torque already dominates the evolution in these phases.

\subsection{Influence of planetary mass and magnetic field on planet migration}\label{sec:planetparams}

We now aim to unravel the influence of planetary intrinsic properties on the fate of the system. We focus in this work on the influence of the mass of the planet as well as its magnetic field. The assessment of the former has a direct influence on the mass dependency of the different torques. Indeed, from the \citet{murray} and the \citet{strugarek16} formulations we have
\begin{equation}\label{eqn:Gtidemp}
\Gamma_\text{tide}\propto M_p^2
\end{equation}
\begin{equation}
\Gamma_\text{mag}\propto R_p^2 B_p^{0.56}.
\end{equation}

In view of the relative dependence on \(M_p\) and \(R_p\) of the two torques, the mass--radius relationship used to describe the planet will have a significant influence on the relative contributions of magnetic and tidal interactions.

We now consider different scenarios to estimate the magnetic field of the planet. First, if we assume a planetary magnetic field independent of all the other quantities of our model, by relying on Eq. \eqref{eqn:RMrel} the magnetic torque scales as
\begin{equation}\label{eqn:Bconst}
\Gamma_\text{mag} \propto
\begin{cases}
M_p^{0.56}, &M_p < 2.0\ M_\oplus\ (6.29\times10^{-3}\ M_\text{Jup})\\
M_p^{1.18}, &2.0\ M_\oplus \leq M_p < 0.4\ M_\text{Jup}\\
M_p^{-0.08}, &M_p \geq 0.4\ M_\text{Jup.}
\end{cases}
\end{equation}
The tidal torque (Eq. \eqref{eqn:Gtidemp}) in the scenario we consider (Eq. \eqref{eqn:Bconst}) is therefore more sensitive to the planetary mass than the magnetic torque.

\begin{figure}[!h]
   \begin{center}
    \includegraphics[scale=0.4]{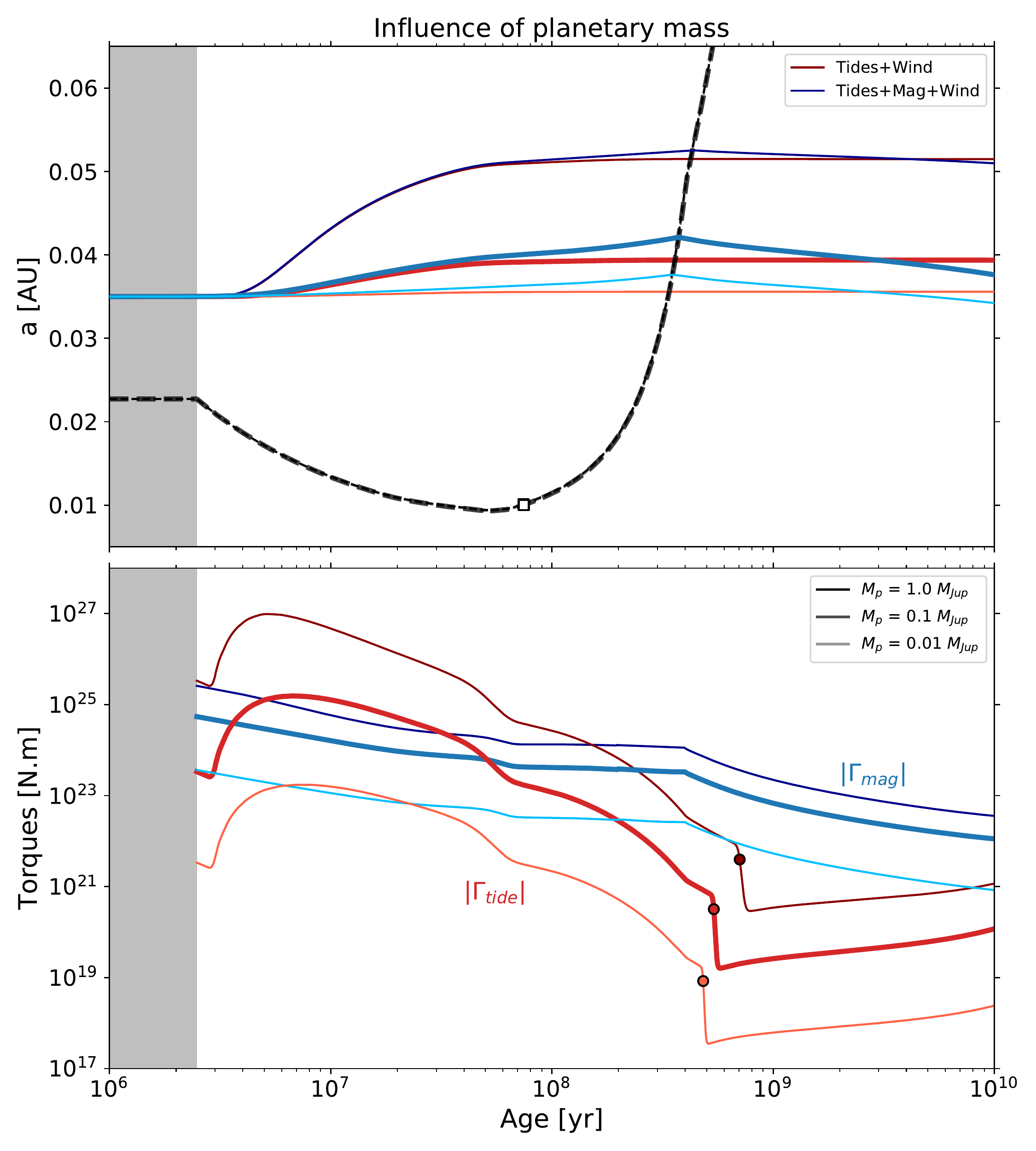}
   \end{center}
   \caption{\label{mp} Secular evolution of a star–planet system  formed by a fast rotating K star (\(M_\star = 0.8\ M_\odot,\ P_\text{rot,ini} = 1.4\text{ d}\)) orbited by a strongly magnetized planet (\(a_\text{ini} = 0.035\text{ AU},\ B_p = 10\text{ G}\)) for three different planetary masses: \(M_p = \{0.01,\ 0.1,\ 1\}\ M_\text{Jup}\) (dark to light colors). The thick curves correspond to our reference case discussed in \S 3.1.1.  \textit{Top panel}: Semi-major axis (solid lines), co-rotation radius of the star (black dashed lines). Wind braking + tides are shown in shades of red ; wind braking + tides + magnetic effects are shown in shades of blue. The gray bands on the left correspond to the disk-locking phase. 
   \textit{Bottom panel}: Tidal (shades of red) and magnetic (shades of blue) torques in the case of evolution with all the combined interactions. The white squares correspond to the ZAMS and the red circles to the crossing of the dynamical tide excitation limit by the planet.}
\end{figure}   

To investigate these different sensitivities to planetary mass  in more detail we consider a lower and a higher planetary mass with respect to our reference case, that is: \(M_p = 0.01\) and \(1\ M_\text{Jup}\). As shown in Fig. \ref{mp}, in the tidal case alone, the migration of massive planets is more efficient (red curves in upper panel) than for their less massive counterparts, as the tidal torque is stronger. In the presence of magnetic interactions (in blue), the evolution of the super-Earths differs significantly from the tidal case alone (upper panel). Indeed, the tidal torque decreases more substantially than the magnetic torque for a decreasing planetary mass (see the bottom panel of Fig. \ref{mp}). Hence, in the case of super-Earths, the magnetic torque (see the blue curves in Fig. \ref{mp}) is more likely to drive the evolution of the star--planet system. By taking magnetic interactions into account, a typical variation of \(M_p\) (from \(10^{-2}\) to \(1\ M_\text{Jup}\)) leads to a delay of around 100
Myr in the crossing of the co-rotation radius. Thus, a change of two orders of magnitude in planetary mass has a similar effect on secular evolution, manifesting as an increase in the initial semi-major axis  by a factor of two and an increase in the initial stellar rotation period  by a factor of five.

One can assess the robustness of the previous results by considering other hypotheses for the planetary magnetic field. For instance, one can assume that the magnetic field of the planet we consider behaves in the same way as what is observed in the Solar System. Then, according to \citet{shkolnik}, in the case of close-in planets with a synchronous rotation and a dipolar magnetosphere, we obtain 
\begin{equation}
B_p \propto \left(\frac{M_p}{P_\text{orb}}\right)^{1.21}R_p^{-3}.
\end{equation}
In such a scenario, according to the \citet{chen} \(M_p-R_p\) relationships (Eq. \eqref{eqn:RMrel}), \(B_p\) increases for an increasing planetary mass if \(M_p < 2.0\ M_\oplus\) and \(M_p \geq 0.4\ M_\text{Jup}\). This leads to the following mass dependencies for the magnetic torque:
\begin{equation}
\Gamma_\text{mag} \propto 
\begin{cases}
M_p^{0.77}, &M_p < 2.0\ M_\oplus\ (6.29\times10^{-3}\ M_\text{Jup})\\
M_p^{0.87}, &2.0\ M_\oplus \leq M_p < 0.4\ M_\text{Jup}\\
M_p^{0.66}, &M_p \geq 0.4\ M_\text{Jup}.
\end{cases}
\end{equation}
This leads to a stronger mass dependency of the magnetic torque for jovian planets as well as super-Earths and a weaker \(M_p\)-dependency in the case of neptunian planets. However, as in the constant surface magnetic field scenario, the tidal torque is more sensitive to planetary mass than the SPMI torque. Hence, we find this effect to be negligible and choose to consider a constant planetary field in the remainder of this work for the sake of simplicity.\\

\begin{figure}[!h]
   \begin{center}
    \includegraphics[scale=0.4]{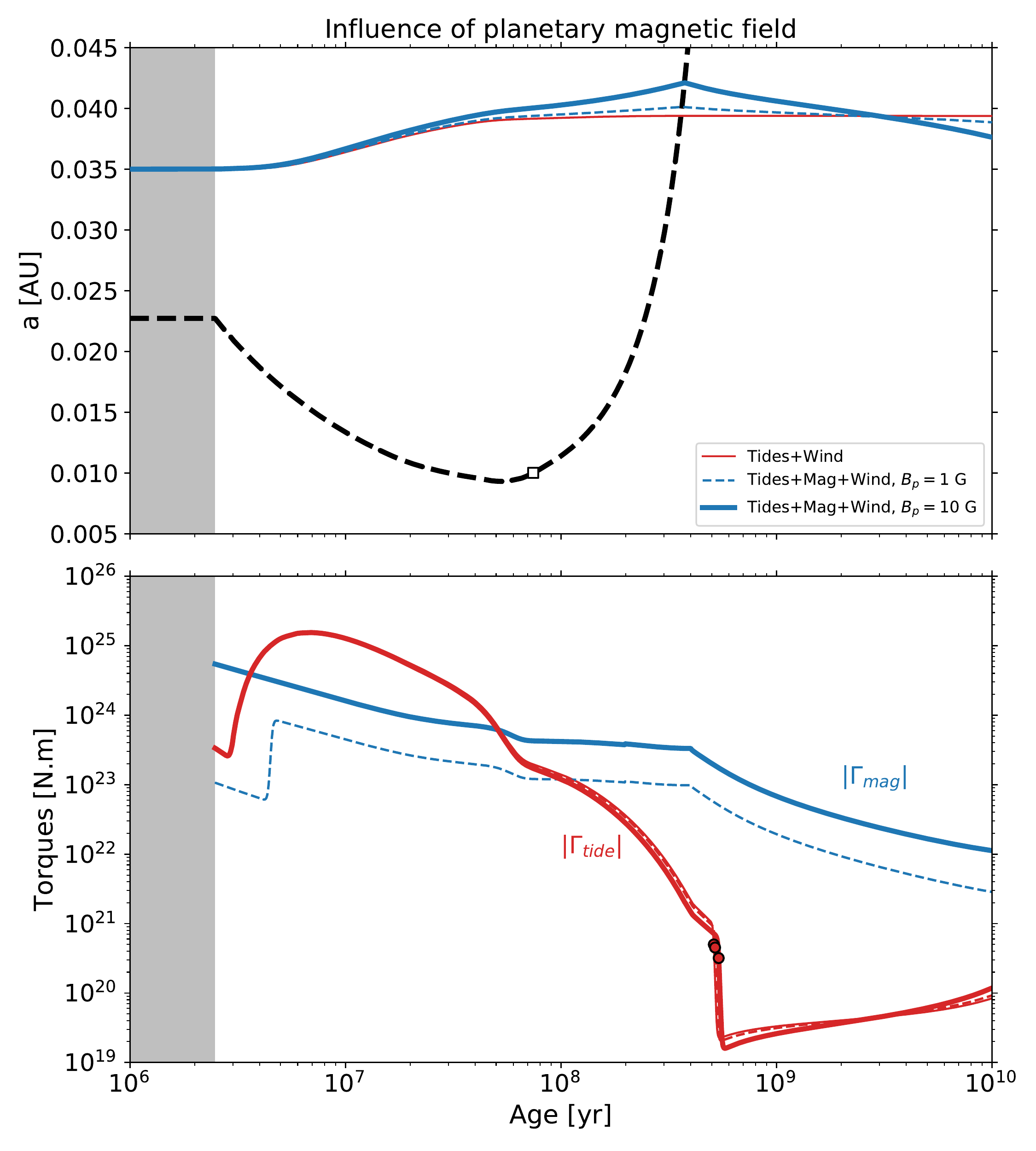}
   \end{center}
   \caption{\label{Bp} Secular evolution of a star–planet system  formed by a fast-rotating K star (\(M_\star = 0.8\ M_\odot,\ P_\text{rot,ini} = 1.4\text{ d}\)) orbited by a magnetized hot Neptune (\(M_p = 0.1\ M_\text{Jup},\ a_\text{ini} = 0.035\text{ AU}\)) for two different values of the planetary magnetic field: \(B_p = \{1,10\}\ \text{G}\) (in dashed and solid line, respectively). The thick curves correspond to our reference case discussed in \S 3.1.1.  \textit{Top panel}: Semi-major axis (solid lines) and co-rotation radius of the star (black dashed lines). Wind braking + tides are shown in shades of red ; wind braking + tides + magnetic effects are shown in shades of blue. The gray bands on the left correspond to the disk-locking phase. 
   \textit{Bottom panel}: Tidal (shades of red) and magnetic (shades of blue) torques in the case of evolution with all the combined interactions. The white square corresponds to the ZAMS and the red circles to the crossing of the dynamical tide excitation limit by the planet.}
\end{figure}  
Let us now assess the sensitivity of our results to the amplitude of the planetary field. In the case of a magnetized planet, as \(\Gamma_\text{mag}\propto R_p^2 B_p^{0.56}\), a higher planetary magnetic field leads to more efficient star--planet interactions and therefore to a more significant migration. Such a trend is illustrated in Fig. \ref{Bp} by varying the surface magnetic field of the planet in our reference case. In this configuration, a factor of ten in the magnetic field leads to a increase in the dipolar torque by a factor of 3.6, which has a significant influence on the semi-major axis of the planet during the main sequence. Such a variation plays a critical role if the planet is likely to be engulfed. However, a change of one order of magnitude in the planetary magnetic field only leads to a delay in exceeding the co-rotation radius of the order of 10 Myr, making \(B_p\) the least sensitive free parameter of our model. 

When a magnetic cavity is formed around the planet, as the geometrical cross-section intervenes in the magnetic torque, the latter is independent of \(B_p\). The magnetic torque then scales as \(\Gamma_\text{mag} \propto R_p^2\). From the \citet{chen} relations, the magnetic torque is therefore less sensitive to the planetary mass than its tidal counterpart in the magnetic cavity regime, but is always negligible compared to the dynamical tide because it decreases by one order of magnitude.\\

Even though planetary magnetism can affect the secular evolution of the star--planet system  significantly, the magnetic field strength of extrasolar planets remains  poorly constrained. For instance, tidal effects may heat the planetary core, hence affecting its dynamics. Those processes may alter the planetary dynamo in a manner that is not yet fully understood.
However, we can gain initial insight into the possible values in various ways. For instance, \citet{mcintyre} estimated the magnetic moment of rocky planets from dynamo models, which leads to a surface magnetic field of between \(1.5 \times 10^{-2}\) and 1.45 G, assuming a dipolar topology. We can also rely on what is observed in the Solar System to assess \(B_p\). While the maximal field strength in this system is observed in Jupiter with a value of around 4 G, the \citet{shkolnik} scaling law applied to a super-Jupiter of mass \(M_p = 10\ M_\text{Jup}\) and an orbital period of 0.5 d (situated near the Roche limit of a \(1\ M_\odot\) star for example) leads to a planetary magnetic field that can reach 10 G. Hence, we adopt in the rest of this work two different values for the planetary magnetic field: \(B_p = 1\) G, and \(B_p = 10\) G, the latter acting as an upper bound compared to the values measured in the Solar System.
Such values of $B_p$ seem to be in agreement with the possible detections of exoplanet radio emission \citep[e.g., $\tau$ Boötis b, we refer the reader to][]{turner21}. However, one has to keep in mind that it is possible for some planets to have an even stronger magnetic field. Indeed, values as high as 28 G or even hundreds of Gauss have been estimated for hot Jupiters using observed abnormal stellar activity correlated to the planet orbital motion \citep{cauley15,cauley19}.

\subsection{Impact on stellar rotation}

We now investigate the impact of the planet on the rotation rate of its host star. Indeed, the transfer of angular momentum between the planet and the star can affect stellar rotation, especially in the case of a planet spiralling inwards, which may spin up the host star by transferring its orbital angular momentum into stellar spin \citep{yee20}.

\begin{figure}[!h]
   \begin{center}
    \includegraphics[scale=0.45]{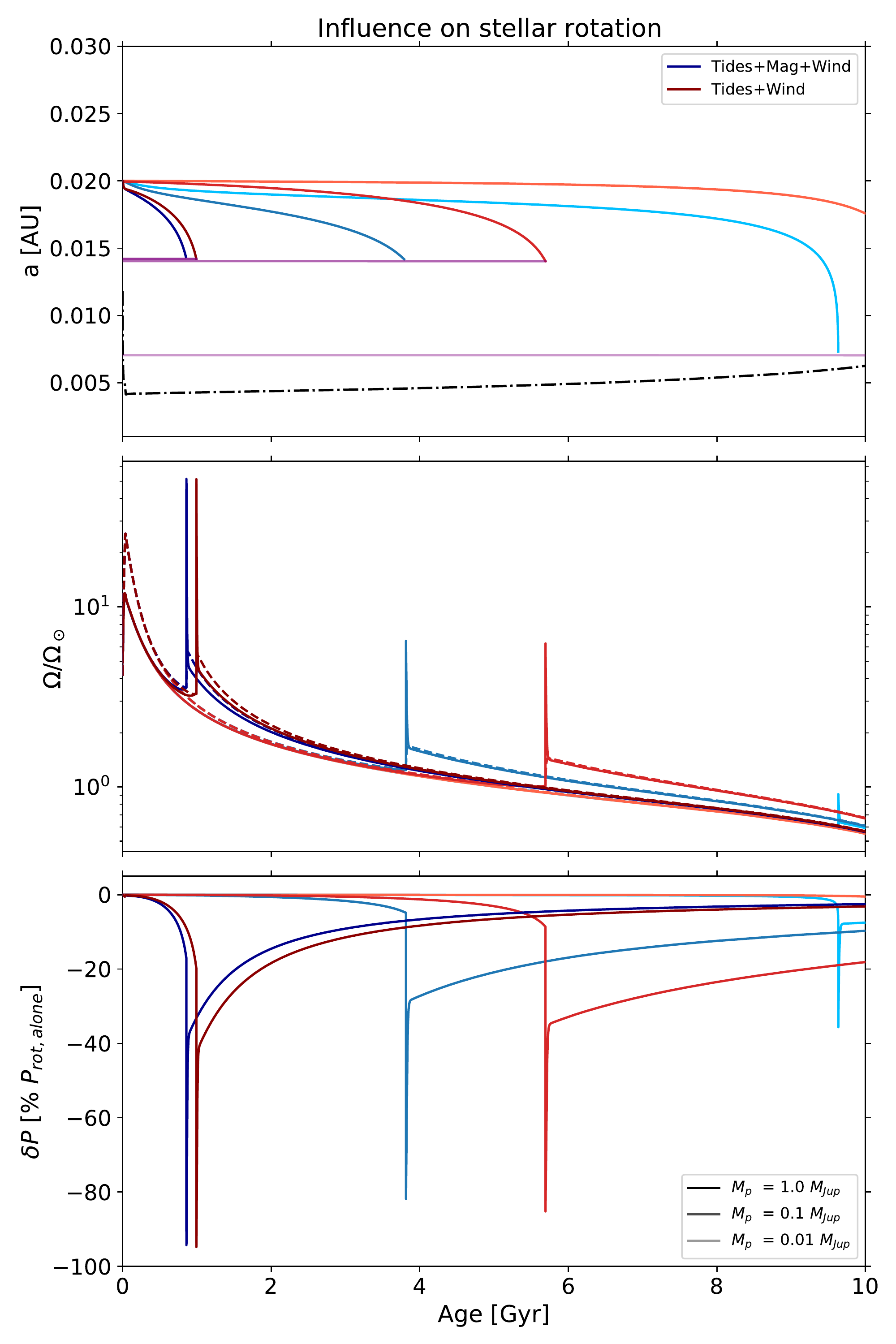}
   \end{center}
   \caption{\label{dP} Secular evolution of a star–planet system formed by a median rotating solar twin (\(M_\star = 1\ M_\odot,\ P_\text{rot,ini} = 5\text{ d}\)) orbited by a magnetized planet (\(a_\text{ini} = 0.02\text{ AU},\ B_p = 1\text{ G}\)) for three different planetary masses: \(M_p = \{0.01,\ 0.1,\ 1\}\ M_\text{Jup}\) (dark to light colors). \textit{Top panel}: Semi-major axis (solid lines). In shades of red: wind braking + tides, in shades of blue: wind braking + tides + magnetic effects, in shades of purple: Roche limit, defined as in \citet{benbakoura}. In dashed-dotted black: stellar radius. \textit{Middle panel}: Time evolution of the rotation rate of the stellar envelope (solid line) and of the stellar core (dashed line). \textit{Bottom panel}: Departure of the rotation period of the planet-hosting star compared to an isolated star.} 
\end{figure}   
We can assess the upper bound of the angular momentum transferred from the orbit to the star by considering the case of a planet engulfment. In this configuration, the initial orbital angular momentum \(L_\text{orb,ini}\) is entirely transferred to the stellar envelope over the migration timescale. Hence, a planet-hosting star with a rotation period \(P_\text{rot}\) and an angular momentum \(L_\star\) has a modified rotational state compared to an isolated star with a rotation period of \(P_\text{rot, al}\) and an angular momentum \(L_{\star,al}\). In this context, the difference in angular momenta  \(\delta L = L_\star-L_{\star,al}\) can be written as
\begin{equation}
\frac{\delta L}{L_{\star,al}} =\frac{P_\text{rot,al}}{P_\text{rot}}-1 =  \frac{L_\text{orb,ini}}{L_{\star,al}}.
\end{equation}
The corresponding difference in rotation periods \(\delta P = P_\text{rot}-P_\text{rot,al}\) then becomes
\begin{equation}\label{deltaP}
\frac{\delta P}{P_\text{rot,al}} = \frac{ -L_\text{orb,ini}/L_{\star,al}}{1+ L_\text{orb,ini}/L_{\star,al}}\sim -\frac{M_p\sqrt{G M_\star a_\text{ini}}P_\text{rot,al}}{2\pi I_\star},
\end{equation}
where \(I_\star\) is the total moment of inertia of the star. Hence, stellar rotation is the most impacted by the star--planet interaction when the engulfed planet is massive or when it has a high initial semi-major axis. Furthermore, all things equal, a later engulfment leads to a larger relative over-rotation of the star because older stars rotate slower. As an example, a \(1\ M_\odot\) star engulfing a planet at \(t = 1\) Gyr (corresponding to \(P_\text{rot,al}\) =  9.33 d) undergoes a three-times-smaller spin-up  than if it had destroyed the same planet at solar age (for which \(P_\text{rot,al}\) = 28 d). More details on this process and on the influence of the different physical quantities can be found in \citet{gallet18} and \citet{benbakoura}. Here we simply illustrate the influence of the magnetic torque on stellar rotation by varying the planetary mass.

To this end, we consider a star–planet system formed by a median rotating solar twin (\(M_\star = 1\ M_\odot,\ P_\text{rot,ini} = 5\text{ d},\ a_\text{corot,ini} = 0.057\text{ AU}\)) orbited by a magnetized planet (\(a_\text{ini} = 0.02\text{ AU},\ B_p = 1\text{ G}\)) for three different planetary masses: \(M_p = \{0.01,\ 0.1,\ 1\}\ M_\text{Jup}\). As visible in Fig. \ref{dP}, the planet can have a strong impact on stellar rotation. Indeed, as the planet migrates inwards, angular momentum is transferred from the orbit to the star, resulting in an increase of the stellar spin until planetary engulfment. More details on the modeling of such a phenomenon in ESPEM can be found in \citet{benbakoura}. After the destruction of the planet, the \(\Omega_\star
^3\) dependency of wind braking makes the star spin down efficiently, hence causing the two rotational histories (single star and host star engulfing a planet) to converge again. Depending on its lifetime, the star may not come exactly back to an unperturbed rotational state. Furthermore, for low-mass planets, their migration is less efficient, which results in a later engulfment, which in turn leads to a later and smaller peak in stellar rotation, because of the lower initial orbital angular momentum. The magnetic effects accentuate the behavior obtained based on tidal effects only: here the planet experiences an earlier engulfment compared to the pure tidal case (for \(M_p = \{1,\ 0.1\}\ M_\text{Jup}\); see the dark blue curves in Fig. \ref{dP}). The destruction of these planets leads to a spin-up of their host star of around \(\Delta P_\text{rot}\) = 8.9 d and 22.4 d (for \(M_p = \{1,\ 0.1\}\ M_\text{Jup}\) respectively). These values are slightly lower than the over-rotations obtained through tidal effects only, as planetary engulfment happens earlier. It is also worth noting that a stellar spin-up of around 20\% lasts between 4 and 6 Gyr in this configuration. Thus, by relying on gyrochronology, stellar age could be underestimated by about 10 \%. For  \(M_p = 0.01\ M_\text{Jup}\), tidal effects alone are not strong enough to disrupt the planet. The addition of the magnetic torque allows an engulfment at the end of the evolution (see the light blue curves in Fig. \ref{dP}), leading to a stellar spin-up of around \(\Delta P_\text{rot}\) = 18.5 d. Hence, star--planet magnetic interactions may actually lead to a late destruction of hot super-Earths.

\section{Star--planet interactions and classification of planetary populations}
\subsection{ESPEM sample and migration timescales}

\begin{figure*}[!h]
   \begin{center}
    \includegraphics[scale=0.25]{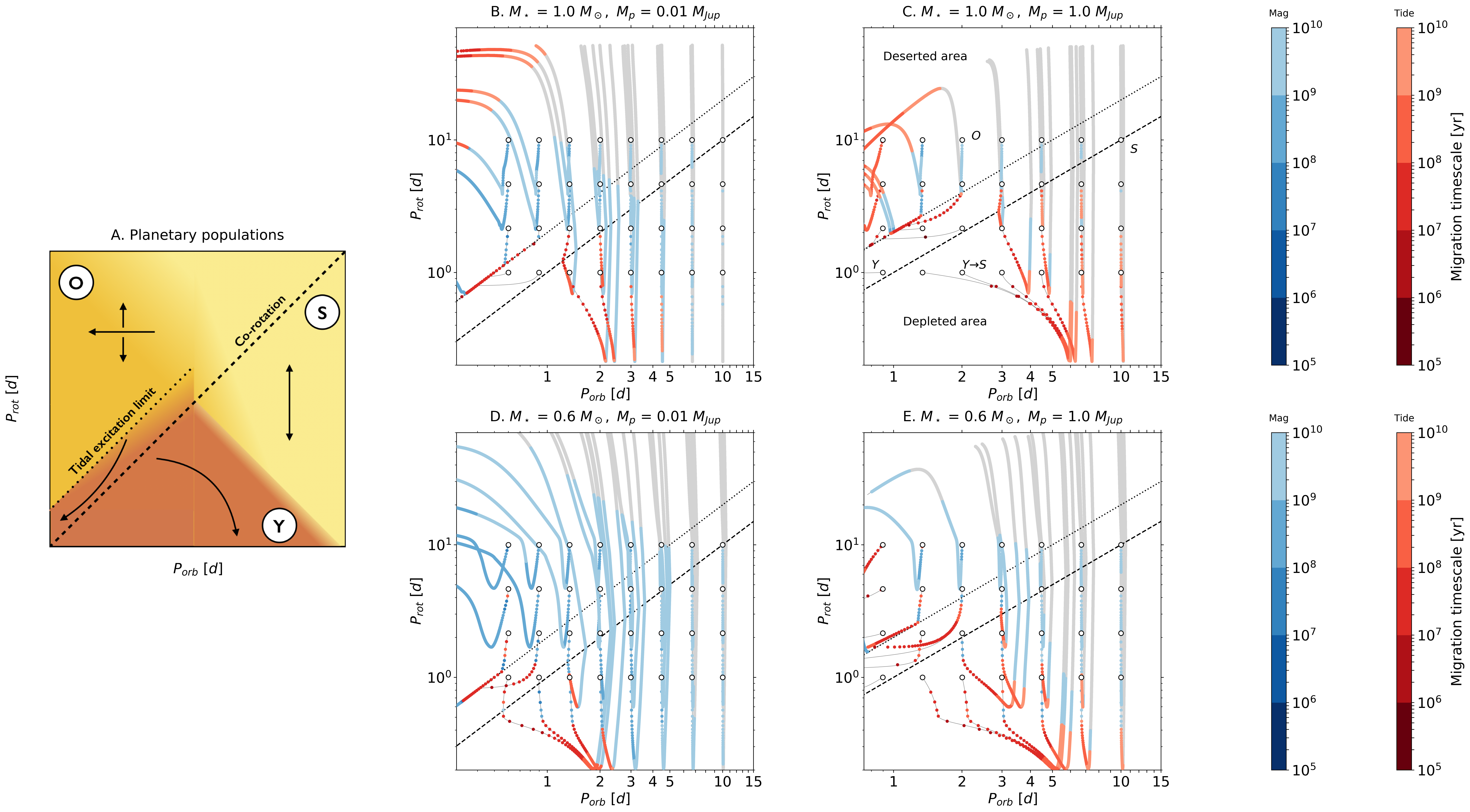}
   \end{center}
   \caption{\label{Tracks} Evolution between 1 Myr and 10 Gyr of a planetary population designed from eight orbital periods \(P_\text{orb,ini}\), equally spaced in logarithm between 0.6 and 10 days, and four initial stellar rotation periods \(P_\text{rot,ini}\) between 1 and 10 days. Those initial parameters correspond to the white circles. Panel A: Sketch representing the evolution of the Steady population (S, in yellow), the Young Migrators (Y, in dark orange), and the Old Migrators (O, in light orange) in the (\(P_\text{orb},P_\text{rot}\)) plane. The black arrows correspond to the possible paths of each population. Panel B: Evolution of the planetary population for \(M_\star = 1\ M_\odot\) and \(M_p = 0.01\ M_\text{Jup}\). Panel C: Evolution of the planetary population for \(M_\star = 1\ M_\odot\) and \(M_p = 1\ M_\text{Jup}\). Panel D: Evolution of the planetary population for \(M_\star = 0.6\ M_\odot\) and \(M_p = 0.01\ M_\text{Jup}\). Panel E: Evolution of the planetary population for \(M_\star = 0.6\ M_\odot\) and \(M_p = 1\ M_\text{Jup}\). Magnetic and tidal dominance are shown in blue and red, respectively. The shades of blue and red correspond to the overall migration timescale of the system, displayed in logarithmic scale (see key on the right). Regions where the overall migration timescale is greater than the age of the Universe are shown in gray. The black dashed line corresponds to co-rotation and the black dotted line to the tidal excitation limit.}
\end{figure*}   

Knowing the influence of each parameter of our model on the fate of a given star--planet system, we now adopt a more global approach by simultaneously varying them. To this end we design a sample including 7000 ESPEM simulations which browses the parameter space. The range for each free parameter we consider as well as their distribution are presented in Table \ref{tab:ranges}. To highlight the behavior of star--planet systems through magnetic and tidal interactions, we first focus on a subsample comprising eight initial orbital periods \(P_\text{orb,ini}\) and four initial stellar rotation periods \(P_\text{rot,ini}\), which are evenly spaced in logarithm, two planetary masses \(M_p = \{0.01,\ 1\}\ M_\text{Jup}\), and two stellar masses \(M_\star = \{0.6,\ 1\}\ M_\odot\).
\begin{table}[!h]
\centering 
      \caption{\label{tab:ranges} Range and distribution of star--planet parameters considered in the ESPEM sample.}
      \begin{tabu}{lll}
            \hline
             \noalign{\smallskip}
             Parameter & Range & Sample distribution \\
             \noalign{\smallskip}
            \hline
            \noalign{\smallskip}
            $M_\star\ [\text{M}_\odot]$ & 0.5 -- 1.1 & Uniform\\
            $P_\text{rot,ini}\ [\text{d}]$ & 1 -- 10 & Uniform\\
            $a_\text{ini}\ [\text{AU}]$ & $5 \times 10^{-3}$ -- 0.2 & Uniform in logarithm\\
            $M_p\ [\text{M}_\oplus]$ & 0.5 -- 1589 (= 5 $\text{M}_\text{Jup}$) & Uniform in logarithm\\
            $B_p\ [\text{G}]$ & 0\tablefootmark{a}, 1, 10 & - \\
            \noalign{\smallskip}
            \hline
         \end{tabu}
         \tablefoottext{a}{This case corresponds to an evolution of the system through tidal effects only.}
   \end{table}
   
We show in Fig. \ref{Tracks}  the evolutive track of our representative subsample of star--planet models in the (\(P_\text{orb},\ P_\text{rot}\)) plane. The efficiency of planetary migration at each time-step is assessed through the characteristic timescale
\begin{equation}
\tau_\text{mig} = \left|\frac{P_\text{orb}}{\dot{P}_\text{orb}}\right|= \frac{1}{3}\left|\frac{L_\text{orb}}{\dot{L}_\text{orb}}\right|.
\end{equation}
As \(P_\text{orb} \propto a^{3/2}\) and \(L_\text{orb} \propto a^{1/2}\), for a given migration, the orbital period is more impacted than the semi-major axis, which is itself more sensitive than the orbital angular momentum. Hence, our expression of \(\tau_\text{mig}\) gives the shortest timescale characterizing planet migration. The overall migration timescale due to the sum of tidal and magnetic
torques can be split into two contributions
\begin{equation}
\tau_\text{mig} = \left(\frac{1}{\tau_M}+\frac{1}{\tau_T}\right)^{-1},
\end{equation}
where \(\tau_M\) and \(\tau_T\) are the magnetic and tidal migration timescales respectively. From the expressions of \(\Gamma_\text{mag}\) and \(\Gamma_\text{tide}\), those timescales can be written as
\begin{equation}
\tau_T\ [\text{Myr}]= 89.31\times\left(\frac{Q'}{10^6}\right)\left(\frac{M_\star}{M_\odot}\right)^\frac{8}{3}\left(\frac{M_\text{Jup}}{M_p}\right)\left(\frac{R_\odot}{R_\star}\right)^{5}\left(\frac{P_\text{orb}}{1\ \text{d}}\right)^{\frac{13}{3}}
\end{equation}
\begin{equation}
\begin{split}
\tau_M=\tau_{M,0}\left(\frac{R_J}{R_p}\right)^{2}\left(\frac{M_p}{M_\text{Jup}}\right)\left(\frac{M_\star}{M_\odot}\right)^\frac{5}{3}\left(\frac{R_\star}{R_\odot}\right)^{-4}\left(\frac{B_\star}{10 \text{ G}}\right)^{-2}\left(\frac{P_\text{orb}}{1\ \text{d}}\right)^\frac{7}{3}\!\!,
\end{split}
\end{equation}
with
\begin{equation}
\tau_{M,0}\ [\text{Myr}] = 
\begin{cases}
\displaystyle 405.3\times\left(\frac{10\ R_p}{R_m}\right)^{1.68}M_a^{-0.44},\ R_m \geq R_p\\
2.095\times 10^{5},\text{ otherwise},
\end{cases}
\end{equation}
where \(R_m = R_p\Lambda_P^\frac{1}{6}\) is  the size of the planet magnetosphere if we assume a dipolar magnetic field. As we only consider subaflvenic interactions between the star and the planet, the \(M_a\) dependency only increases the migration timescale through magnetic torques. Therefore, one can neglect the influence of this quantity to provide a lower bound for \(\tau_M\).

The instantaneous overall migration timescale is shown for each system in Fig. \ref{Tracks}. As already presented in \S 3 for each parameter of the model independently, we can see from Fig. 11 that tidal effects tend to dominate star--planet magnetic interactions (shown in red) for high stellar and planetary masses, long rotation periods (beyond several dozen days in the case of the equilibrium tide) or short rotation periods (below two approximately days  in the case of the dynamical tide), and small semi-major axis.

\subsection{Classification of planetary populations}

Overall, we find that three populations arise from the action of star--planet interactions. The Steady population is composed of planets for which the mean migration timescale is greater than the age of the Universe along most of their evolution because of negligible tidal and magnetic torques. This population is situated at long orbital periods (from 1 to 10 days depending on stellar rotation period; see gray areas in Fig. \ref{Tracks}). The Young Migrators are exoplanets that experience a significant migration during the PMS of their host stars. Situated at short orbital and rotation periods (less than two days approximately), these latter are subject to extreme star--planet interactions. Two fates are possible for those systems: If the planet is initially below the co-rotation radius, it is engulfed by the star during the early stages of its evolution. Conversely, if the planet is initially beyond the co-rotation radius, it migrates outward efficiently and may join the Steady population during the MS. However, if magnetic and tidal interactions are efficient enough later on (in the case of low stellar and planetary masses; see panel D in Fig. \ref{Tracks}), these can still undergo a significant migration during the MS of their host star, thus becoming Old Migrators (our third migration population; see Fig. \ref{Tracks}). Situated at high rotation periods and low orbital periods, the latter population presents the highest sensitivity to the physical parameters of our model, such as the planetary mass and the stellar mass. Their location in the (\(P_\text{orb}, P_\text{rot}\)) plane is that of the most efficient angular momentum transfer from the planet to the star, as is made more explicit below.\\

The evolution of the three aforementioned populations allows us to define different regions in the (\(P_\text{orb}, P_\text{rot}\)) plane. First, a depleted area is visible at low orbital and rotation periods \citep{teitler,benbakoura}. Such a region becomes larger for high stellar and planetary masses, up to orbital periods of 6 days and rotation periods of 2 days in our sample (see e.g., panel C). The depopulation of this region is due to the rapid engulfment of Young Migrators within the co-rotation radius, as well as the efficient outward migration of individuals from the same population initially beyond \(r_\text{corot}\). As the dynamical tide dominates the evolution of the systems in this region, the efficiency of this interaction has a direct influence on the extension of the depleted area.

Furthermore, Old Migrators are able to efficiently transfer their orbital angular momentum to their host star. As seen in panel C of Fig. \ref{Tracks}, such planets, spiraling inwards, spin up the star and lead to a break in the gyrochronology \citep[e.g.,][]{gallet18,benbakoura}, the highest rotation periods remaining unreached \citep{gallet19}. Such a deserted area at high rotation periods and low orbital periods, only visible for the most massive planets, is extended for high stellar masses (panel C). Then, star--planet systems with an orbital period of less than 3 days and a stellar rotation period of more than 20 days cannot appear, because of the spin-up of the host star. Such a process is driven by magnetic interactions (in blue) for the least massive stars of our sample and by the equilibrium tide (in red) for their massive counterparts. 

In addition, the region in the (\(P_\text{orb},\ P_\text{rot}\)) plane populated by Young and Old Migrators defines an area of influence of star--planet interactions. Such a region is the most extended in the case of less massive planets orbiting K-type stars (panel D). There, planets with orbital periods up to 10 days may undergo some migration through the magnetic torque. Indeed, the enhancement of stellar magnetism and stellar wind leads to strong star--planet magnetic interactions. In addition, for giant planets orbiting K-type stars (panel E) and super-Earths orbiting G-type stars (panel B), the range in orbital periods of star--planet interactions is entirely defined by the magnetic torque, as it favors the migration of more distant planets around slower rotators (cf. \S 3.1.2 and \S 3.2.1).

Therefore, a larger number of planets from our sample undergoes slow migration, as the associated timescale is of the order of 1 Gyr. Furthermore, in the case of super-Earths orbiting slowly rotating G-type stars (panel B), the area of influence of star--planet interaction is dictated by the equilibrium tide, as the closest planets are engulfed by the star. If the planet and the star are both massive (panel C), the most distant planets undergoing star--planet interactions are Young Migrators, moving away from the star under the action of the dynamical tide. The area of influence strongly depends on the initial stellar rotation in this case. It can extend up to an orbital period of 2 days for \(P_\text{rot} = 10\) d, and beyond 10 days for \(P_\text{rot} = 0.2\) d. Magnetic effects then expand the area of influence of star--planet interactions to slower rotators, for which the dynamical tide is less efficient. It is therefore necessary to consider magnetic interactions to determine if a planet is likely to undergo a migration, because in most cases they affect the area of influence of star--planet interactions.

\subsection{Influence on the global distributions}

\begin{figure}[!h]
   \begin{center}
    \includegraphics[scale=0.4]{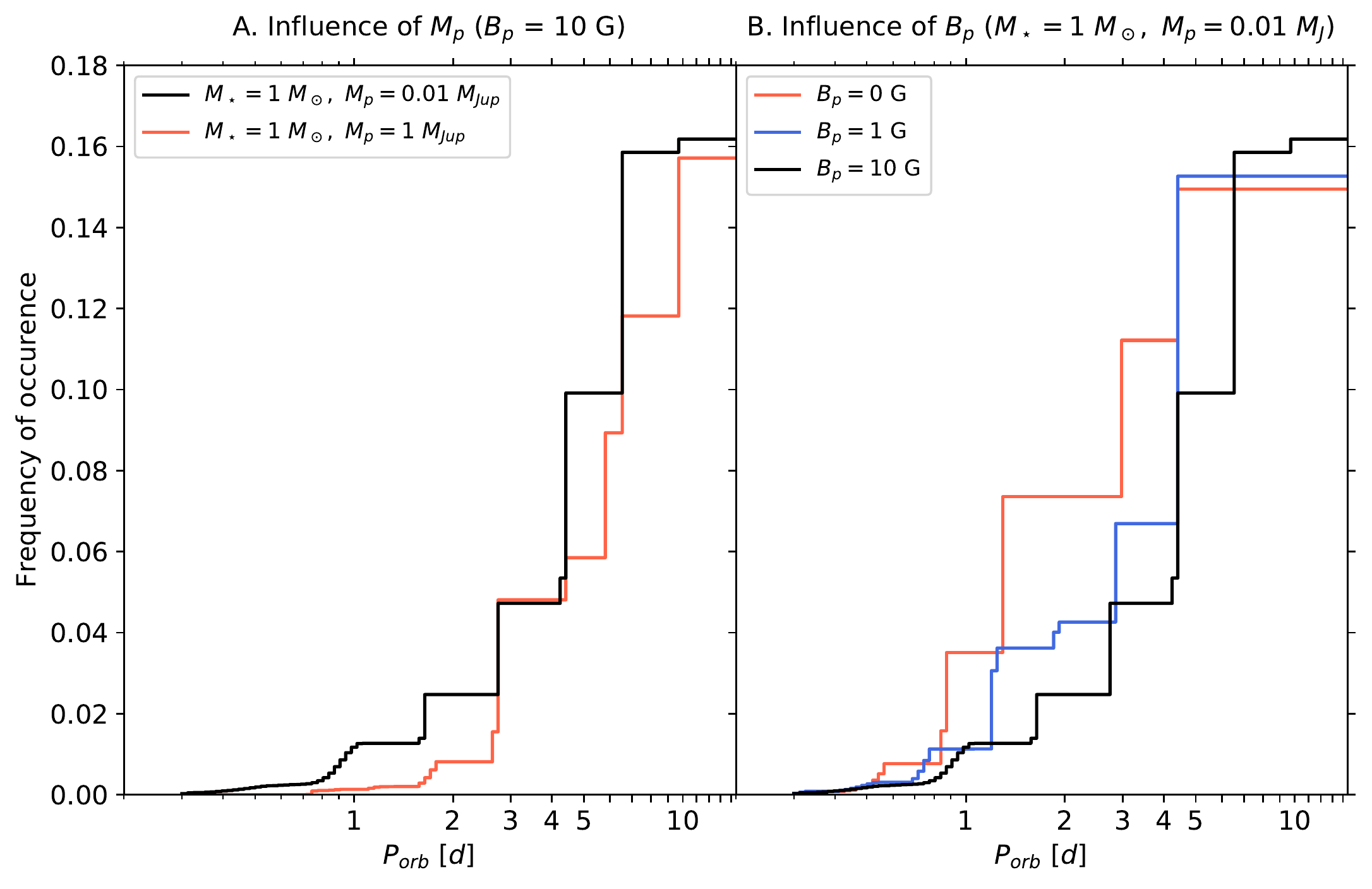}
   \end{center}
   \caption{\label{Tracks_Porb} Panel A: Global distribution in orbital periods for \(M_p =  0.01\ M_\text{Jup}\) in black, and \(M_p = 1\ M_\text{Jup})\) in red. The planetary magnetic field is fixed at 10 G and the stellar mass at \(1\ M_\odot\). Panel B: Global distribution in orbital periods for \(B_p = 0\ \text{G}\) (in red), \(B_p = 1\ \text{G}\) (in blue), and \(B_p = 10\ \text{G}\) (in black). A star of mass \(1\ M_\odot\) and a planet of mass \(0.01\ M_\text{Jup}\) are considered here.}
\end{figure}

The existence of different planetary populations with diverse evolution paths is likely to influence the global distributions of stellar rotation periods and orbital periods in our ESPEM sample. In order to highlight the role played by the different physical parameters of our model, we represent in Fig. \ref{Tracks_Porb} the global distribution in \(P_\text{orb}\) for two different planetary masses (see panel A in Fig. \ref{Tracks_Porb}) and three different planetary magnetic fields (see panel B in Fig. \ref{Tracks_Porb}). These distributions are obtained by considering 2,000 evolutionary states between the dissipation of the disk and the terminal-age main sequence (TAMS) of our solar-mass models and five initial stellar rotations. \\

Planetary mass may affect the distribution in orbital periods of the sample significantly. Indeed, as \(M_p\) increases, a steeper \(P_\text{orb}\) distribution is observed (in red in the panel A of Fig. \ref{Tracks_Porb}), as planetary migration becomes more efficient with higher $M_p$ (see \S \ref{sec:planetparams}). Therefore, less close-in planets are likely to be detected due to higher engulfment rate.
\begin{figure}[!h]
   \begin{center}
    \includegraphics[scale=0.5]{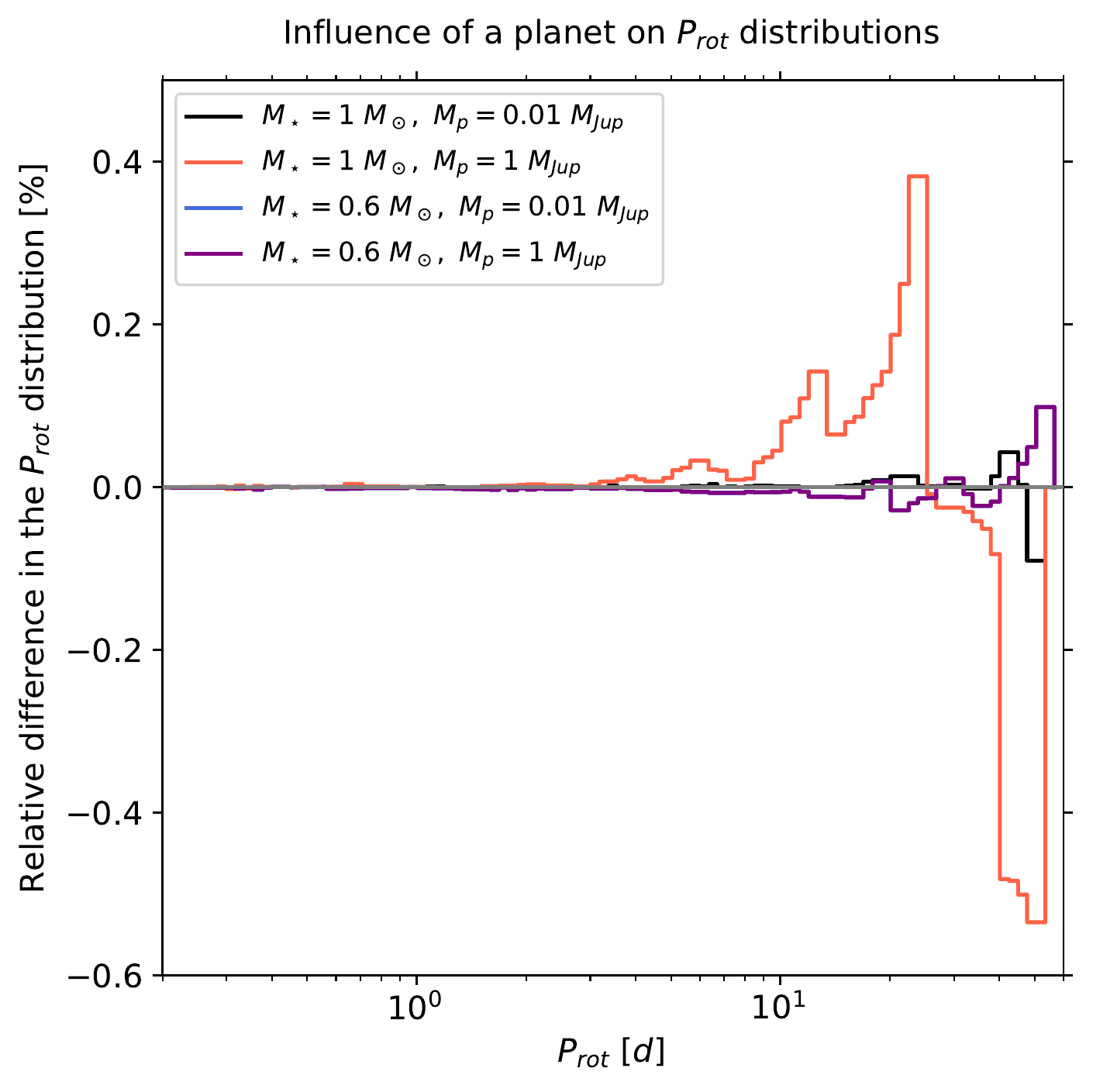}
   \end{center}
   \caption{\label{Tracks_Prot} Relative difference in the $P_\text{rot}$ distribution due to the presence of a planet, for \((M_\star,M_p) = (1\ M_\odot,\ 0.01\ M_\text{Jup})\) in black, \((M_\star,M_p) = (1\ M_\odot,\ 1\ M_\text{Jup})\) in red, \((M_\star,M_p) = (0.6\ M_\odot,\ 0.01\ M_\text{Jup})\) in blue, and \((M_\star,M_p) = (0.6\ M_\odot,\ 1\ M_\text{Jup})\) in violet.}
\end{figure}  
A similar effect is observed by increasing the planetary magnetic field (panel B). Indeed, this directly increases the intensity of the magnetic torque, which reinforces the migration of close-in planets. As seen in panel B of Fig. \ref{Tracks_Porb}, if \(M_\star = 1\ M_\odot\) and \(M_p = 0.01\ M_J\), an increase of one order of magnitude of \(B_p\) induces a stronger depopulation only at low orbital periods. However, under more favorable conditions, a stronger planetary magnetic field may favor the repopulation of the same region by more distant planets, as is the case for the lowest stellar masses (see panel D in Fig. \ref{Tracks}).\\

Star--planet interactions may also influence the distribution of \(P_\text{rot}\) in our sample. Following the same analysis, Fig. \ref{Tracks_Prot} shows the percentage of stars with modified rotation as a function of stellar rotation period. In accordance with the results of the previous section, the transfer of angular momentum from the planetary orbit to the star is favored for giant planets orbiting the most massive stars. Thus, the most significant influence of star--planet interactions on the \(P_\text{rot}\) distribution is due to the spin-up of the more massive stars, which depopulates rotation periods longer than 20 days in favor of \(P_\text{rot}\) values between 8 and 20 days (in red in Fig. \ref{Tracks_Prot}), corresponding to stellar ages ranging between 80 Myr and 2 Gyr. However, such effects affect at most 0.5 \% of the (\(1\ M_\odot,\ 1\ M_\text{Jup}\)) ESPEM subsample, which represents 0.07 \% of the whole sample. Indeed, such a configuration requires the engulfment of massive planets, thus reducing its probability of occurrence. For \(M_p = 0.6\ M_\odot\), only 0.1 \% of the corresponding subsample (i.e., 0.01 \% of the total ESPEM sample) undergoes modified rotation for \(P_\text{rot} \geq 30\) d (corresponding to systems older than 5 Gyr). We can therefore consider that star--planet interactions essentially impact the distribution in orbital periods, the distribution in stellar rotation periods being marginally affected here.

\section{Building a synthetic population of exoplanets}
\subsection{Observational data}
Knowing the possible evolutions of planetary populations as well as the influence of our model physical parameters on the distributions of orbital and stellar rotation periods, we now aim to confront our results with the statistics of observed star--planet systems. To do so, we focus on data from the \textit{Kepler} mission, and more specifically on two studies performed by \citet[][hereafter MMA13]{MMA13} and \citet[][hereafter MMA14]{MMA14}.

The MMA14 study provides the largest homogeneous rotation dataset in the Kepler field to date, involving F-type to M-type stars; it includes 34,030 Kepler MS stars whose rotation period has been measured through an autocorrelation-based method, which represents 25.6\% of the 133,030 MS Kepler targets detected at that time (excluding known eclipsing binaries and Kepler objects of interest). In this sample, only effective temperatures lower than 6500 K are considered in order to keep only solar-type stars with convective envelopes.
\begin{figure}[!h]
   \begin{center}
    \includegraphics[scale=0.6]{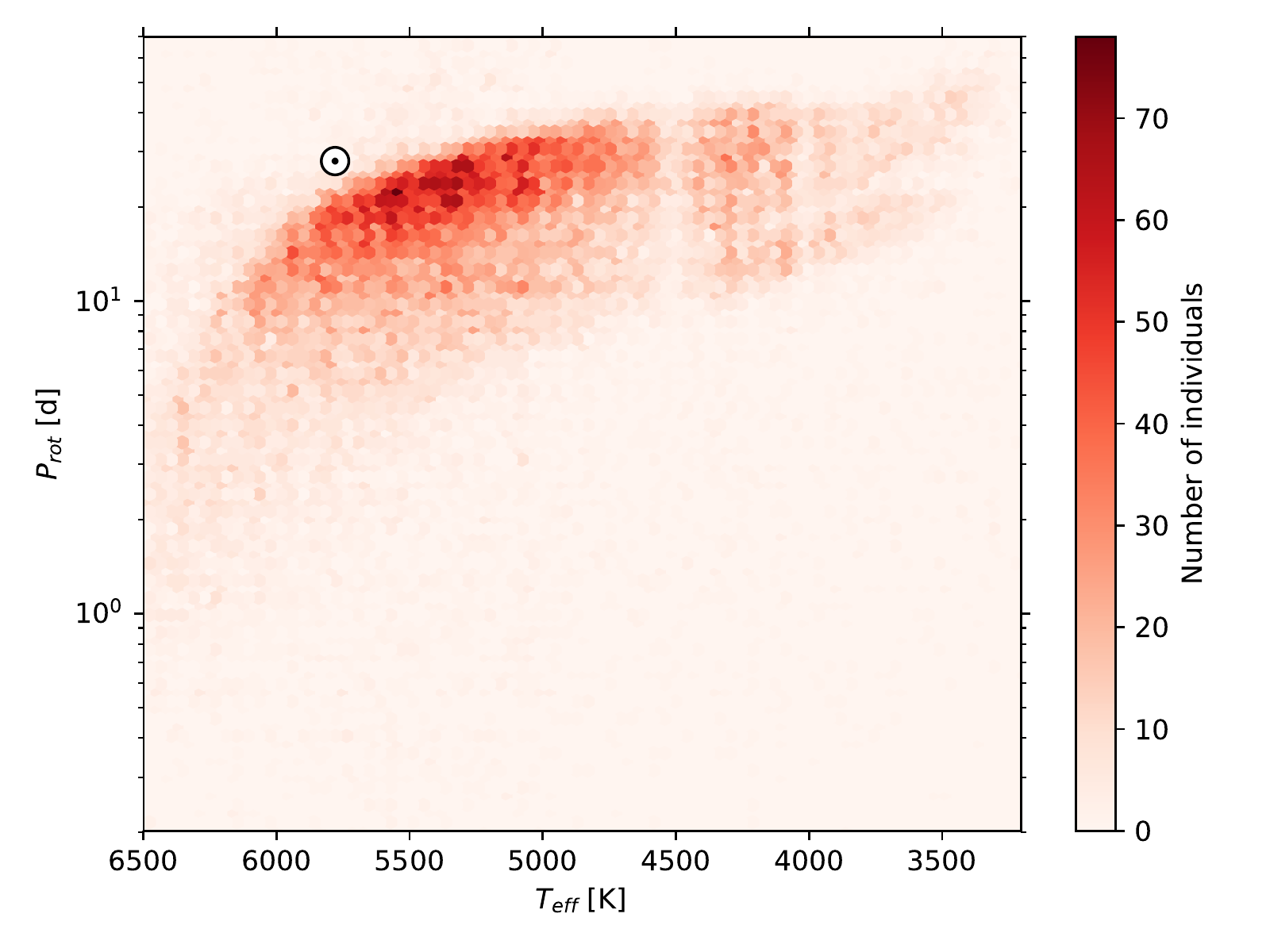}
   \end{center}
   \caption{\label{ProtMMA14} Observed distribution of rotation periods from MMA14 as a function of the effective temperature. The black circle correspond to the solar values.}
\end{figure} 
We reproduce the \(P_\text{rot}\) distribution of the MMA14 targets as a function of their effective temperature in Fig. \ref{ProtMMA14}. The MMA14 sample can be seen to present a concentration of stars with rotation periods ranging between 10 and 40 days and effective temperatures between 4000 and 6000 K. Thus, slow-rotating F-type and G-type stars are the most represented in this sample. Furthermore, the distribution of rotation periods presents an upper envelope that increases with decreasing temperature. The Sun is situated close to this upper envelope, which means that few stars older than the Sun have been detected. Indeed, more slowly rotating stars in the MS would not have been removed by exclusion processes of evolved stars. This dataset is used in the remainder of this study to design a synthetic stellar population taking into account the observational biases of the Kepler field as well as the potential influences of the stellar distribution in the galaxy.

\begin{figure}[!h]
   \begin{center}
    \includegraphics[scale=0.45]{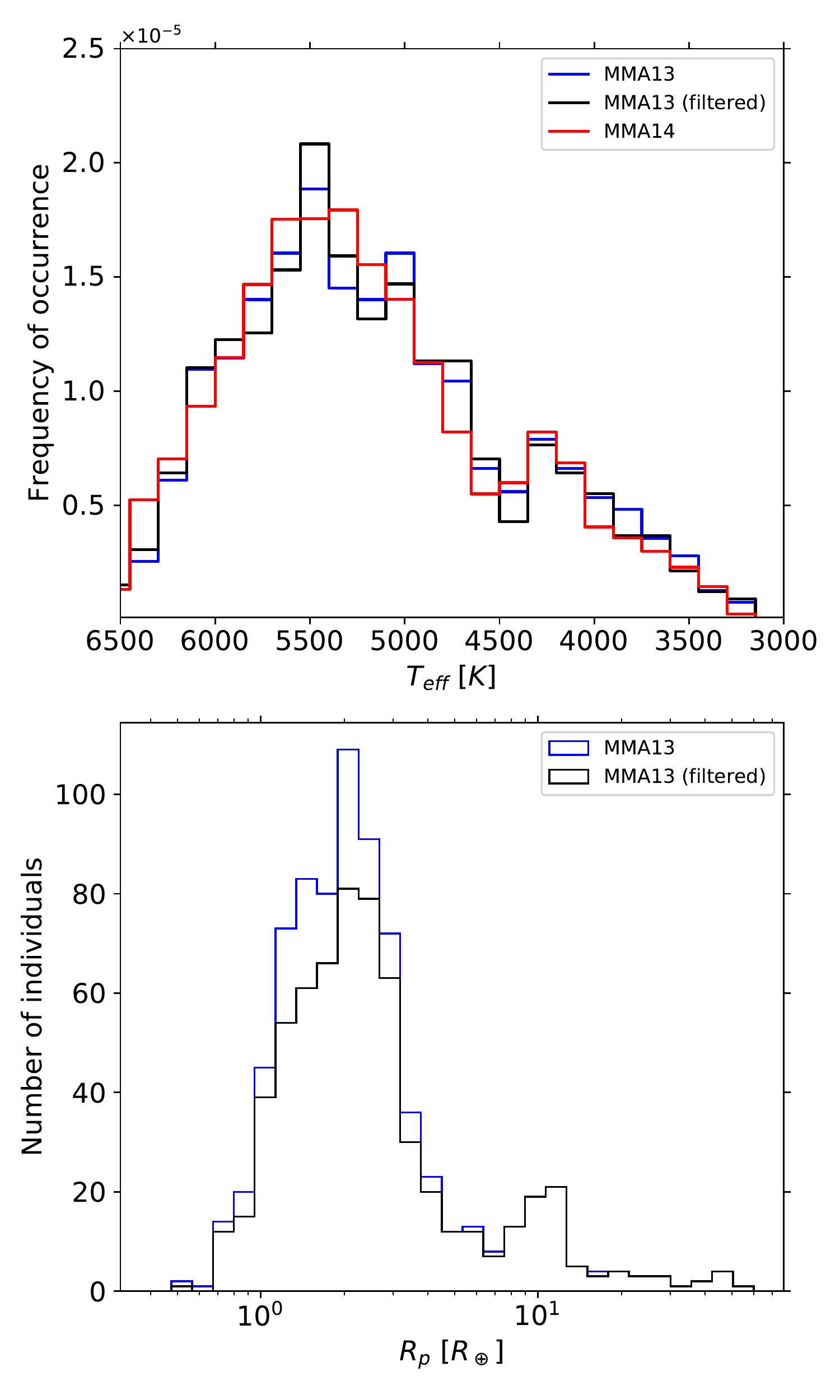}
   \end{center}
   \caption{\label{MMA1314} Top: Distribution in effective temperature from MMA14 (in red), MMA13 (in blue), and a subsample of MMA13 excluding multiplanetary systems (in black). Bottom: Distribution in planetary radii from MMA13 (in blue) and a subsample of MMA13 excluding multiplanetary systems (in black).}
\end{figure}

MMA13 is one of very few studies to date that combine the orbital period of detected exoplanets and the stellar rotation of their host star, making the associated dataset a valuable tool in the study of star--planet interactions. The main highlight of the study was the observed lack of close-in planets around fast rotators \citep[we also refer the reader to][]{pont}. The MMA13 dataset includes 737 KOI with detected orbital and stellar rotation periods. The targets are all in the MS, and were selected using the same processes as in the MMA14 sample.  Thus, comparing these two studies allows us to keep identical target selection processes as well as detection methods between the populations of isolated and planet-hosting stars. Moreover, for the sake of consistency, as our model does not deal with interactions between several planets, we do not take into account multiplanetary systems in the MMA13 dataset. Therefore, in the remainder of this work we aim to compare ESPEM results with the  MMA13 sample which has been filtered to exclude detected multiplanetary systems. As seen in the top panel of Fig. \ref{MMA1314}, the distribution in stellar effective temperatures is marginally affected by the choice of dataset (MMA13 vs. MMA14) as well as the removal of multiplanetary systems (black line). Therefore, we can rely on the \(T_\text{eff}\) distribution from the MMA14 study to design a stellar population and compare the ESPEM results to the filtered MMA13 sample without adding significant biases. In addition, as shown in the bottom panel of Fig. \ref{MMA1314}, the initial MMA13 sample as well as its filtered analog mostly contain planets with radii between 0.6 and 8 \(R_\oplus\), corresponding to planetary masses ranging between 0.2 and 72 \(M_\oplus\) according to \citet{chen} conversion laws (cf. Eq. \eqref{eqn:RMrel}). Thus, low-mass planets are extensively represented in these samples.

\subsection{Synthetic populations and \textit{Kepler} observations: comparison of global distributions}

In the remainder of this work, we aim to generate a synthetic population of star--planet systems out of our whole sample of ESPEM simulations, and then to compare the distributions we obtain with the filtered MMA13 dataset. To do so, we first design a realistic population of stars based on the MMA14 study. For the sake of consistency, as we assume a two-layer structure for the star, we consider stellar masses ranging between 0.5 and 1.1 \(M_\odot\) with a bin size of 0.1 \(M_\odot\). This corresponds to effective temperatures of between 3700 and 6000 K, which allows us to cover most of the MMA14 sample. The associated surface gravity $\log g$ ranges between 4.25 and 5.07, which is consistent with the selection processes used in MMA13 and MMA14 to exclude likely giants \citep{ciardi}. Furthermore, in order to account for the rotational evolution of stars in open clusters \citep{gallet15}, in our synthetic populations generated with ESPEM we consider  five initial rotation periods evenly distributed between 1 and 10 days. We then rely on a common sampling of stellar age for all evolutionary tracks in order to remove biases due to the adaptative time-step used in the ESPEM integrator \citep[for more details, we refer the reader to][]{benbakoura} as well as to account for the relative duration of the different stages of stellar evolution for each stellar mass. More precisely, stellar age is sampled between 1 Myr and 10 Gyr to obtain  2,000 equally spaced instants. Then, for a given star, only stellar ages that fall in the MS are taken into account. To make sure that, in the absence of planets, our stellar sample reproduces the MMA14 stellar distribution, we introduce a coefficient \(C_\text{stellar}(P_\text{rot},T_\text{eff})\) to weight a given star--planet system at a given age according to the rotation period of the host star as well as its effective temperature (for more details, we refer the reader to Appendix A). This allows us to indirectly  bias the stellar masses and ages of the systems according to the observations of the \textit{Kepler} mission.

We then include planets by considering 40 initial semi-major axes ranging between \(5 \times 10^{-3}\) and 0.2 AU, evenly spaced in logarithm. The corresponding orbital periods are then situated between 0.12 and 46 days, which allows us to initially populate all the orbital distances for which star--planet interactions may act efficiently. Five planetary masses between 0.5 Earth masses and 5 Jupiter masses, uniformly spaced in logarithm, are considered in this planetary population. This corresponds to planetary radii of between 0.8 and 12.7 \(R_\oplus\), thus covering the majority of the planets considered by the MMA13 study, as seen in Fig. \ref{MMA1314}. Each planetary mass is weighted in our population in order to reproduce the \(R_p\) distribution of the filtered MMA13 subsample. Moreover, three configurations are investigated for a given population of planetary systems: we make it evolve through tidal effects only, or along with magnetic effects by considering \(B_p = 1\ \text{G}\) and \(10\ \text{G}\). As the initial configuration of the systems and the engulfment ratio are unknown in the MMA13 study, we only take  into account the planets that survived. This allows us to define a density of probability of presence \(\text{DPP}_\text{ESPEM}(P_\text{orb},P_\text{rot})\). The probability \(\text{d}\mathbb{P}_\text{ESPEM}\) of finding a star--planet system with an orbital period included within the interval \([P_\text{orb},P_\text{orb}+\text{d}P_\text{orb}]\) and a stellar rotation period in \([P_\text{rot},P_\text{rot}+\text{d}P_\text{rot}]\) is then defined as
\begin{equation}
\text{d}\mathbb{P}_\text{ESPEM}(P_\text{orb},P_\text{rot}) = \text{DPP}_\text{ESPEM}(P_\text{orb},P_\text{rot}) \text{d}P_\text{orb}\text{d}P_\text{rot}.
\end{equation}

To compare these results with the observed distributions, we assume that each individual from the ESPEM sample corresponds to a detected star--planet system. This way, the probability $\mathbb{P}_\text{ESPEM}$ can be interpreted as a distribution of detected systems assuming that only star--planet interactions affect the initial population. However, several biases may be involved in the MMA13 distributions (such as the probability of transit, which is equal to $R_\star/a$, and also biases linked to the selection of KOIs as well as the detectability of stellar rotation period). Thus, synthetic distributions may show a larger number of systems with increasing orbital period when compared with the corresponding observed planet population. Hence, to compare consistently synthetic and observed distributions, we first consider a subsample of the filtered MMA13 sample, for which orbital periods are shorter than 20 days. This way, we only focus on close-in planets, for which our study is relevant. Each system is then weighted by the inverse of the probability of transit $a/R_\star$ to remove the associated bias.  In the remainder of \S 5.2, both synthetic and observed distributions are normalized to 1. Finally, we systematically quantify the differences between the planetary global distributions by performing a two-sample Kolmogorov-Smirnov \citep[KS,][]{kolmogorov,smirnov} test . The corresponding statistics of the test and $p$-values are presented in Table \ref{tab:KS_global} of Appendix C.

\begin{figure}[!h]
   \begin{center}
    \includegraphics[scale=0.35]{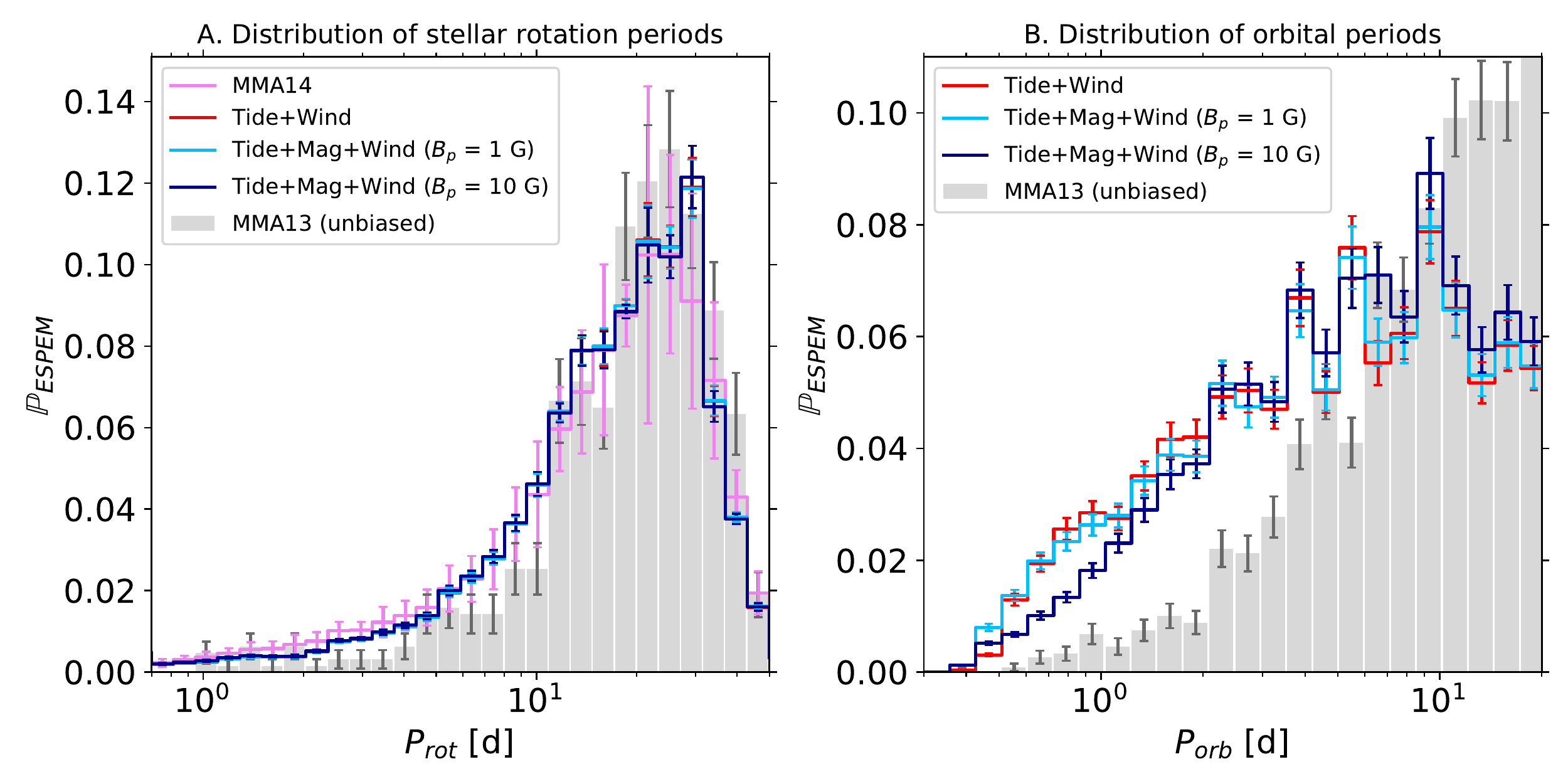}
   \end{center}
   \caption{\label{globaldis} Distribution of stellar rotation periods (left) and orbital periods (right). The \(P_\text{rot}\) distribution from the MMA14 study is shown in violet. The gray bars correspond to the distributions obtained with the unbiased MMA13 sample. The ESPEM distributions with \(B_p = 0,1,10\) G are shown in red, light blue and dark blue respectively. The calculation of the error bars for the synthetic distributions is presented in Appendix A.}
\end{figure}  
The probability of finding a planet-hosting star at a rotation period \(P_\text{rot}\), regardless of the position of the planet, is presented in panel A of Fig. \ref{globaldis}. By construction, the synthetic populations produced by ESPEM for isolated stars reproduce the MMA14 distribution (in violet in panel A of Fig. \ref{globaldis}). In the presence of a planet, the ESPEM distributions show fewer fast rotators (for which \(P_\text{rot} < 10\) d) and more stars whose rotation period ranges between 10 and 30 days. Hence, they tend to reproduce the MMA13 distribution (shown in gray in panel A of Fig. \ref{globaldis}). Moreover, taking into account magnetic torques for a planetary magnetic field \(B_p = 1,\ 10\) G marginally affects the \(P_\text{rot}\) distributions. This implies that only tidal effects are likely to play a role in the distribution of stellar rotation periods. Furthermore, no significant difference in the KS statistic and the corresponding $p$-value is observed between the synthetic populations (we refer the reader to Appendix C for more details). Such a discrepancy between the \(P_\text{rot}\) distribution of isolated stars and planet-hosting stars cannot be explained by the alteration of stellar rotation through star--planet interactions, whether the planet is detected or not. Indeed, as seen in \S 4.3, planetary migration marginally affects the distribution in stellar rotation periods; the most favorable case of a massive star and a massive planet emphasizes the lowest periods of rotation (see the red curve in Fig. \ref{Tracks_Prot}). We can therefore conclude that the engulfment of planets orbiting fast rotators through tidal effects makes the detection of these stars less likely, favoring slower rotators in the \(P_\text{rot}\) distribution of MMA13.

The distributions in orbital periods, regardless of stellar rotation, are shown in panel B of Fig. \ref{globaldis}. Taking into account star--planet magnetic interactions, in particular for \(B_p\) = 10 G (see the dark blue curve in panel B), significantly affects the \(P_\text{orb}\)  distributions by depopulating orbital periods shorter than 2 days. Such a modification then makes it possible to approach the MMA13 distribution for \(P_\text{orb} < 1\) d (see the gray histogram in panel B) with a confidence level of 1$\sigma$ (we refer the reader to Appendix A for more details about the calculation of the error bars), where the populations for \(B_p = 0\) and 1 G show a more significant excess of planets at these orbital periods compared to observations (red and light blue curves in panel B). More precisely, taking into account star--planet magnetic interactions with a stronger planetary magnetic field tends to improve the goodness of fit by lowering the value of the test statistic from 0.32 to 0.27 and increasing the corresponding $p$-value from 0.17 to 0.34 (we refer the reader to Table \ref{tab:KS_global} for more details). For orbital periods of greater than 10 days, the ESPEM distributions become uniform, as star--planet interactions are negligible for remote planets. In any case, taking the magnetic torque for large values of \(B_p\) into account is necessary to account for the distribution of close-in planets. Such a scenario seems to be supported by planetary dynamo considerations. Indeed, as planetary rotation is synchronized in our model, close-in planets may rotate with periods shorter than 1 day. They are therefore likely to generate high magnetic fields. Moreover, \citet{cauley19} inferred values of \(B_p\) of between 20 G and 120 G for planets whose orbital period ranges between 2 and 4 days. Such a configuration would lead to even stronger modifications of the \(P_\text{orb}\) distribution than what is presented in this work. Furthermore, as we consider a sufficiently wide range of initial semi-major axes, corresponding to orbital periods of between 0.12 and 46 days, the initial conditions adopted to generate our synthetic populations have no influence on the planetary distributions at low orbital periods.  Despite a better goodness of fit at high planetary magnetic fields, we see an excess of exoplanets for all the synthetic populations at low orbital periods compared to the observed distribution. This may highlight that other star--planet interactions not considered in this work, or the initial distribution of orbital periods resulting from planetary formation, may have played a role in shaping the observed population from MMA13.

\subsection{Synthetic populations and \textit{Kepler} observations: comparison of young, middle-aged, and old star--planet systems}

\begin{figure}[!h]
   \begin{center}
    \includegraphics[scale=0.35]{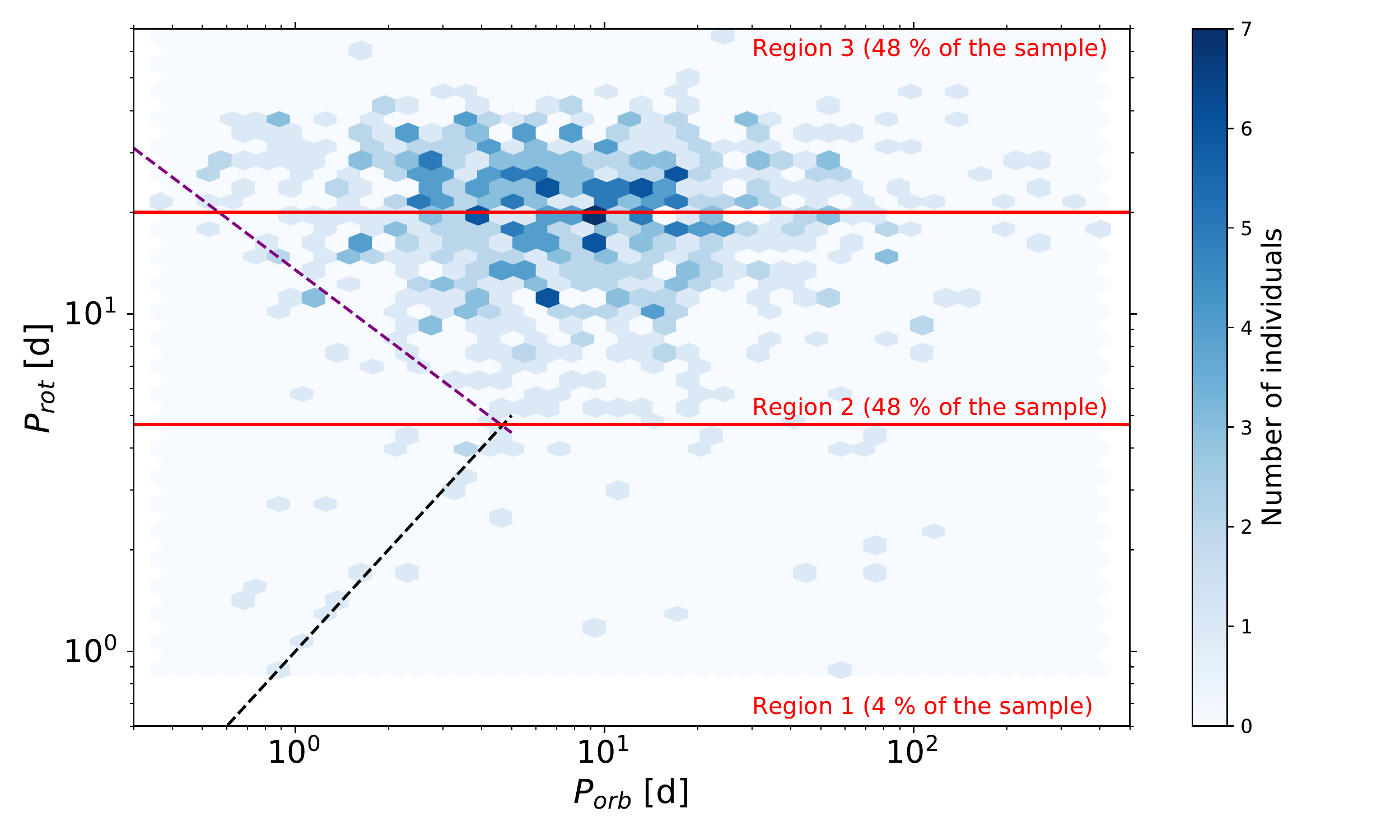}
   \end{center}
   \caption{\label{MMA13_2D} Observed distribution of stellar rotation periods from the whole MMA13 sample as a function of the orbital period of the planet. The purple dashed line corresponds to the lower edge of
the distribution fitted in MMA13. The black dashed line corresponds to the synchronization between \(P_\text{orb}\) and \(P_\text{rot}\).}
\end{figure}

We now turn to the detailed distributions of star--planet systems at different ages. To do so, we rely on the \((P_\text{orb},P_\text{rot})\) distribution of the whole MMA13 sample shown in Fig. \ref{MMA13_2D}, which allows us to define three regions. For stars rotating with a period shorter than around 4.7 days (the so-called Region 1 in Fig. \ref{MMA13_2D}), few planets are detected. Moreover, in Region 2, corresponding to stellar rotation periods ranging between 4.7 and 20 days, the detected exoplanets have an orbital period greater than around 1 day. Finally, a wide range of orbital periods, namely between 0.4 and 500 days, is observed for exoplanets orbiting around  the slowest rotators (\(P_\text{rot} > 20\) d; see the Region 3 in Fig. \ref{MMA13_2D}).

\begin{figure*}[!h]
   \begin{center}
    \includegraphics[scale=0.38]{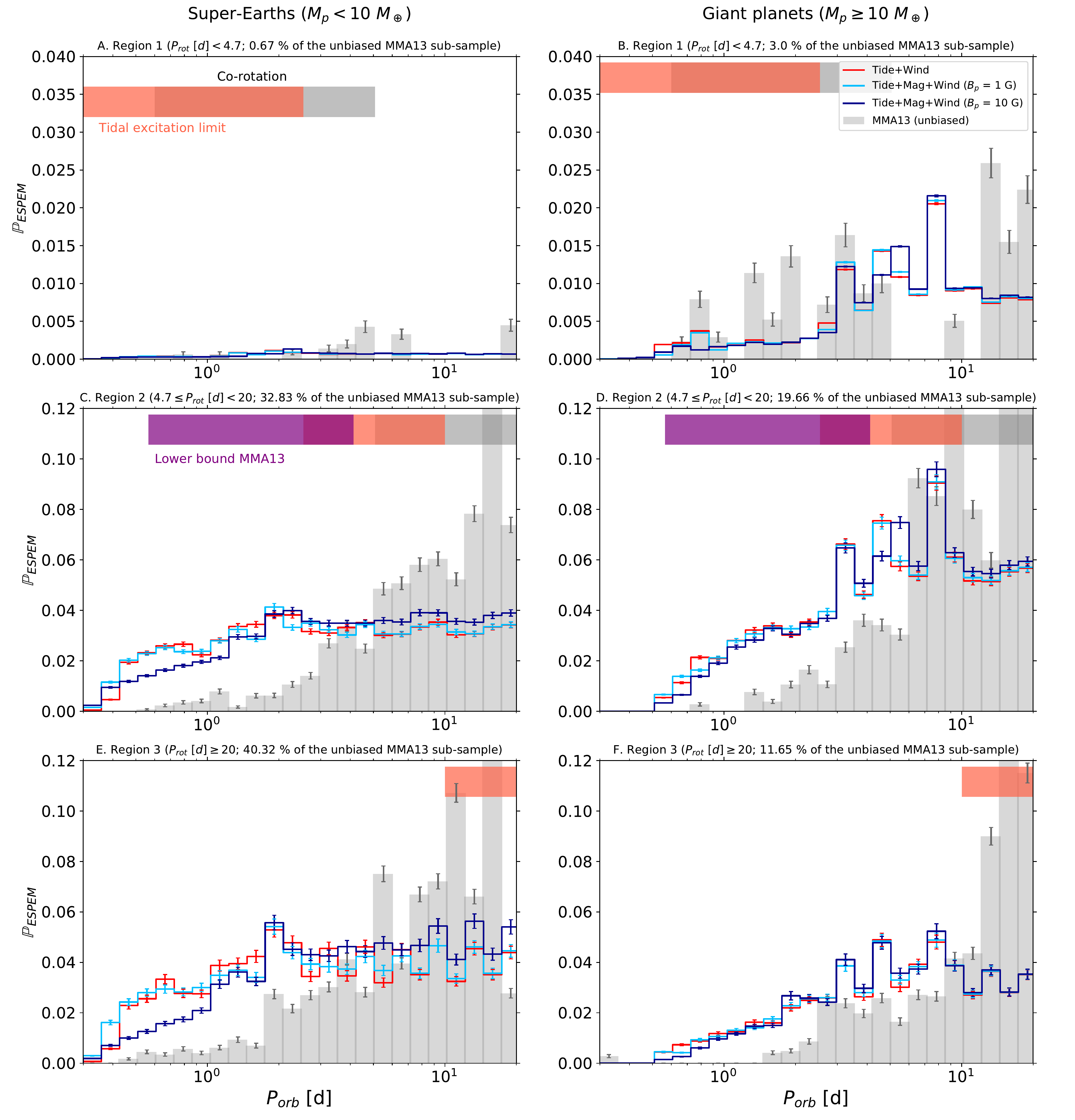}
   \end{center}
   \caption{\label{stat} Distribution of orbital periods for super-Earths (left column) and giant planets (right column). Each row corresponds to a population of young (Region 1, for which \(P_\text{rot} < 4.7\) d), middle-aged (Region 2, for which \(4.7 \leq P_\text{rot}\ [\text{d}] < 20\) ), and old (Region 3, \(P_\text{rot} \geq 20\) d) star--planet systems. The gray histogram corresponds to the distributions obtained with the unbiased MMA13 sample. The ESPEM distributions with \(B_p = 0,1,10\) G are shown in red, light blue and dark blue respectively. The gray, red, and purple horizontal bands at the top of each panel correspond to the range of values in each region of the co-rotation, the dynamical tide excitation limit, and the lower edge of the MMA13 distribution, respectively. The calculation of the error bars for the synthetic distributions is presented in Appendix A.}
\end{figure*}

For each of these regions, each corresponding to a row in Fig. \ref{stat}, we compare the distributions of the synthetic populations to their MMA13 counterparts in the case of super-Earths (\(M_p < 10\ M_\oplus\), see the first column of Fig. \ref{stat} ) and giant planets (\(M_p \geq 10\ M_\oplus\), see the second column of Fig. \ref{stat}). The observed star--planet systems are selected and unbiased with the same method as in \S 5.2. Both synthetic and observed distributions of super-Earths and giant planets for all the stellar rotation periods considered are then normalized to 1.

In the case of fast rotators (panels A and B), Young Migrators have already migrated outwards or been engulfed, depending on their initial position relative to the co-rotation orbit (see the horizontal gray bands at the top of each panel in Fig. \ref{stat}). The frequency of occurrence then may be higher for orbital periods longer than 3 days for the most massive planets (panel B), and longer than around 1 day in the case of super-Earths, as star--planet interactions are less efficient (panel A). Taking into account star--planet magnetic interactions has a marginal influence on the distribution of giant planets, as tidal interactions dominate secular evolution. However, a slight shift towards higher orbital periods is observed if \(M_p< 10\ M_\oplus\) (see red and dark blue curves in panel A), with more planets being engulfed. In those regions, the small number of observations and the low probability of occurrence in the case of synthetic populations prevent significant discrimination between the different distributions (see upper rows in Table \ref{tab:KS_local}).

In the case of giant planets located in Region 2 (panel D), no significant migration takes place for orbital periods longer than 3 days. The associated distribution is then almost uniform. When \(P_\text{orb} < 3\) d, Old Migrators get closer to the star through the action of the equilibrium tide and to a lesser extent star--planet magnetic interactions. This induces a depopulation of the lowest orbital periods, slightly enhanced by the adjunction of the magnetic torque. More precisely, the case $B_p$ = 10 G leads to slightly better KS statistic compared to the other synthetic populations. In the case of the slowest rotators (Region 3 ; see panel F in Fig. \ref{stat}), the frequency of occurrence decreases with decreasing orbital period. This can be attributed to planetary engulfment, which is at the origin of the deserted area presented in \S 4.2. Moreover, as the extension of this region decreases with stellar mass \citep[see \S 4.2; we also refer the reader to][]{gallet19}, the frequency of the occurrence of massive planets decreases smoothly with their orbital period. Such a behavior results from the action of tidal effects, with magnetic torque playing a negligible role. Moreover, we see that the probability of detecting a massive planet around slow rotators in the MMA13 sample is generally lower than what can be expected from synthetic populations. As giant planets are more likely to be detected than super-Earths, such a discrepancy might not be entirely attributed to the low number of systems detected in this region. Therefore, other migration mechanisms may be responsible for the engulfment of giant planets, thus reducing their frequency of occurrence around slowly rotating stars. For instance, the excitation of tidal gravity waves in the stellar radiative zone may affect the evolution of the system \citep{barker10,guillot14,ahuir21}. Such a contribution will be studied in future work. Furthermore, magnetic fields of strength greater than 10 G \citep{cauley15,cauley19} may also significantly affect the distributions of the giant planets by depopulating the shortest orbital periods.

Star--planet magnetic interactions significantly affect \(P_\text{orb}\) distributions in the case of low-mass planets orbiting slower rotators (panels C and E). Indeed, Older Migrators efficiently get closer to the star at the beginning of the MS thanks to the magnetic torque. The closest planets are then likely to be engulfed through the combined action of the equilibrium tide and magnetic torque. In this configuration, we recover the characteristics of the global distribution in orbital periods. Therefore, taking into account a strong magnetic torque (for \(B_p\) = 10 G; see the dark blue curve on panels C and E of Fig. \ref{stat}) significantly modifies the distributions by depopulating the orbital periods shorter than two days, and these distributions then show a better agreement with the observations of MMA13 at low orbital periods with a confidence level of 1$\sigma$. Indeed, the cases \(B_p = 0\) G and \(B_p = 1\) G present an excess of planets at periods of between 1 and 2 days. More precisely, taking into account star--planet magnetic interactions with a planetary magnetic field of 10 G  significantly improves  the KS statistic and the corresponding $p$-value compared to the other synthetic populations (we refer the reader to Table \ref{tab:KS_local} for more details). However, as in the global $P_\text{orb}$ distribution, an excess of planets at orbital periods shorter than 3 days is still noticeable.

Thus, it is possible to approach the \(P_\text{orb}-P_\text{rot}\) distribution observed in \textit{Kepler} systems (e.g., from the MMA13 study) by relying on a realistic stellar population (such as the MMA14 distribution) and star--planet interactions after the dissipation of the protoplanetary disk. More precisely, the magnetic torque tends to modify the distribution of super-Earths around slower rotators, which improves the agreement between synthetic populations and observations.\\

\section{Summary and discussions}
\subsection{Results}

In this paper, we focus on the secular evolution of star--planet systems by taking stellar magnetic braking and star--planet magnetic and
tidal interactions into account simultaneously. The two latter effects act together on planetary migration and stellar rotation. Furthermore, both interactions may dominate the other throughout secular evolution depending on the initial configuration of the system and the evolutionary phase considered. More precisely, tidal effects are found to dominate star--planet magnetic interactions for high stellar and planetary masses as well as low semi-major axis. This implies in particular that super-Earths close to their host star essentially migrate through magnetic torques. Those interactions may actually lead to a late destruction of hot super-Earths. The dynamical tide governs the evolution of planets orbiting fast rotators while slower rotators evolve through magnetic interactions. However, at high rotation periods, the equilibrium tide may become the dominating contribution during planetary engulfment. Moreover, stellar rotation may be significantly impacted at any age due to the engulfment of a close-in planet. In this configuration, an alteration of the stellar rotation period of several dozens of days may last more than a few million years. However, such events are scarce, as they require high stellar and planetary masses, or sufficiently high planetary magnetic fields in the case of super-Earths, to be truly effective.

Three populations of star--planet systems emerge from the combined action of magnetic and tidal torques. Systems undergoing negligible migration define an area of influence of star--planet interactions, which can extend for instance up to orbital periods of 10 days for super-Earths orbiting K-type stars. Furthermore, for sufficiently large planetary magnetic fields, the magnetic torque determines the extension of this region. Planets outside this influence region form our first population, which we dub a ``Steady'' population. 

The second population we identified, which we call the ``Young Migrators'', is composed of planets initially close to fast rotators that migrate efficiently during the PMS. They may either be expelled away from the star or be rapidly engulfed, which engenders a depleted region at low rotation and orbital periods. 

Finally, we identified a third population of ``Old Migrators'', composed of planets migrating inward around slower rotators, which happens during the MS. This population is more sensitive to the physical parameters involved in our modeling. They can also lead to an efficient angular momentum transfer from the planetary orbit to stellar rotation, like Young Migrators. They could induce a break of gyrochronology for high stellar and planetary masses, as the star can efficiently spin up. This population finally creates a region at high stellar rotation periods and low orbital periods not populated by star--planet systems.

Furthermore, star--planet interactions significantly impact the global distribution in orbital periods.
Indeed, for higher planetary mass and planetary magnetic fields, magnetic and tidal torques are more efficient. Low orbital periods are then more likely to be depopulated. However the global distribution in stellar rotation periods is marginally affected. Indeed, around 0.5 \% of G-type stars and 0.1 \% of K-type stars may spin up because of planetary engulfment. As a significant stellar over-rotation requires the destruction of massive planets, its probability of occurrence in a given planetary population is found to be relatively low. 

Finally, we designed synthetic populations based on observed stellar distributions (such as the MMA14 distribution). We found that star--planet magnetic interactions, after the dissipation of the disk, significantly affect the distribution of super-Earths around slower rotators, which improves the agreement between synthetic populations and observations, while tidal effects shape the distribution of giant planets. Such a result is obtained without relying on specific prescriptions for the initial semi-major axis distribution of planets after the disk dissipation. However, we found that all these populations present an excess of exoplanets at low orbital periods compared to the distribution observed in \textit{Kepler} systems (e.g., from the MMA13 study). This may indicate that additional star--planet interactions not taken into account in this work, such as the dynamical tide in the stellar radiative zone \citep[e.g.,][]{terquem,goodman,barker10} or magnetic interactions in the unipolar regime \citep{laine,lainelin}, are at play in shaping the observed close-in planet population in the Kepler field. Another possibility is that part of the observed distribution is actually already present immediately after the disk dissipation, and differs from the initial uniform distribution in $P_\text{orb}$ we assume here. A combination of the two aspects could be shaping the observed distribution, and a detailed investigation is left for future work.

\subsection{Discussions}

The results presented in this work may have strong implications on exoplanet detection. For instance, they imply that  the detection of jovian planets with orbital periods shorter than 6 days around solar twins rotating with a period shorter than 2 days is very unlikely because of the presence of a depleted area. Moreover, similar systems having an orbital period of shorter than 3 days and a stellar rotation period of longer than 20 days could hardly be detected because of potential planetary engulfment. Such a drop in the frequency of occurrence is less significant when considering weaker stellar and planetary masses. More generally, planets with orbital periods of up to 10 days orbiting stars with a rotation period shorter than 10 days are most likely to undergo migration through magnetic (and to a lesser extent tidal) interactions. In the case of slower rotators, stellar mass, initial rotation, and planetary mass have a strong impact on the possibility of planetary migration (cf. \S 4.1.2).

Moreover, regarding the influence of star--planet interactions on stellar rotation,  with ESPEM we recover the results from previous studies \citep[e.g.,][]{zhang, gallet18, benbakoura, gallet19}. Indeed, as seen in Fig. \ref{dP}, the engulfment of a jovian planet by a solar twin may lead to an alteration of the stellar rotation period of more than 90\%, which corresponds to an error of 45\% in the estimation of stellar age if we assume the Skumanich law. It is worth noting that an over-rotation of around 20\% is likely to last for a few billion years. A deviation from gyrochronology is also possible for stars hosting super-Earths as star--planet magnetic interactions may lead to their engulfment later on during the main sequence. The induced spin-up is less significant is this case. We show in this work that these effects are barely noticeable in global \(P_\text{rot}\) distributions, as at most 0.07 \% of a population of star--planet systems may see its stellar rotation significantly altered. A signature of planetary migration may nevertheless become apparent at a given age, in particular in stellar open clusters \citep{teitler,gallet18}, and confronted with alternative scenarios that have been proposed to account for the observed \(P_\text{rot}\) distribution of low-mass stars \citep{brown,vansaders}.

Theoretical models predict that planetary migration within the protoplanetary disk is more efficient than the effects considered in this
work \citep[e.g.,][]{baruteau14,bouviercebron,heller}. This leads to the idea that the planetary population achieved immediately after disk dissipation is close to being steady until later phases of stellar evolution after the main sequence. Nonetheless, we show that the magnetic and tidal interactions occurring after the dissipation of the disk have a significant influence on planetary populations during the PMS and MS. As a result, our work shows that the observed populations need to be carefully studied to derive constraints on planet formations, as both disk (not taken into account here) and post-disk migration come into play for the population on short-period orbits. 

In addition, studying planetary populations for orbital periods of less than 2 days is essential for understanding planetary systems and their evolution. Indeed, such orbital periods seem to highlight the presence of strong planetary magnetic fields (cf. \S 5.3). They may also highlight the presence of some star--planet interactions not taken into account in this work (see \S \ref{sec:perspectives}). Uncertainties remain regarding the value of the planetary magnetic field $B_p$. Indeed, this latter could reach values much higher than those assumed in this work \citep[e.g.,][]{cauley15,cauley19}. In particular, from dynamo considerations, magnetic fields as high as 4000 G are expected in hot Jupiters during the PMS \citep{hori}. This would significantly affect the planetary distributions by enhancing the depopulation of star--planet systems at short orbital periods, whether in the case of super-Earths or even giant planets. Finally, although star--planet interactions may be necessary to account for planetary populations, they only concern planets close to their host star, which are thus located below the habitable zone \citep[for an extensive study of the secular evolution of this region, we refer the reader to][]{gallet17b}.

\subsection{Perspectives}\label{sec:perspectives}
The present work is a first attempt to study the secular evolution of star--planet systems through magnetic and tidal interaction with a fully dynamical approach. We also attempt to investigate their role in shaping the distributions of planetary populations. In order to best account for the observations, we need to take these types of interactions into account in a comprehensive manner. For instance, to get a complete picture of tidal dissipation in stars, the dynamical tide in the stellar radiative
zone needs to be taken into account \citep[e.g.,][]{zahn75,goldreich89, goodman, terquem}, which is likely to compete with the dissipation of inertial waves
in convective layers \citep{ivanov13} and to affect secular evolution of star--planet systems \citep{barker10,barker11,guillot14,barker20}. Moreover, in differentially rotating stellar convective zones, tidal inertial waves may interact with mean flows at  critical layers; they can therefore either deposit or extract angular momentum from or to the surrounding fluid, which leads to exchanges of angular momentum between the star and the planet \citep{astoul20}. Finally, tides can be affected by stellar and planetary magnetic fields \citep{wei,lin18,astoul19}.

Regarding star--planet magnetic interactions, the unipolar interaction may occur for a weakly magnetized planet with a low magnetic diffusivity \citep{laine,lainelin}. Taking such a regime into account is likely to increase the magnetic torque by several orders of magnitude and to make the subsequent planetary migration even more efficient \citep{strugarek17}. New behaviors could then be expected. As an example, if the planet gets closer to its host star, the latter may significantly spin up, which may enhance the stellar magnetic field, favoring a transition from the dipolar to the unipolar regime, which has been found to be at the origin of a strong increase in magnetic torque. Planetary migration is then likely to enter a runaway regime.

Here we only considered isolated, circularized star--planet
systems. More complex geometries, for example including eccentricities, inclinations \citep{kaula61}, and dynamical interactions in multi-planet systems \citep[e.g.,][]{laskar12,bolmont15}, have to be implemented in a future work to fully characterize the PMS and MS evolution of multi-planetary systems.

\begin{acknowledgements}
We would like to thank the anonymous referee for helpful comments and suggestions regarding our work. The authors thank Emeline Bolmont for the help and information she has generously provided. We also thank Cillia Damiani and Noë Brucy-Ciaramella for fruitful discussions regarding our work. The authors acknowledge funding from the European Union’s Horizon-2020 research and innovation programme (Grant Agreement no. 776403 ExoplANETS-A) as well as the PLATO CNES funding at CEA/IRFU/DAp. A.S. and A.S.B. acknowledge funding by ERC WHOLESUN 810218 grant, INSU/PNST and CNES Solar Orbiter. A.S. and J.A. acknowledges funding from the Programme National de Plan\'etologie (PNP). S.M. and J.A. acknowledge funding by the European Research Council through the ERC grant SPIRE 647383. 
\end{acknowledgements}


\begin{appendix}
\section{Synthesis of a population of star--planet systems}
\subsection{Stellar population}
To confront the MMA13 planetary distribution with the one obtained through the action of star--planet interactions, we rely on a synthetic population of star--planet systems generated with our ESPEM code. The goal of this section is to explain in more detail the methods to generate this population. 

We first need to design a stellar population. To do so, we consider stellar masses ranging between 0.5 and 1.1 \(M_\odot\) with a bin of 0.1 \(M_\odot\). Five initial rotation periods evenly distributed between 1 and 10 days are chosen to account for the rotational evolution of stars in open clusters \citep{gallet15}. Stellar age is sampled between 1 Myr and 10 Gyr to obtain 2,000 dates equally spaced, regardless of the simulation being processed. Then, for a given star, only the stellar ages falling inside the MS are taken into account. The stellar sample is then generated thanks to ESPEM by making each star evolve without a planet. This way, the stellar rotation period evolves due to changes in stellar structure, redistribution of angular momentum inside the star, and magnetic braking. We then determine the frequency of occurrence \(f_\text{ESPEM,al}(T_\text{eff},P_\text{rot})\) of a star at a given pair \((T_\text{eff},P_\text{orb})\). By defining the same frequency of occurrence \(f_\text{MMA14}(T_\text{eff},P_\text{rot})\) for the MMA14 sample, we introduce the coefficient
\begin{equation}\label{eqn:Cstel}
C_\text{stellar} (T_\text{eff},P_\text{rot})= \frac{f_\text{MMA14}(T_\text{eff},P_\text{rot})}{f_\text{ESPEM,al}(T_\text{eff},P_\text{rot})}.
\end{equation}
Such a ratio quantifies the difference between the \(P_\text{rot}\) distribution obtained by ESPEM for isolated stars and the one observed in the MMA14 study. This difference may be due to the limitations of our rotational evolution model, to the distribution of stars in the galaxy, to observational biases, or to uncertainties on the detection of rotation periods in MMA14 \citep[for an assessment on the reliability of the MMA14 results, see][]{santos19}.

\begin{figure}[!h]
   \begin{center}
    \includegraphics[scale=0.5]{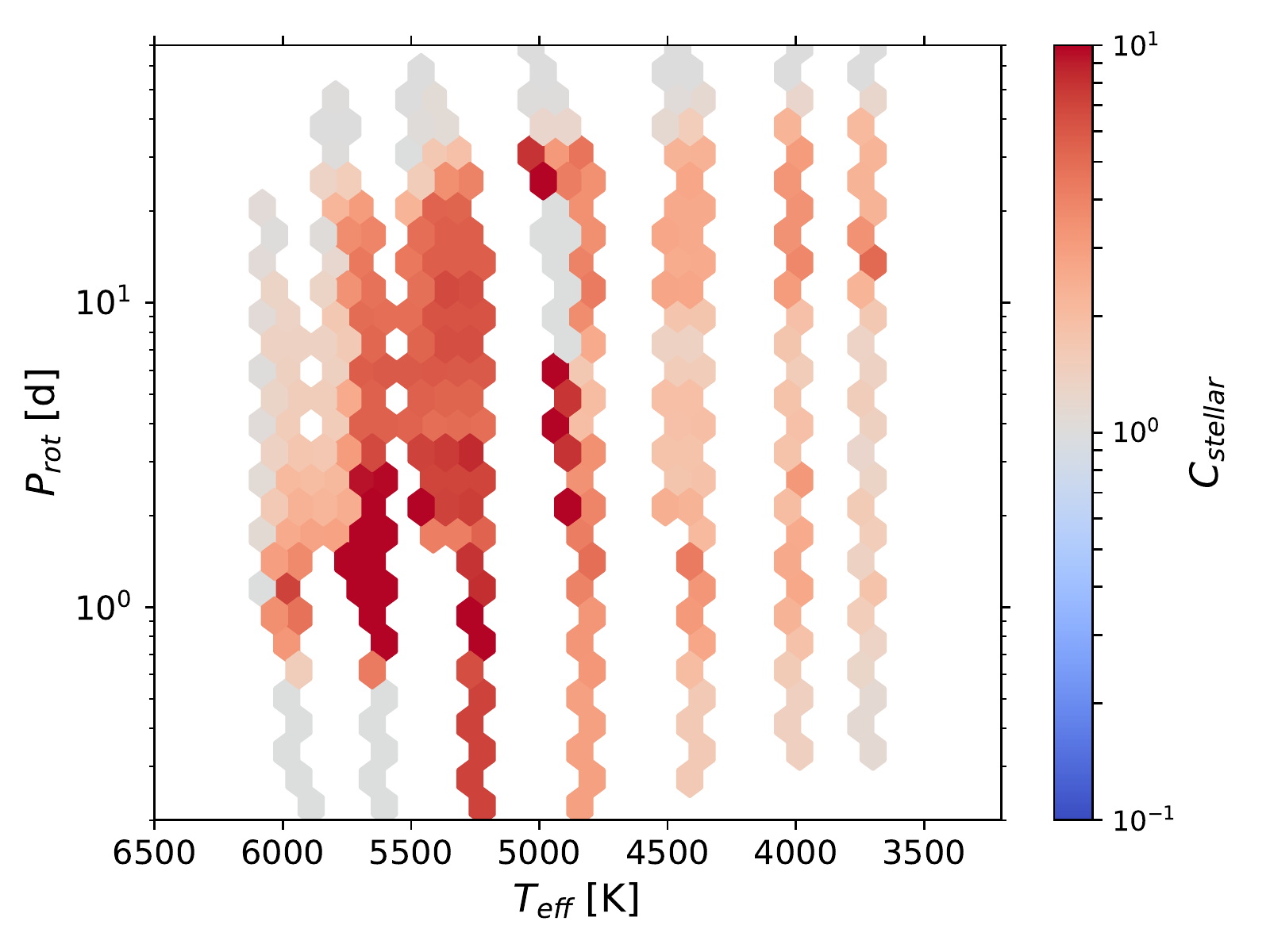}
   \end{center}
   \caption{\label{Cstel} Values of the \(C_\text{stellar}\) coefficient as a function of the stellar rotation period \(P_\text{rot}\) and the effective temperature \(T_\text{eff}\).}
\end{figure}  
As shown in Fig. \ref{Cstel}, high values of \(C_\text{stellar}(P_\text{rot},T_\text{eff})\) indicate an under-representation of the corresponding stars in the ESPEM population compared to the MMA14 distribution. In order to reproduce the latter, it is necessary to give greater weight to stars with effective temperatures between 5000 and 6000 K and rotation periods between 1 and 30 days. This then amounts to favoring the presence of G-type stars younger than the Sun within the synthetic stellar population.

We now aim to assess in a first approach the influence of the uncertainties in effective temperature and stellar rotation period on the $C_\text{stellar}$ coefficient. To this end, we consider that each star from the MMA14 sample has an effective temperature $T_\text{eff}$ and a stellar rotation period $P_\text{rot}$. We define the associated uncertainties $\sigma[P_\text{rot}]$, given in MMA14, and $\sigma[T_\text{eff}]$. For the latter quantity, we assume a typical value of 200 K \citep[we refer the reader to][for more details]{huber}.
We assume as a first approximation that each observation from MMA14 follows a 2D normal distribution $\mathcal{N}(T_\text{eff},\sigma[T_\text{eff}])\times \mathcal{N}(P_\text{rot},\sigma[P_\text{rot}])$.

We now introduce a bin in effective temperature and stellar rotation period $(T_k,P_k)$, with dimensions ($\Delta T_k, \Delta P_k$), and $p_{ik}$ the probability that the observation $i$ falls into the bin $k$. Hence we have:
\begin{equation}
\begin{split}
p_{ik} = &\left[\Phi_{T_\text{eff},\ \sigma[T_\text{eff}]}\left(T_k+\frac{\Delta T_k}{2}\right)-\Phi_{T_\text{eff},\ \sigma[T_\text{eff}]}\left(T_k-\frac{\Delta T_k}{2}\right)\right]\times\\
&\left[\Phi_{P_\text{rot},\ \sigma[P_\text{rot}]}\left(P_k+\frac{\Delta P_k}{2}\right)-\Phi_{P_\text{rot},\ \sigma[P_\text{rot}]}\left(P_k-\frac{\Delta P_k}{2}\right)\right],
\end{split}
\end{equation}
where
\begin{equation}
\Phi_{\mu,\sigma} (x) = \frac{1}{2}\left[1+\text{erf}\left(\frac{x-\mu}{\sigma\sqrt{2}}\right)\right]
\end{equation}
is the cumulative distribution function of a normal distribution with a mean $\mu$ and a standard deviation $\sigma$. The number $N_k$ of observations in the bin $k$ can now be interpreted as a sum of Bernoulli variables $B_{ik}$, equal to 1 if the observation $i$ is in the bin $k$, and 0 otherwise. By assuming that all the observations from the MMA14 sample have been performed independently, the standard deviation associated to $N_k$ becomes
\begin{equation}
\sigma [N_k] = \sqrt{\sum_i p_{ik} (1 - p_{ik})}.
\end{equation}
By introducing the number of stars $N_\text{MMA14}$ in the MMA14 sample, we obtain
\begin{equation}
\sigma[f_\text{MMA14}] = \frac{\sigma [N_k]}{\sqrt{N_\text{MMA14}}}.
\end{equation}
Next, from Eq. \eqref{eqn:Cstel}, one can assess the associated standard deviation $\sigma[C_\text{stellar}]$ as
\begin{equation}
\frac{\sigma[C_\text{stellar}]}{C_\text{stellar}} = \frac{\sigma[f_\text{MMA14}]}{f_\text{MMA14}}.
\end{equation}

\begin{figure}[!h]
   \begin{center}
    \includegraphics[scale=0.5]{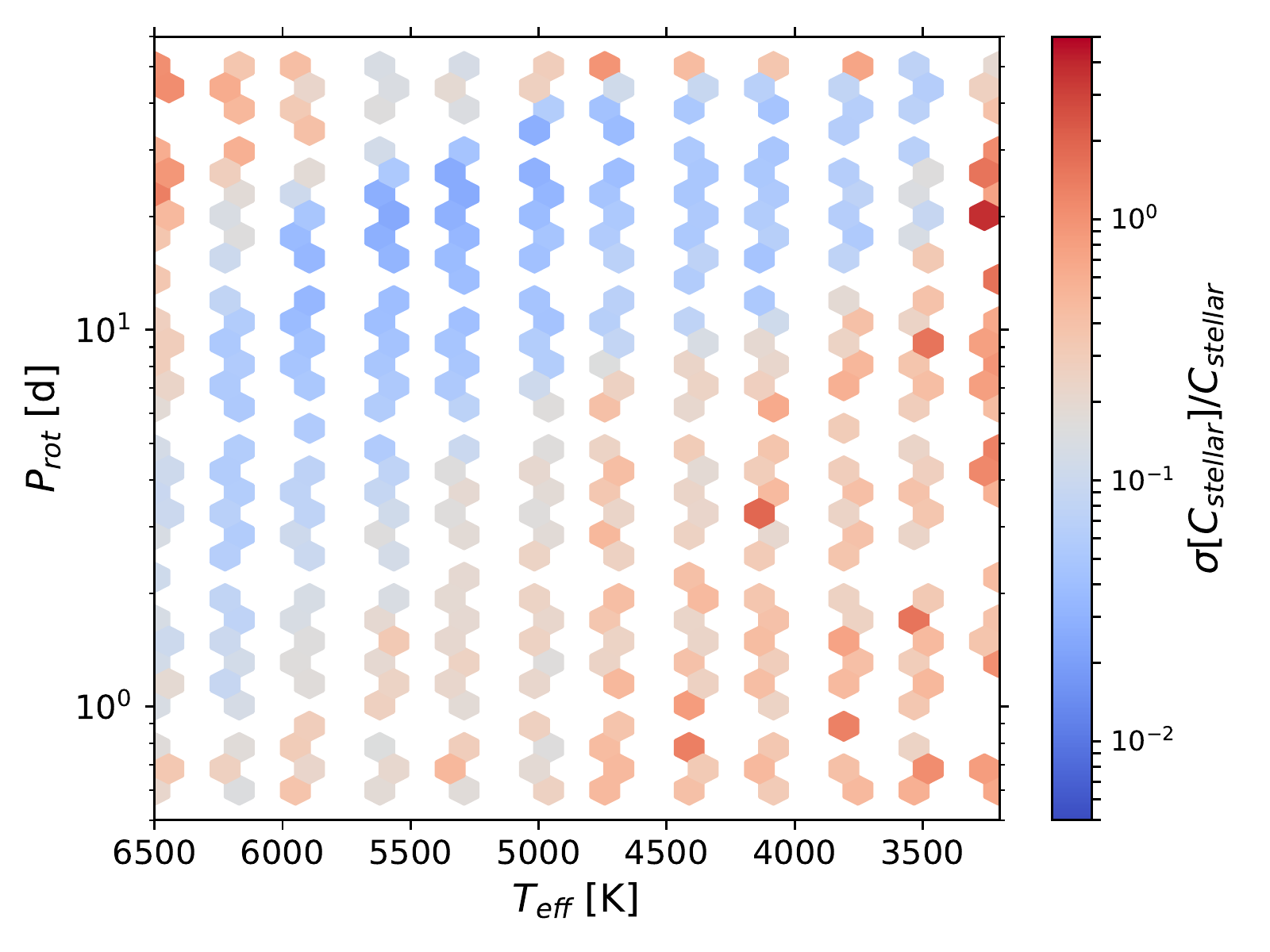}
   \end{center}
   \caption{\label{Cstel_sig} Values of \(\sigma[C_\text{stellar}]/C_\text{stellar}\) as a function of the stellar rotation period \(P_\text{rot}\) and the effective temperature \(T_\text{eff}\).}
\end{figure}  
As shown in Fig. \ref{Cstel_sig}, the relative uncertainty of $C_\text{stellar}$ is higher at low effective temperatures and low rotation periods, reaching values as high as 100 \%. However, it marginally affects synthetic populations, as few systems from the MMA14 sample are located in this region. In the regions favored by the MMA14 sample, $\sigma[C_\text{stellar}]/C_\text{stellar}$ is less than 1 \%. A more extensive study of those uncertainties is left for future work.

\subsection{Star--planet population}

Once the stellar parameters have been chosen and consistently biased, we include the planets by choosing 40 initial semi-major axes ranging between \(5 \times 10^{-3}\) and 0.2 AU, uniformly spaced in logarithm, and five planetary masses between 0.5 Earth masses and 5 Jupiter masses, uniformly spaced in logarithm. From these initial conditions, we can define three different star--planet populations by making them evolve through tidal effects only, or along with magnetic effects by considering a planet magnetic field equal to 1 and 10 G. We then compute the frequency of occurrence \(f_\text{ESPEM}(P_\text{orb},P_\text{rot})\) of star--planet systems at a given pair \((P_\text{orb},P_\text{rot})\). This frequency is then biased as in the case of isolated stars by relying on the \(C_\text{stellar}\) factor. This way, the stellar distribution is consistent with the MMA14 study. Furthermore, each planetary mass is weighted to reproduce the MMA13 distribution of planetary radii. This leads to the following biased frequency of occurrence, accounting for both the MMA13 and MMA14 distributions:
\begin{equation}\label{eqn:f_ESPEM}
f_\text{ESPEM}^\text{biased} =  f_\text{ESPEM}(P_\text{orb},P_\text{rot})C_\text{stellar} (T_\text{eff},P_\text{rot})f_\text{MMA13}(M_p).
\end{equation}
The associated distribution function \(\text{DF}_\text{ESPEM}\) is defined as
\begin{equation}
\text{d}n_\text{ESPEM}(P_\text{orb},P_\text{rot}) = \text{DF}_\text{ESPEM}(P_\text{orb},P_\text{rot}) \text{d}P_\text{orb}\text{d}P_\text{rot},
\end{equation}
where \(\text{d}n_\text{ESPEM}\) is the fraction of systems with an orbital period included within the interval \([P_\text{orb},P_\text{orb}+\text{d}P_\text{orb}]\) and a stellar rotation period in \([P_\text{rot},P_\text{rot}+\text{d}P_\text{rot}]\). As the rotation periods as well as the orbital periods are sampled, we calculate this distribution by considering the frequency of occurrence at each bin of surface \(\Delta P_\text{orb}\Delta P_\text{rot}\) as
\begin{equation}
\text{DF}_\text{ESPEM}(P_\text{rot},P_\text{orb}) = \frac{f_\text{ESPEM}^\text{biased}(P_\text{rot},P_\text{orb})}{\Delta P_\text{orb}\Delta P_\text{rot}}.
\end{equation}
The function obtained is then normalized such as
\begin{equation}
\iint \text{DF}_\text{ESPEM}(P_\text{orb},P_\text{rot})\text{d} P_\text{orb}\text{d} P_\text{rot} = 1-\mathcal{D},
\end{equation}
with \(\mathcal{D}\) the ratio of the planets from the initial population that have been engulfed. Such an approach makes it possible to compare synthetic populations with different values of \(B_p\), as the rate of planet engulfment may change because of the variable efficiency of the magnetic torque (see Table \ref{tab:Destruction}).

\begin{table}[!h]
\centering 
      \caption{\label{tab:Destruction} Engulfment ratio of the synthetic star--planet population for different planetary magnetic fields.}
      \begin{tabu}{cc}
            \hline
             \noalign{\smallskip}
            $B_p$ [G] & $\mathcal{D}$\\
            \noalign{\smallskip}
            \hline
            \noalign{\smallskip}
            0\tablefootmark{a} & 10.05 \%\\
            1 & 10.78 \%\\
            10 & 12.68 \%\\
            \noalign{\smallskip}
            \hline
         \end{tabu}
         \tablefoottext{a}{Tidal effects only.}
   \end{table}

However, in the case of planetary populations observed by the \textit{Kepler} mission we have no idea of the associated initial population. In particular, the rate of planets engulfed by their host star is unknown. To compare our synthetic populations with the observations, we only take into account the planets in our sample that have survived. To do so, we define a probability density of presence \(\text{DPP}_\text{ESPEM}(P_\text{orb},P_\text{rot})\), obtained by normalizing the distribution function \(\text{DF}_\text{ESPEM}(P_\text{orb},P_\text{rot})\) so that
\begin{equation}
\iint \text{DPP}_\text{ESPEM}(P_\text{orb},P_\text{rot})\text{d} P_\text{orb} \text{d} P_\text{rot} = 1.
\end{equation}

\begin{figure}[!h]
   \begin{center}
    \includegraphics[scale=0.5]{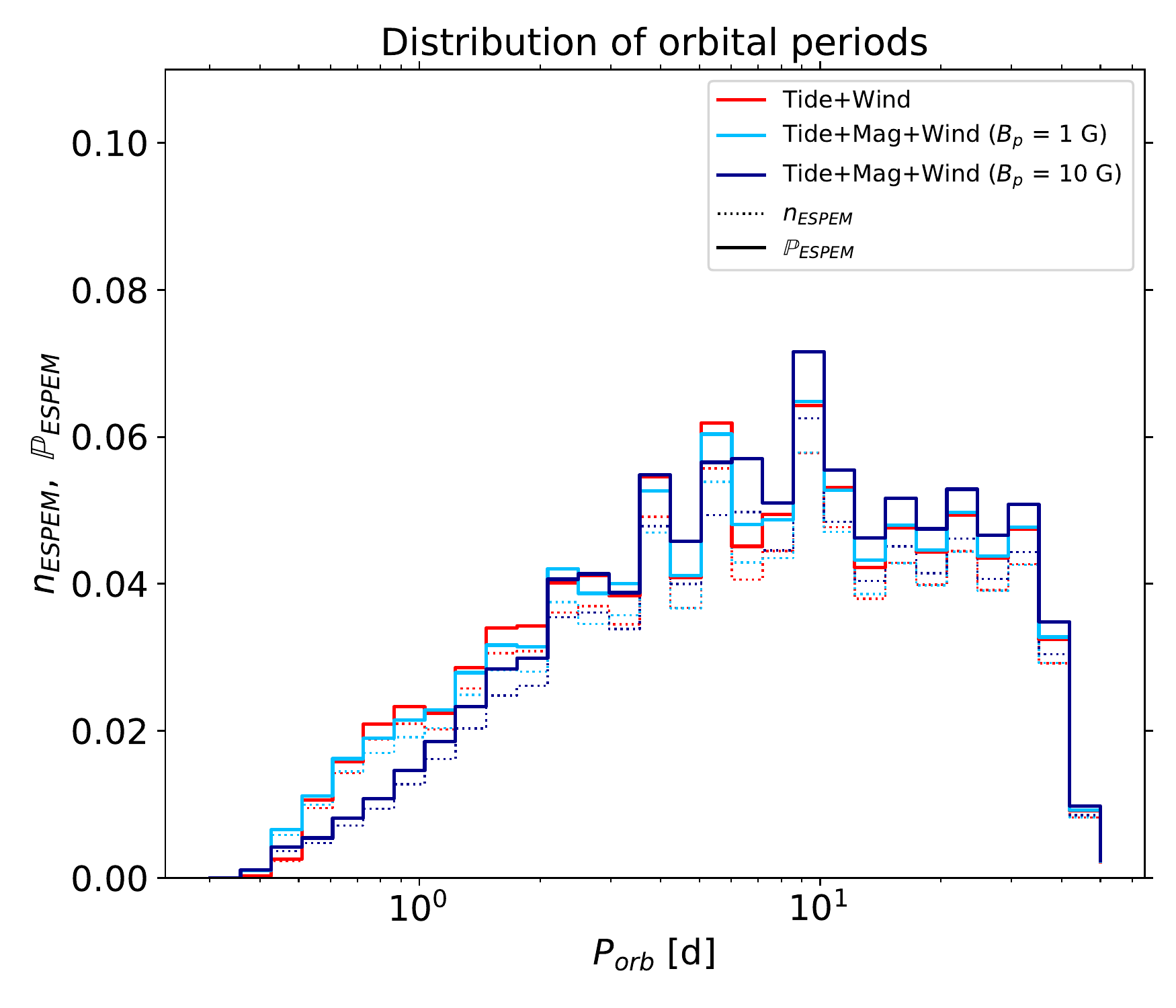}
   \end{center}
   \caption{\label{DFDPP} Probability of presence \(\mathbb{P}_\text{ESPEM}\) (solid lines) and fraction \(n_\text{ESPEM}\) of systems present in the synthetic population (dashed lines) as a function of orbital period. In red: ESPEM distributions with \(B_p = 0\) G. In light blue: ESPEM distributions with \(B_p = 1\) G. In dark blue: ESPEM distributions with \(B_p = 10\) G.}
\end{figure}  
Thus, as shown in Fig. \ref{DFDPP}, only the relative shape of the distributions is considered with such a normalization. More precisely, a higher concentration of planets at a given orbital period induces higher probabilities of presence, regardless of the number of planets destroyed in the sample. The probability of presence \(\mathbb{P}_\text{ESPEM}\) (solid lines in Fig. \ref{DFDPP}) is thus higher than the fraction \(n_\text{ESPEM}\) (dashed lines in Fig. \ref{DFDPP}) of systems present in the sample. This is particularly true in case \(B_p = 10\) G, because the associated planet destruction rate is the highest.

To assess the influence of uncertainties of the MMA14 sample on these distributions, we consider for the sake of simplicity the relative uncertainty of $C_\text{stellar}$ averaged in effective temperatures. Furthermore, to decorrelate the contributions from the MMA13 and MMA14 samples, we do not consider any uncertainty in planetary radii, as the same distribution is used in both observed and synthetic populations. Hence, from Eq. \eqref{eqn:f_ESPEM}, the standard deviation $\sigma\left[\text{DPP}_\text{ESPEM}\right]$ linked to the density of probability of presence can be assessed as
\begin{equation}
\frac{\sigma[\text{DPP}_\text{ESPEM}]}{\text{DPP}_\text{ESPEM}(P_\text{orb},P_\text{rot})} = \left(\frac{\sigma[C_\text{stellar}]}{C_\text{stellar}}\right) (P_\text{rot}).
\end{equation}
We then rely on this prescription to build the error bars associated to the distributions presented in \S 5.2 and \S 5.3.

\section{Influence of instantaneous stellar rotation}

\begin{figure}[!h]
   \begin{center}
   \includegraphics[scale=0.53]{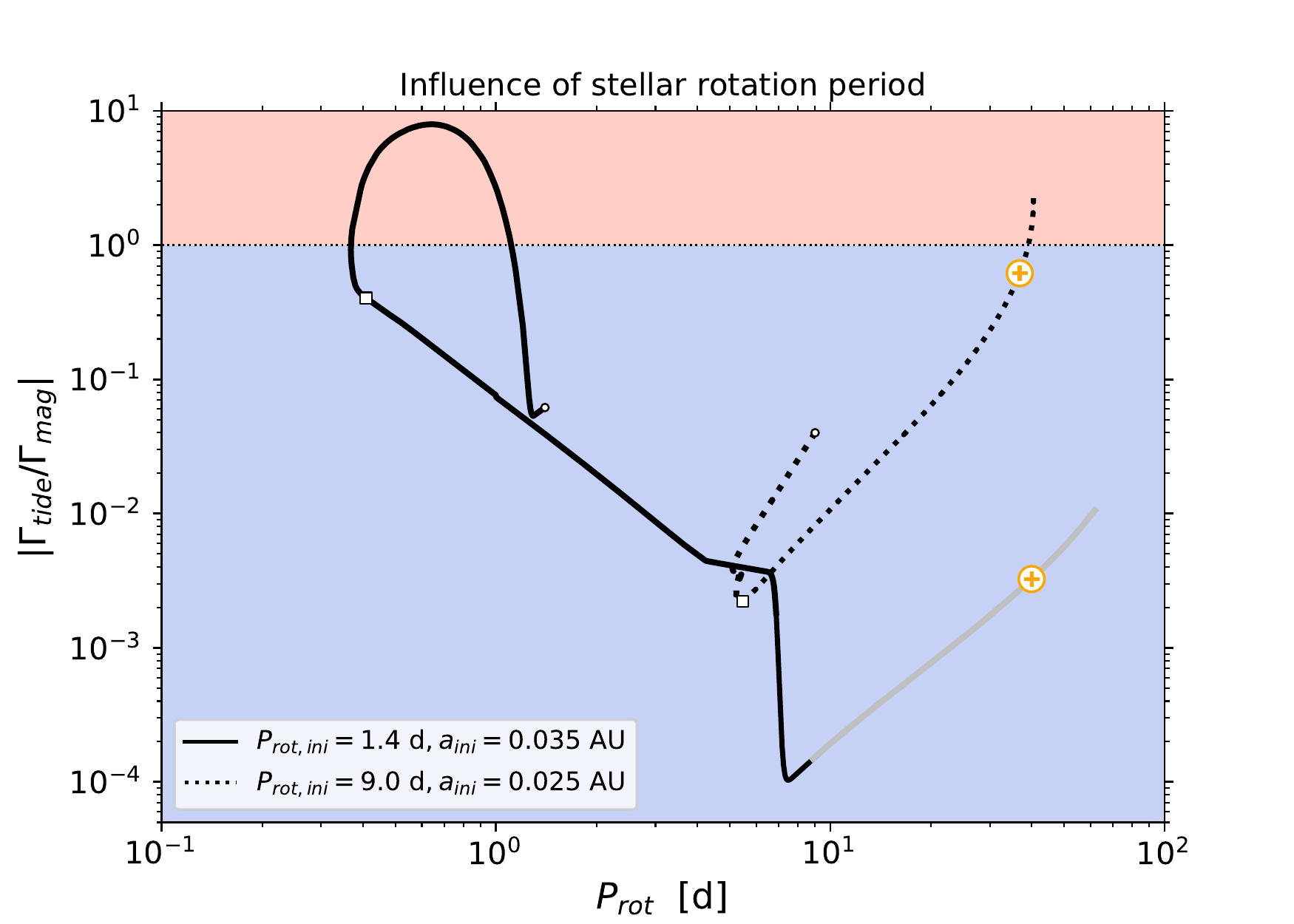}
   \end{center}
   \caption{\label{Prot} Ratio between the tidal and magnetic torques as a function of the stellar rotation period for our reference case (solid line) and for a similar star–planet system with \(P_\text{rot,ini} = 9\ \text{d} \text{ and } a_\text{ini} = 0.025\text{ AU}\) (dotted line). In gray: regions where the overall migration timescale is greater than the age of the universe. Blue background: magnetic dominance. Red background: tidal dominance. The white circles correspond to the beginning of the evolution. The white squares correspond to the ZAMS. The orange circles correspond to solar age.}
\end{figure}

To highlight the role of instantaneous stellar rotation on the fate of the system, we consider the ratio between the tidal and magnetic torques as a function of the stellar rotation period for the reference case presented in \S 3.1 and a similar star–planet system with an initial stellar rotation period \(P_\text{rot,ini} = 9\ \text{d}\) and an initial semi-major axis \(a_\text{ini} = 0.025\text{ AU}\). In both cases the planet is able to sustain a magnetosphere. As seen in Fig. \ref{Prot}, when inertial waves are excited in the stellar envelope, the associated torque tends to dominate for lower rotation periods. Thus, the dynamical tide generally dominates the evolution during most of the PMS. If the inertial waves cannot be excited by the tidal potential, only the equilibrium tide contributes to the evolution of the system. In this case the ratio between the tidal and magnetic torques increases for higher rotation periods, as the tidal torque is less sensitive to stellar rotation than the magnetic torque. Hence, at the end of the MS (see the highest rotation periods in Fig. \ref{Prot}) the tidal torque can be of the same order of magnitude as the magnetic torque or even dominate the latter, depending on the initial configuration of the system.\\
\newpage
\section{Results of the Kolmogorov-Smirnov test}
\begin{figure}[!h]
   \begin{center}
    \includegraphics[scale=0.35]{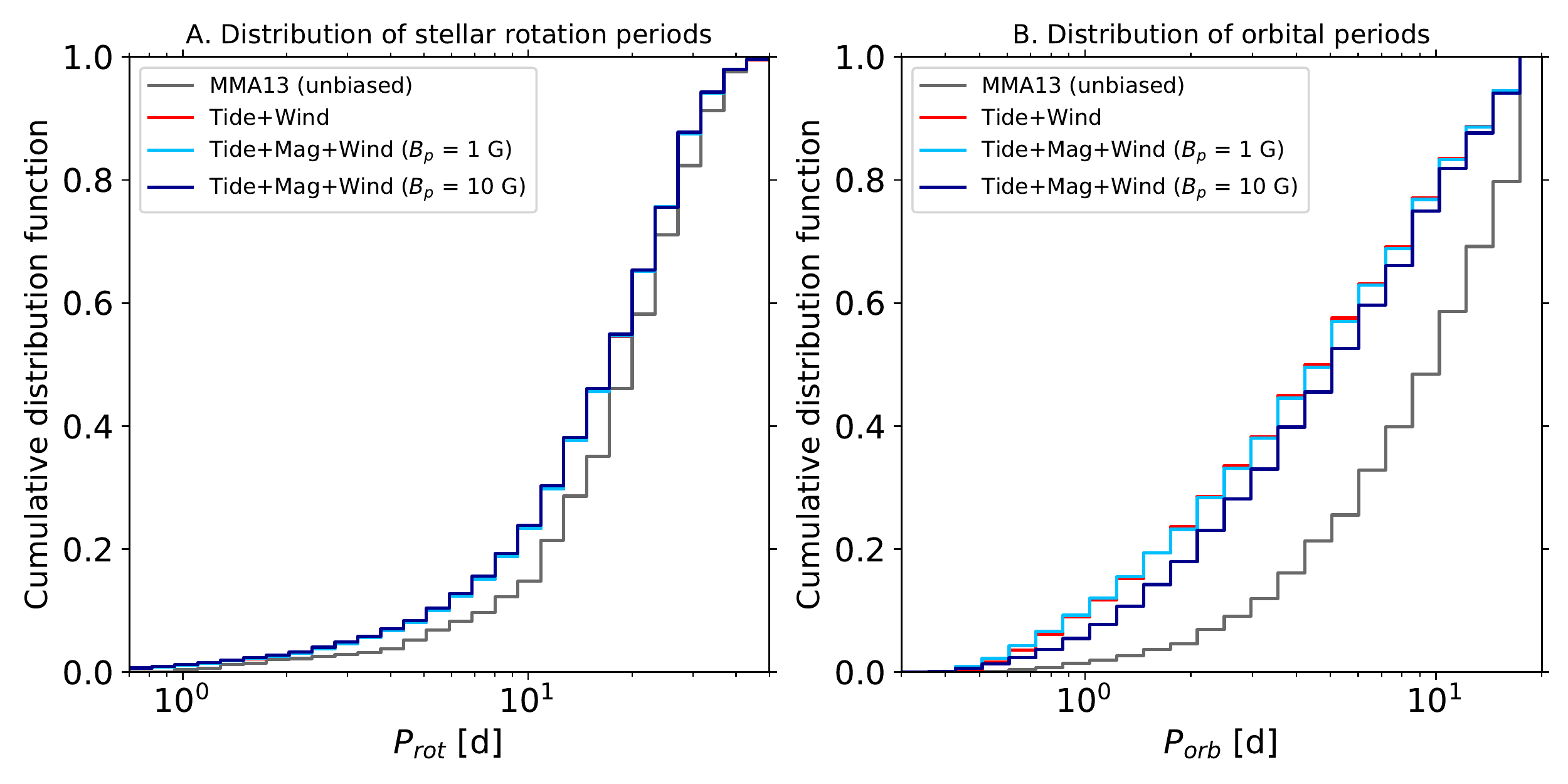}
   \end{center}
   \caption{\label{globalCDF} Cumulative distributions of stellar rotation periods (left) and orbitals periods (right). In gray: unbiased MMA13 sample. In red: ESPEM distributions with \(B_p = 0\) G. In light blue: ESPEM distributions with \(B_p = 1\) G. In dark blue: ESPEM distributions with \(B_p = 10\) G.}
\end{figure}

\begin{figure*}[!h]
   \begin{center}
    \includegraphics[scale=0.38]{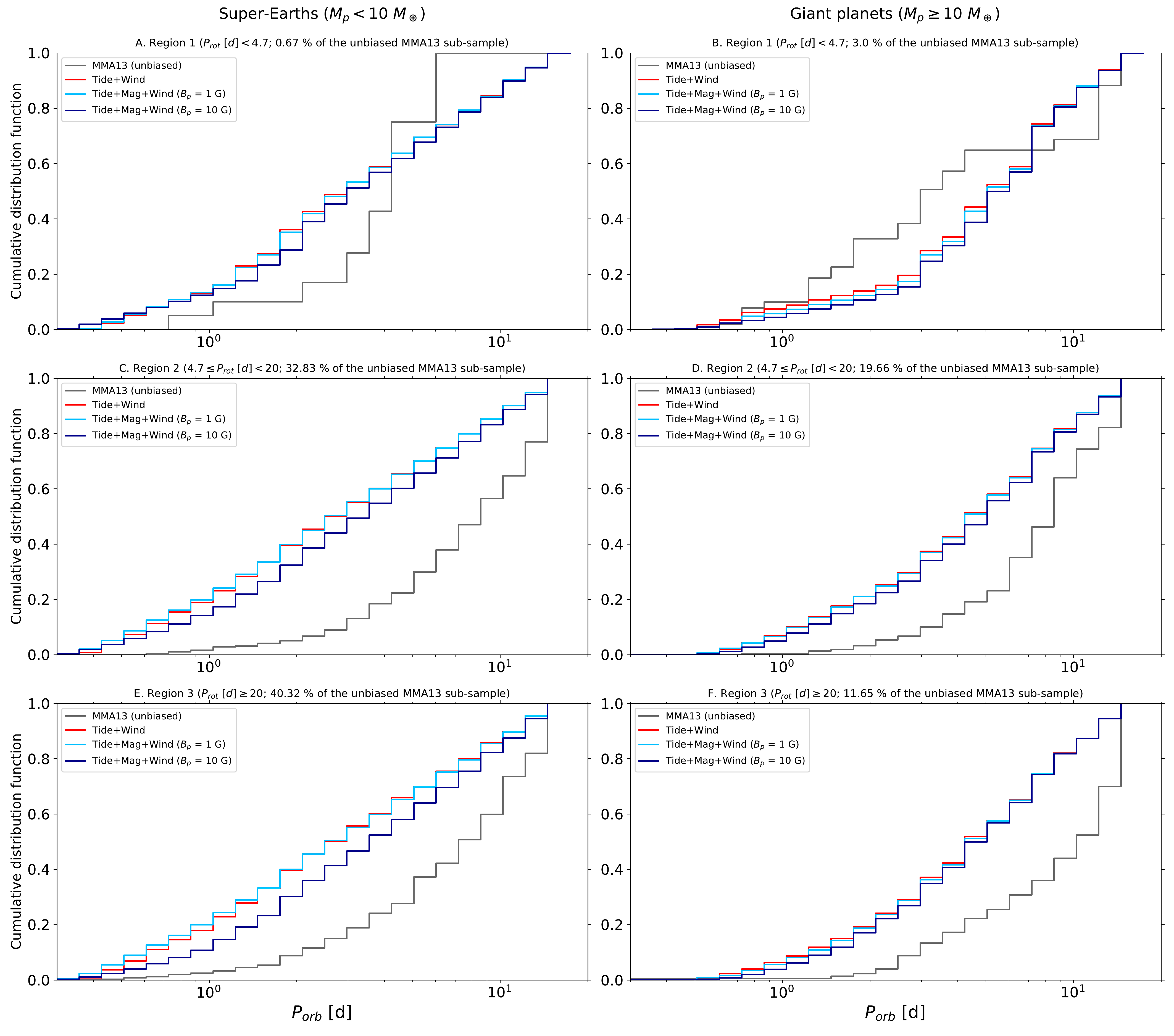}
   \end{center}
   \caption{\label{localCDF} Cumulative distributions of orbital periods for super-Earths (left column) and giant planets (right column). Each row corresponds to a population of young (Region 1, for which \(P_\text{rot} < 4.7\) d), middle-aged (Region 2, for which \(4.7 \leq P_\text{rot}\ [\text{d}] < 20\)), and old (Region 3, \(P_\text{rot} \geq 20\) d) star-planet systems. In gray:  unbiased MMA13 sample. In red: ESPEM distributions with \(B_p = 0\) G. In light blue: ESPEM distributions with \(B_p = 1\) G. In dark blue: ESPEM distributions with \(B_p = 10\) G.}
\end{figure*}

We perform a two-sample Kolmogorov-Smirnov test to assess whether the observed and synthetic samples show the same underlying probability distribution in orbital and stellar rotation periods, which constitutes the null hypothesis. To perform the KS test, the probability distributions are first normalized to 1. We then compute the KS statistic $\mathcal{D}_\text{KS}$, defined as
\begin{equation}
\mathcal{D}_\text{KS} = \text{sup}_{P_\text{orb}}\left|F_\text{ESPEM} (P_\text{orb})-F_\text{MMA13,unbiased} (P_\text{orb})\right|,
\end{equation}
where $F$ corresponds to the cumulative distribution function. We then rely on the Kolmogorov distribution $K$ to estimate the $p$-value as follows:
\begin{equation}
\mathbb{P}\left(K \geq \sqrt{\frac{N}{2}}\mathcal{D}_\text{KS}\right) = 2\sum_{k=1}^{+\infty}(-1)^{k-1} e^{-2k^2\left(\sqrt{\frac{N}{2}}\mathcal{D}_\text{KS}\right)^2},
\end{equation}
where $N$ is the size of both observed and synthetic samples. We present the global cumulative distribution functions in $P_\text{orb}$ and $P_\text{rot}$, for both the observed and synthetic samples, in Fig. \ref{globalCDF}. The corresponding KS statistics and $p$-values, mentioned in \S 5.2, are presented in Table \ref{tab:KS_global}.
We also present the cumulative distribution functions for super-Earths and giant planets, for both the observed and synthetic samples, in Fig. \ref{localCDF}. The corresponding KS statistics and $p$-values, mentioned in \S 5.3, are presented in Table \ref{tab:KS_local}. 

\begin{table}[!h]
\centering 
      \caption{\label{tab:KS_global} Kolmogorov-Smirnov test statistic $\mathcal{D}_\text{KS}$ and $p$-values for the synthetic and observed global distributions.}
      \begin{tabu}{ccc}
            \hline
             \noalign{\smallskip}
            $B_p$ [G] & $\mathcal{D}_\text{KS}$, $p$-value\\
            \noalign{\smallskip}
            \hline
            \noalign{\smallskip}
            \textit{$P_\text{rot}$ distribution.}\\
            0\textsuperscript{*} & 0.11, 0.99\\
            1 & 0.11, 0.99\\
            10 & 0.11, 0.99\\
            \\
            \hline
            \textit{$P_\text{orb}$ distribution.}\\
            0\textsuperscript{*} & 0.32,0.17\\
            1 & 0.31,0.19\\
            10 & 0.27,0.34\\
            \noalign{\smallskip}
            \hline
         \end{tabu}
         \textsuperscript{*}\footnotesize{Tidal effects only}
   \end{table}
   
\begin{table}[!h]
\centering 
      \caption{\label{tab:KS_local} Kolmogorov-Smirnov test statistic $\mathcal{D}_\text{KS}$ \text{and $p$-values} for the synthetic and observed sub-populations.}
      \begin{tabu}{ccc}
            \hline
             \noalign{\smallskip}
            $B_p$ [G] & Super-Earths& Giant planets\\
            \noalign{\smallskip}
            \hline
            \noalign{\smallskip}
            \textit{Region 1.}& $\mathcal{D}_\text{KS}$, $p$-value & $\mathcal{D}_\text{KS}$, $p$-value \\
            \\
            0\textsuperscript{*} & 0.32, 0.18 & 0.24, 0.50\\
            1 & 0.31, 0.20 & 0.25, 0.42 \\
            10 & 0.28, 0.29 & 0.27, 0.35\\
            \\
            \hline
            \textit{Region 2.}& \\
            \\
            0\textsuperscript{*} & 0.43, 0.023 & 0.35, 0.11\\
            1 & 0.43, 0.023 & 0.35, 0.11\\
            10 & 0.38, 0.064 & 0.33, 0.16\\
            \\
            \hline
            \textit{Region 3.}& \\
            \\
            0\textsuperscript{*} & 0.38,0.060 & 0.39, 0.054 \\
            1 & 0.38, 0.068 & 0.39, 0.056\\
            10 & 0.30, 0.22 & 0.38, 0.058\\
            \noalign{\smallskip}
            \hline
         \end{tabu}
         \textsuperscript{*}\footnotesize{Tidal effects only}
   \end{table}
\end{appendix}
\end{document}